\documentclass[a4paper,11pt]{article}

\usepackage[normalem]{ulem}
\usepackage{jheppub} 

\usepackage{bbm}
\usepackage{graphicx}  
\usepackage{dcolumn}   
\usepackage{bm}        
\usepackage{amssymb}   
\usepackage{braket}
\usepackage{amsmath}
\usepackage{color}  	
\usepackage{slashed}
\usepackage{subfigure}
\usepackage[utf8]{inputenc}
\DeclareGraphicsExtensions{.pdf,.png,.jpg,.eps}
\graphicspath{ {./images/} }

\newcommand{\Tr}{\mathrm{Tr}}
\newcommand{\Pexp}{\mathrm{Pexp}}
\newcommand{\td}{\mathrm{d}}

\newcommand{\TT}[1]{\mathrm{#1}}

\newcommand{\pdf}[2]{\dfrac{\partial #1}{\partial #2}}

\newcommand{\tdf}[2]{\dfrac{\td #1}{\td #2}}

\renewcommand{\v}[1]{\mathbf{#1}}

\renewcommand{\>}{\right\rangle}
\newcommand{\<}{\left\langle}
\newcommand{\lkl}{\left|}
\newcommand{\rkl}{\right|}

\newcommand{\qqquad}{\qquad\qquad}
\newcommand{\qqqquad}{\qquad\qquad\qquad}
\newcommand{\qqqqquad}{\qquad\qquad\qquad\qquad}
\newcommand{\Ldg}{\textrm{Leading}}
\newcommand{\Nc}{N_{\mathrm{c}}}
\newcommand{\As}{\alpha_{\mathrm{s}}}
\usepackage{breakurl} 

\hypersetup{urlcolor=black}

\title{Building a consistent parton shower}

\preprint{\begin{flushright}MAN/HEP/2020/002 \\
		UWTHPH-2020-8\end{flushright}
}    
	
\author[a,b]{Jeffrey R. Forshaw,}
\author[a,b]{Jack Holguin,} 
\author[b,c]{Simon Pl\"{a}tzer.}

\affiliation[a]{Consortium for Fundamental Physics, School of Physics \& Astronomy, \\
	University of Manchester, Manchester M13 9PL, United Kingdom}
\emailAdd{jeffrey.forshaw@manchester.ac.uk}
\emailAdd{jack.holguin@manchester.ac.uk}
\emailAdd{simon.plaetzer@univie.ac.at}
\affiliation[b]{Erwin Schr\"{o}dinger Int. Institute for Mathematics and Physics, \\ University of Vienna, 1090 Wien, Austria}
\affiliation[c]{Particle Physics, Faculty of Physics, \\
	University of Vienna, 1090 Wien, Austria}
\date{\today}

\abstract{Modern parton showers are built using one of two models:
  dipole showers or angular ordered showers. Both have distinct
  strengths and weaknesses. Dipole showers correctly account for
  wide-angle, soft gluon emissions and track the leading flows in QCD
  colour charge but they are known to mishandle partonic recoil. Angular
  ordered showers keep better track of partonic recoil and correctly
  include large amounts of wide-angle, soft physics but azimuthal
  averaging means they are known to mishandle some correlations. In this paper, we derive both approaches from the same starting point; linking our understanding
  of the two showers.  This insight allows
  us to construct a new dipole shower that has all the strengths of a
  standard dipole shower together with the collinear evolution of an angular-ordered shower. We show that this new approach corrects the
  next-to-leading-log errors previously observed in parton showers and
  improves their sub-leading-colour accuracy.}

\begin{document} 

\maketitle
\flushbottom

\section{Introduction}
\label{sec:Intro}

Parton showers simulate the particle content of scattering events at collider experiments and provide the backbone to modern experimental analyses \cite{Pythia,Pythia8,Herwig_dipole_shower,DIRE,Herwig_shower,Gleisberg:2008ta,Giele:2007di}.  Yet questions over their accuracy and on how best to improve them remain. 
In this paper we present a unified analysis of the two main approaches to formulating parton showers: dipole showers \cite{Lonnblad:1992tz,Pythia8,Herwig_dipole_shower,DIRE} and angular ordered showers \cite{Marchesini:1983bm,Herwig_shower,Gleisberg:2008ta}. As a result, we are able to construct a new dipole shower that does not suffer from the next-to-leading logarithm (NLL) problems suffered by existing parton showers and has increased next-to-leading colour (NLC) accuracy \cite{Dasgupta:2018nvj}.

In our previous papers \cite{SoftEvolutionAlgorithm,Forshaw:2019ver} we introduced an algorithm for amplitude-level parton branching (the PB algorithm). The PB algorithm was designed to capture both the soft and collinear logarithms associated with the leading infra-red singularities of scattering amplitudes without making any approximations on the spin and colour. In \cite{Forshaw:2019ver} we showed how the PB algorithm can be used to derive the resummation of observables at leading-logarithmic accuracy (it has the capacity to be extended to include next-to-leading-logarithms) and we showed that it gives rise to the collinear factorisation of parton density and fragmentation functions. In \cite{SoftEvolutionAlgorithm} we showed that the colour evolution is equivalent to that of other approaches \cite{Becher:2016mmh,Caron-Huot:2015bja,Caron-Huot:2016tzz,BMSEquation}. The PB algorithm is the starting point for the analysis presented here.

In the next section, we present a brief overview of the algorithm before going on to use it to derive both dipole and angular ordered showers. In these derivations we keep close track of the approximations made, with the goal of gaining a solid understanding of the sources for errors in these showers. We focus on deriving showers in $e^+e^-$, though much of the machinery necessary to derive showers for hadron-hadron processes is also present in this paper. The full discussion of our derivations is technical and largely handled in Appendix \ref{sec:Supp}. 

More specifically, in Section \ref{sec:CBEvo}, we derive an angular ordered shower starting from the PB algorithm. In doing so we are able to constrain the recoil functions in the original PB algorithm, since angular ordered showers provide clear constraints on how momentum longitudinal to a jet must be conserved in order to get NLL physics correct.  In Section \ref{sec:DipoleEvo} we then derive a dipole shower from the PB algorithm, taking particular care over the constraints observed from our angular ordered derivation. The result is a dipole shower that reduces the doubly-logarithmic NLC errors noted in \cite{Dasgupta:2018nvj} (complete removal of NLC errors at a given logarithmic accuracy generally requires amplitude-level evolution). Having pinned down longitudinal recoil, in Section \ref{sec:Correcting} we present a scheme (inspired by \cite{Bewick:2019rbu}) for the transverse recoil. This completes the specification of our shower. We then go on to recreate the fixed order analysis of \cite{Dasgupta:2018nvj} and show that our shower corrects the NLL errors from incorrect transverse recoil previously observed in dipole showers. In Appendix \ref{sec:Checks} we go further and show that our new shower is sufficient for the correct leading-colour NLL resummations of thrust and the generating functions for jet multiplicity.

\section{Evolution equations}
\label{sec:Evo}

\subsection{Amplitude evolution overview}
\label{sec:AmpEvo}

\begin{figure}[b]
	\centering
	\includegraphics[trim=0 80 20 0,clip,width=0.95\textwidth]{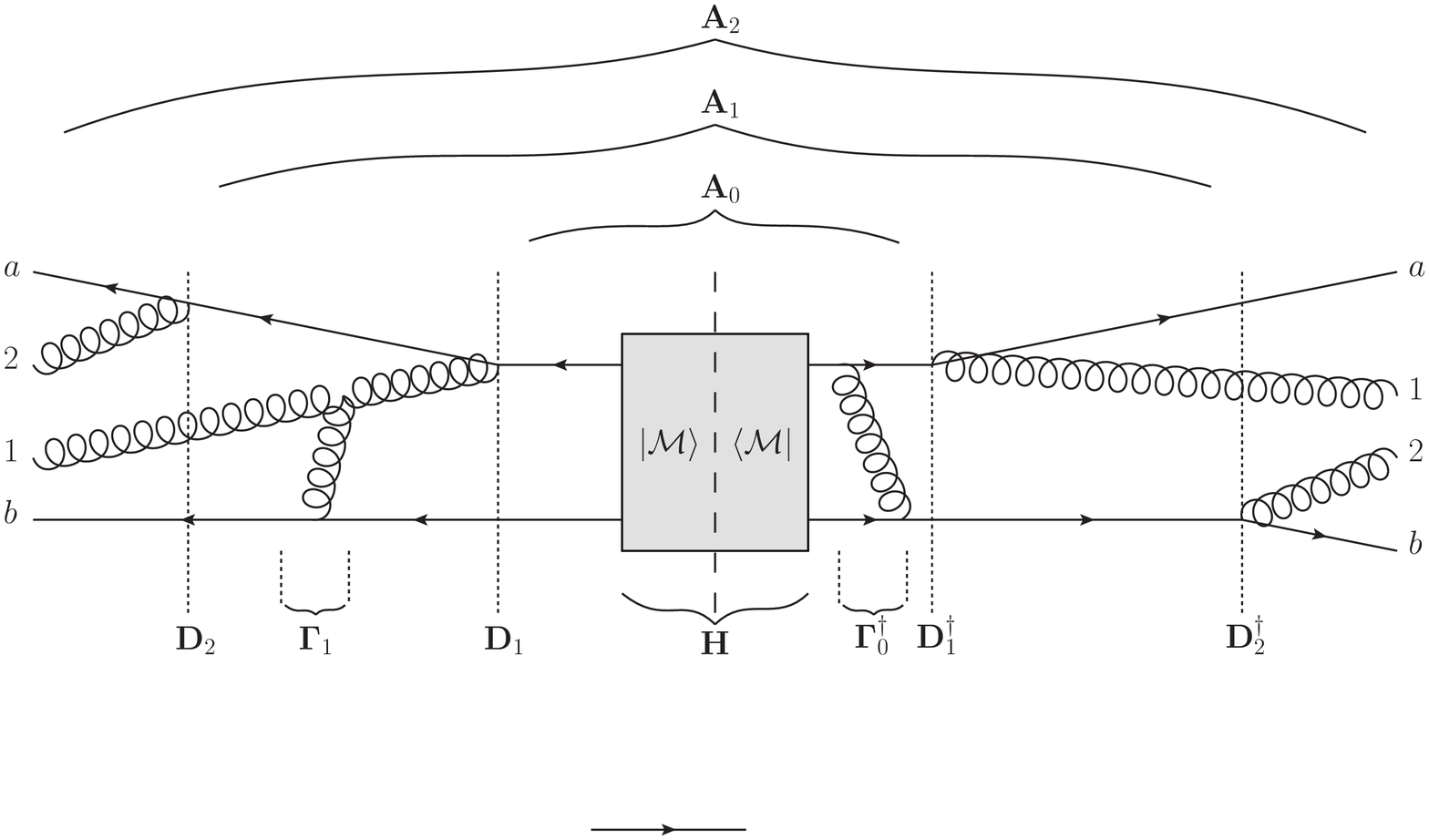}
	\caption{A general term in the Markov chain of amplitude density matrices, $\v{A}_{n}$, constructed by the PB algorithm. $\v{H} \equiv \lkl \mathcal{M} \> \< \mathcal{M} \rkl$ is the initial hard process; in this case it has two hard coloured legs, $a$ and $b$. $\v{D}_{n}$ dresses an amplitude with the $n$th emission that is either soft or collinear. Collinear emissions are emitted symmetrically from the amplitude and conjugate amplitude, such as gluon $1$. Soft emissions appear as interference terms, such as gluon $2$. $\v{\Gamma}_{n}$ dresses the amplitude after $n$ soft or collinear emissions with a loop. }
	\label{fig:Evolution}
\end{figure}

The PB algorithm defines a sequence of transitions in a Markov chain
of amplitude density matrices: $\v{A}_{0}(q_{0\, \bot}; \{p\}_{0})
\mapsto \v{A}_{1}(q_{1\, \bot}; \{p\}_{1}) \mapsto ... \mapsto
\v{A}_{n}(q_{n\, \bot}; \{p\}_{n})$. The sequence is illustrated
in Figure \ref{fig:Evolution}. We use $n$ to index the number of
partons dressing the hard process. Each amplitude is defined at a
given scale (parametrised by an ordering variable), this is its first
argument. The second argument, after a semi-colon,
specifies its full dependence on the relevant parton momenta (which we often choose to omit). The
Markov chain uses the initial condition $\v{A}_{0}(Q; \{p\}_{0}) =
\v{H}(Q; P_{1},...,P_{n_{\TT{H}}})$, where $\v{H}\equiv \lkl
\mathcal{M} \> \< \mathcal{M} \rkl$ is the hard process density matrix
for a process of hard scale $Q$ and with $n_\text{H}$ hard
partons. The hard partons' momenta form the set
$\{P_{1},...,P_{n_{\TT{H}}}\} \equiv \{p\}_{0}$. The Markov chain
terminates on the amplitudes $\v{A}_{n}(\mu; \{p\}_{n})$; $\mu$
is an infra-red cut-off and $\{p\}_{n} = \{P_{1},...,P_{n_{\TT{H}}},
q_{1},...,q_{n}\}$ where $q_{1},...,q_{n}$ are the momenta of the $n$
partons that dress the hard process. Steps in the Markov chain are
constructed from the action of two operators, $\v{D}_{n}$ and
$\v{\Gamma}_{n}$. The $\v{D}_{n}$ operators are emission operators;
they act as maps from a state $\v{A}_{n-1}(q_{\bot}; \{p\}_{n-1})$ to
a state $\v{A}_{n}(q_{\bot}; \{p\}_{n})$, and they describe the
emission of the $n$th parton. Operators $\v{\Gamma}_{n}$ provide a map
from a state $\v{A}_{n}(q_{\bot}; \{p\}_{n})$ onto some other
$\tilde{\v{A}}_{n}(q_{\bot}; \{p\}_{n})$. Physically, they dress the
density operator with (iterated) virtual corrections. The
path-ordered exponent of $\v{\Gamma}_{n}$ is an amplitude level
Sudakov factor/operator which we call $\v{V}_{a,b}$:
\begin{align}
\v{V}_{a,b} = \Pexp \left(- \int^{b}_{a}\frac{\td q_{\bot}}{q_{\bot}} \, \v{\Gamma}_{n}(q_{\bot})\right).
\end{align}
$\v{V}_{a,b}$ evolves a state $\v{A}_{n}(b; \{p\}_{n})$ to a state at a lower scale $\tilde{\v{A}}_{n}(a; \{p\}_{n})$; for a complete discussion of $\v{V}_{a,b}$ see \cite{Forshaw:2019ver}. In \cite{Forshaw:2019ver} we presented the PB algorithm in the following form:
\begin{equation}
\v{A}_{n}(q_{\bot}; \{p\}_{n}) = \int \td R_{n} \v{V}_{q_{\bot},q_{n \, \bot}} \v{D}_{n} \v{A}_{n-1}(q_{n \, \bot}; \{p\}_{n-1}) \v{D}^{\dagger}_{n} \v{V}^{\dagger}_{q_{\bot},q_{n \, \bot}} \Theta(q_{\bot} \leq q_{n \, \bot}). \label{eq:Aevo}
\end{equation}
The algorithm maps the set of partonic momenta prior to the $n$th emission ($\{ p_{n-1} \}$) onto a new set ($\{ p_{n} \}$), by adding a parton ($q_n$). In order to conserve energy-momentum, the set of momenta prior to the emission are adjusted after each emission, i.e. $\{ p_{n-1} \} \to  \{ \tilde{p}_{n-1} \}$ and $\{ p_{n} \} = \{ \tilde{p}_{n-1} \cup q_n \}$.  We achieve this by integrating over delta functions relating the two sets of momenta. This is all hidden inside $\int \td R_{n}$, which we describe in Appendix \ref{sec:AevoDetails} and give examples of in Section \ref{sec:Correcting}. We also provide definitions of each operator involved in the evolution in Appendix \ref{sec:AevoDetails}.

In this paper, it better suits our purposes to work with the PB
algorithm expressed as an evolution equation, i.e. working
differentially in the ordering variable, $q_{\bot}$. Broadly speaking,
$q_{n \, \bot}$ is the transverse momentum of the $n$th parton and it is a function of the $n$-parton phase-space. The precise definition of $q_{n \, \bot}$ is context dependent and is
given in Appendix \ref{sec:AevoDetails}.
The evolution equation is
\begin{align}
q_{\bot}\pdf{\v{A}_{n}(q_{\bot}; \{p\}_{n})}{q_{\bot}} = &- \v{\Gamma}_{n}(q_{\bot}) \, \v{A}_{n}(q_{\bot}; \{p\}_{n}) \nonumber -  \v{A}_{n}(q_{\bot}; \{p\}_{n}) \, \v{\Gamma}^{\dagger}_{n}(q_{\bot}) \\ & + \int \td R_{n} \; \v{D}_{n}(q_{n \, \bot}) \, \v{A}_{n-1} (q_{n \, \bot}; \{p\}_{n-1}) \, \v{D}^{\dagger}_{n}(q_{n \, \bot}) \; q_{\bot} \; \delta(q_{\bot} - q_{n \, \bot}). \label{eqn:evo}
\end{align}
It is from this equation that we will derive generalised dipole and angular ordered showers. 

The phase-space measure for the $n$th parton emitted in the cascade is variously written as
\begin{equation}
\frac{\td^{3} q_{n}}{2E_{q_{n}}} =  \frac{q^{2}_{n \, \bot} \td q_{n \, \bot}}{2q_{n \, \bot}} \, \td S^{(q_{n})}_{2} = \frac{\pi^{2}q^{2}_{n \, \bot}}{2\As} \td \Pi_{n}.
\end{equation}
We typically parametrise the evolution so that real emissions use the phase-space measure $\td \Pi_{n}$ and loops $\td \ln q_{n \, \bot}\td S^{(q_{n})}_{2}$. From each $\v{A}_{n}$ we can compute the differential $n_{\TT{H}}+n$ parton cross section:
\begin{align}
\td \sigma_{n}(\mu) = & \left(\prod^{n}_{i = 1} \td \Pi_{i} \right) \Tr \, \v{A}_{n}(\mu),
\end{align}
where $\mu$ is either an infra-red regulator that should be taken to zero or the shower cut-off scale. We will focus on $e^+e^-$ hard matrix elements, in which case observables are computed using
\begin{align}
\Sigma(\mu; \{p\}_{0}, \{v\}) = & \int \sum_{n} \td \sigma_{n}(\mu) \, u (\{p\}_{n}, \{v\}), \label{eqn:cross-sectionee}
\end{align}
where $u (\{p\}_{n}, \{v\})$ is a measurement function for an observable defined by the set of parameters $\{v\}$.\footnote{$\Sigma(\mu; \{p\}_{0}, \{v\})$ is $\sum_{\delta} \tdf{\sigma_{\delta}}{\mathcal{B}} f_{\mathcal{B}, \delta}(v)$ in \cite{Banfi:2004yd}.} The formula for processes involving incoming hadrons is given in Appendix \ref{sec:hadrons}.

\subsection{Angular ordered shower}
\label{sec:CBEvo}

\begin{figure}[b]
	\centering
	\includegraphics[width=0.33\textwidth]{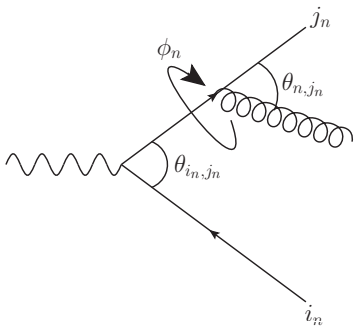}
	\caption{The angles used to derive angular ordering by azimuthal averaging. $\phi_{n}$ is the azimuth that is averaged over. In some equations two azimuths are present, in these situations we give $\phi_{n}$ a second index, e.g. $\phi_{n, j_{n}}$. Angular ordering corresponds to $\theta_{i_{n},j_{n}} > \theta_{n,j_{n}}$.}
	\label{fig:angles}
\end{figure}

In this section we give an overview of the derivation of an angular
ordered shower, starting from Eq.~\eqref{eqn:evo}. The unabridged
derivation is given in Appendix \ref{sec:CBderivation}. Angular
ordering is derived after averaging over the azimuth of each emitted
parton, as measured relative to their parent parton (and neglecting
all subsequent azimuthal correlations). After performing this averaging in
Eq.~\eqref{eqn:evo}, the colour structures can be greatly simplified
(a manifestation of QCD coherence). We exploit this to re-write the
evolution in terms of squared matrix elements,
$|\mathcal{M}_{n}|^{2}$. What follows is a little more detail of the
key steps.
\begin{enumerate}
	\item The  $\v{D}_{n}$ operators in Eq.~\eqref{eqn:evo} describe the emission of soft gluons from dipoles (via eikonal currents) and the emission of hard-collinear partons. The probability for the emission of a soft gluon is partitioned as $$	\frac{n_{i_{n}} \cdot n_{j_{n}}}{n_{i_{n}}\cdot n \; n_{j_{n}} \cdot n} =  P_{i_nj_n} + P_{j_ni_n}, \quad\TT{where} \quad	2 P_{i_nj_n} = \frac{n_{i_{n}} \cdot n_{j_{n}}-n_{i_{n}}\cdot n}{n_{i_{n}}\cdot n \; n_{j_{n}} \cdot n} + \frac{1}{n_{i_{n}}\cdot n},$$ $n_{i_{n}} = p_{i_{n}}/E_{i_{n}}$ and $n = q_{n}/E_{q_{n}}$, and $E$ is an energy in the event zero-momentum frame. Note that $P_{i_nj_n}$ only has a pole when the emission is parallel to $i_n$. When integrated, this term gives rise to a theta function that enforces angular ordering.
	\item We average over the emitted parton's azimuth, $\<...\>_{1,...,n}$, such that (for some quantity $f$) $$ \<f\>_{1,...,n} = \int \frac{\td \phi_{n}}{2\pi} ... \int \frac{\td \phi_{1}}{2\pi} f(\phi_{1},...,\phi_{n}).$$ The relevant angles are defined in Figure \ref{fig:angles}. We use this operation on both sides of Eq.~\eqref{eqn:evo} and spin-average, see Appendices \ref{sec:CBderivation} and \ref{sec:Spin} for details. It is at this point we see that $\<P_{i_nj_n}\>_{n} \propto \Theta(\theta_{j_{n},i_{n}} - \theta_{n,i_{n}})$.
	\item We perform a change of variables, $q_{n \, \bot} \rightarrow \zeta_{n, j_{n}} = 1 - \cos \theta_{n, j_{n}}$, so as to make the angular ordering explicit. We merge the soft and hard-collinear emission kernels; expressing them in terms of collinear splitting functions. We also must sort out recoil so that the longitudinal component of the total momentum in a $1 \rightarrow 2$ splitting is conserved. Finally, using kinematic variables defined in the event zero-momentum frame\footnote{i.e. for $e^+e^- \rightarrow q \bar{q}$,  $z_{n} = \tilde{p}_{i_{n}} \cdot n / p_{i_{n}} \cdot n$ and $n$ is chosen so that $n || P_{\bar{q}}$ for all emissions in the quark jet and vice versa for the anti-quark jet.} allows us to saturate the $\Theta(\theta_{j_{n},i_{n}} - \theta_{n,i_{n}})$ angular ordering constraint for emissions originating from the primary hard partons (which are anti-parallel to each other). For all other emissions, it is necessary to approximate $\Theta(\theta_{j_{n},i_{n}} - \theta_{n,i_{n}}) \approx 1$. This approximation (which corresponds to strong ordering in angles) is equivalent to assuming the angle of the current emission is smaller than the opening angle of every other dipole, not just the opening angle of its parent dipole. This is the familiar angular ordering used in both resummations \cite{CATANI1992419,CATANI1991635} and parton showers when showering from an $e^+e^- \rightarrow q \bar{q}$ hard process \cite{Herwig_shower}. Strong ordering in angles simplifies the colour structures, so that all colour-charge operators can be reduced to Casimir, i.e. $\mathcal{C}_{\TT{F}}$ for a quark and $\mathcal{C}_{\TT{A}}$ for a gluon. The simplified colour reduces the evolution equation to an evolution of matrix elements, $|\mathcal{M}_{n}|^{2}$.
\end{enumerate}
The final result is 
\begin{align}
&\zeta \pdf{\, \<|\mathcal{M}_{n}(\zeta)|^{2}\>_{1,...,n}}{\zeta} \approx \nonumber  \\
 &-  \sum_{j_{n+1}} \sum_{\upsilon} \frac{\As}{\pi} \int \td z \, \mathcal{P}_{\upsilon \upsilon_{j_{n+1}}}(z) \<\Theta_{\TT{on\, shell}}\>_{n+1} \, \<|\mathcal{M}_{n}(\zeta)|^{2}\>_{1,...,n} + \sum_{\upsilon} \frac{\As}{\pi} \mathcal{P}_{\upsilon \upsilon_{j_{n}}}(z_{n}) \nonumber  \\ 
& \times \<\Theta_{\TT{on\, shell}}\>_{n} \int \td^4 p_{j_{n}} \; \delta^{4}(p_{j_{n}} - z^{-1}_{n}\tilde{p}_{j_{n}}) \, \<|\mathcal{M}_{n-1} (\zeta_{n, j_{n}}) |^{2}\>_{1,...,n-1} \; \zeta_{n, j_{n}} \; \delta(\zeta - \zeta_{n, j_{n}}). \label{eqn:CB}
\end{align}
The angular ordering variable $\zeta_{n, j_{n}} = 1 - \cos \theta_{n, j_{n}}$. $\mathcal{P}_{\upsilon \upsilon_{j_{n}}}(z_{n})$ are the usual collinear splitting functions, e.g. $\mathcal{P}_{q q}(z_{n}) = \mathcal{C}_{\TT{F}} \frac{1+ z^{2}_{n}}{1-z_{n}}$. Here we have used $\upsilon_{j_{n}}$ to label the species of parton $j_n$ and $\upsilon$ to label the species $j_n$ transitions to; if $\upsilon_{j_{n}} = q$ then $\upsilon=q$ and if $\upsilon_{j_{n}} = g$ then $\upsilon=q,g$. $z_{n}$ is the momentum fraction of parton $n$, i.e. if we have a collinear splitting that induces $j_{n-1} \rightarrow j_{n} \, n$ then $p_{j_{n}} \approx z_{n} p_{j_{n-1}}$ and $q_{n} \approx (1- z_{n}) p_{j_{n-1}}$. $\Theta_{\TT{on\, shell}}$ is a product of theta functions that ensures each parton is integrated over the phase space corresponding to a real particle (see Section \ref{sec:aawte}). In the first term, $\Theta_{\TT{on\, shell}}$ is a function of $\zeta$, $z$ and the $n$-parton phase space. In the second term $\Theta_{\TT{on\, shell}}$ is a function of $\zeta_{n, j_{n}}$, $z_{n}$ and the $(n-1)$-parton phase space. $\< |\mathcal{M}_{n}(\zeta ; \{P_{1},...,P_{n_{\TT{H}}},  (z_{1}, \zeta_{1, j_{1}}),...,(z_{n},\zeta_{n, j_{n}})\})|^{2}\>_{1,...,n}$ is the azimuthally averaged, squared matrix element for a hard process dressed with $n$ strongly-ordered partons with a unique branching topology; each emitted parton is specified by a pair $(z_{m}, \zeta_{m , j_{m}})$ and parton $j_{m}$ is the corresponding parent. The delta function enforces longitudinal momentum conservation; $|\mathcal{M}_{n}|^{2}$ depends on the momentum after the emission, $\tilde{p}_{j_{n}}$, and $|\mathcal{M}_{n-1}|^{2}$ depends on the momentum before the emission, $p_{j_{n}}$. 

Observables in $e^{+}e^{-}$ are computed after summing over emission topologies:
\begin{align}
\Sigma(\mu; \{p\}_{0}, \{v\}) \approx & \int \sum_{n} \sum_{j_{1},...,j_{n}} \left(\prod_{m = 1}^{n} \frac{\td \zeta_{m, j_{m}}}{\zeta_{m, j_{m}}} \frac{\td z_{i} \td \phi_{i}}{2\pi} \right) \<|\mathcal{M}_{n}(\mu)|^{2}\>_{1,...,n}\, u (\{p\}_{n}, \{v\}), \label{eqn:cross-sectionCB}
\end{align}
where $\mu$ should be taken to zero (or the shower cutoff) and for hadron-hadron collisions see Appendix \ref{sec:hadrons}.\footnote{ In the appendix, we sum over branching topologies: $\sum_{j_{1},...,j_{n}} \<|\mathcal{M}_{n}|^{2}\>_{1,...,n} = \<|M_{n}|^{2}\>_{1,...,n}$.}

There are several noteworthy points involved in this derivation:
\begin{itemize}
	\item In order to reduce the colour structures to being
          diagonal, we made the approximation
          \mbox{$\Theta(\theta_{j_{n},i_{n}} - \theta_{n,i_{n}})
            \approx 1$} for emissions from partons other than the two
          primary hard particles. The approximation is generally only
          good to LL accuracy (though angular ordered showers are able
          to go beyond this when combined with the CMW running of the
          coupling \cite{CATANI1991635}, e.g. to compute thrust at NLL
          \cite{CATANI1992419}). Moreover, modern angular ordered
          showers retain information on the hard-process, leading
          $\Nc$ colour flows by working in the dipole frames of
          initially colour-connected partons. This improves the
          approximation for hard processes with greater than two hard
          jets, since it is then only required to assume
          \mbox{$\Theta(\theta_{j_{n},i_{n}} - \theta_{n,i_{n}})
            \approx 1$} for emissions from partons other than the
          primary hard partons. During the subsequent evolution,
          traditional angular ordered showers lose the information on
          QCD colour flows\footnote{Some azimuthal correlations due
            to colour correlations can be re-instantiated in coherent
            branching algorithms
            \cite{Knowles:1987cu,KNOWLES1990271}.}, while dipole
          showers retain it to all orders at leading $\Nc$. We will
          exploit this in our dipole shower construction. Appendix
          \ref{sec:CBderivation} and \ref{sec:Dipolederivation} give
          more details on this point.
	\item The shower does not yet fully conserve energy and momentum. Rather it only conserves energy-momentum longitudinal to a jet. Accounting fully for energy-momentum conservation is formally sub-leading in many observables. However, it is phenomenologically important and necessary for shower unitarity. Furthermore, if total energy-momentum conservation is handled incorrectly it can spoil the NLL accuracy of a shower for some observables \cite{Dasgupta:2018nvj}. We will return to this in Section \ref{sec:Correcting}.
	\item We averaged the azimuthal dependence of the matrix elements. However, this ignores possible azimuthal dependence of the observable. Really one should compute $\<|\mathcal{M}_{n}|^{2} \, u(\{p\}_{n},\{v\})\>_{1,...,n}$. It is therefore important to ask whether $$\<|\mathcal{M}_{n}|^{2} \, u(\{p\}_{n},\{v\})\>_{1,...,n} \approx \<|\mathcal{M}_{n}|^{2} \>_{1,...,n} \< u(\{p\}_{n},\{v\})\>_{1,...,n}$$ is a good approximation. In other words, are the azimuthal dependencies of the matrix element and the observable correlated? This is clearly an observable dependent statement. Despite this we can make some progress; we can remove the approximation and find
	\begin{align}
	&\<|\mathcal{M}_{n}|^{2} \, u(\{p\}_{n})\>_{1,...,n} = \<|\mathcal{M}_{n}|^{2}\>_{1,...,n} \< u(\{p\}_{n})\>_{1,...,n} \nonumber \\ & \qquad + \sum^{n}_{m = 1} \sigma_{m}(\<|\mathcal{M}_{n}|^{2}\>_{1,...,n}) \, \sigma_{m}(\<u(\{p\}_{n})\>_{1,...,n}) \, \TT{Cor}_{m}(\<|\mathcal{M}_{n}|^{2}\>_{1,...,n},\<u(\{p\}_{n})\>_{1,...,n}) \nonumber \\ 
	& \qquad + \TT{higher} \; \TT{order} \; \TT{correlations}, \label{eqn:cor}
	\end{align}
	where $\sigma_{n}(x) =  \sqrt{\< x^2 \>_{n} - \< x \>^2_{n}}$ and $\TT{Cor}_{n}(x,y) = \frac{\< (x - \< x \>_{n})(y - \< y \>_{n}) \>_{n}}{\sigma_{n}(x)\sigma_{n}(y)}$. The first order correlation term (the second line of Eq.~\eqref{eqn:cor}) acts as a switch. If it is suppressed relative to the uncorrelated term then all higher correlations will be too. If it is not suppressed then higher order correlations may not be. In Appendix \ref{sec:CBEvo} we show that the higher order correlations are subdominant in the computation of NLL thrust. This is because the observable is two-jet dominated\footnote{Observables, such as thrust, for which the leading logarithms quantify small deviations from the two-jet limit or, more generally, the $n$-jet limit in the case of $n$-jettiness \cite{Stewart:2010tn}} and exponentiates, and so at NLL accuracy $\sigma_{m}(\<u(\{p\}_{n})\>_{1,...,n}) \approx 0$. However, we also find that the correlation term can provide a formally leading contribution to non-global logarithms. In Appendix \ref{sec:CBderivation} we observe that the correlation terms contribute leading logarithms to observables like gaps-between-jets, for which $\As^n L^n$ logs are leading. The miscalculation of non-global logarithms by angular ordered showers has previously been subject to numerical study in \cite{Dasgupta:2002bw,Banfi:2006gy}, where it was observed that leading non-global logarithms are incorrectly computed by angular ordered showers. However, \cite{Dasgupta:2002bw,Banfi:2006gy} also observed the error to be a phenomenologically small effect.
\end{itemize}

\subsection{Dipole shower}
\label{sec:DipoleEvo}

In the PB algorithm, the mechanism for energy-momentum conservation is unspecified. This is because interference terms make it difficult to see how recoil should be distributed. There are no such issues in angular ordered showers. In this case, the naive guess for how to conserve momentum longitudinal to a jet is correct and is sufficient for the computation of NLL DGLAP evolution and jet physics \cite{Dokshitzer:1991wu,Dokshitzer:1977sg,Dokshitzer:1992iy,Gribov:1972ri,APSplitting,Forshaw:1999iv}. We can exploit this to constrain the form of the recoil ($\int \td R_{n}$) so that the PB algorithm is consistent with an angular ordered shower. In this section, we will derive a dipole shower with this constraint in place from the outset. The resulting dipole shower is very similar to the dipole showers that are commonplace in event generators \cite{Pythia8,Herwig_dipole_shower}. However, it has a crucial difference: it does not use Catani-Seymour dipole factorisation \cite{Catani:1996vz}.

To derive the dipole shower proceed as follows.
\begin{enumerate}
	\item Expand Eq.~\eqref{eqn:evo} in powers of the number of
          colours $\Nc$ and keep only the leading terms, which go as
          $\As^n\Nc^n$, see
            \cite{Platzer:2013fha,SoftEvolutionAlgorithm}. This is
          necessary as only in the leading colour limit can we write
          evolution equations for $|\mathcal{M}_{n}|^{2}$. For the
          same purpose, spin average the evolution, see Appendix
          \ref{sec:Spin} for details.
	\item The colour expansion reduces the evolution equation so that it only depends on dipoles formed by colour connected partons. We use the form of $\int \td R_{n}$ to partition each dipole into two parts, introducing longitudinal momentum conservation to each part of the dipole in such a way that it is exactly consistent with the angular ordered shower. This is similar to how dipoles are usually partitioned using Catani-Seymour dipole factorisation. This partitioning allows us to exchange the sum over dipoles with a sum over emitting parton colour lines.
	\item Use the dipole partitioning to restore the (full-colour) hard-collinear physics that is correctly computed by an angularly ordered shower. This is uniquely determined by how longitudinal recoil is assigned. The result is a dipole shower that does not suffer the NLC errors in radiation ordered in angle noted in \cite{Dasgupta:2018nvj}.
\end{enumerate}
In Appendix \ref{sec:Dipolederivation} the complete proof is presented. The final result, expressed in the colour flow basis, is
\begin{align}
&q_{\bot} \pdf{|\mathcal{M}^{(\sigma)}_{n}(q_{\bot})|^{2}}{q_{\bot}} \nonumber \\  
&\approx - \, \frac{\As}{\pi}  \sum_{i^c_{n+1}}\int \td q^{(i^c_{n+1}, \overline{i^c}_{n+1})}_{ \bot} \delta (q^{(i^c_{n+1}, \overline{i^c}_{n+1})}_{ \bot} - q_{\bot})\int \td z \, \Theta_{\TT{on}\,\TT{shell}} \;  P_{\upsilon_{i_{n+1}}\upsilon_{i_{n+1}}}(z)  \, |\mathcal{M}^{(\sigma)}_{n}(q_{\bot})|^{2} \nonumber \\ 
& + \frac{\As}{\pi} \int \bigg(\prod_{j_{n}} \td^4 p_{j_{n}} \bigg) \,  \mathfrak{R}^{\TT{dipole}}_{i^c_{n}} \, P_{\upsilon_{i_{n}} \upsilon_{i_{n}}}(z_{n})  \; q_{\bot}\delta(q^{(i^c_{n}, \overline{i^c}_{n})}_{n \, \bot} - q_{\bot} ) |\mathcal{M}^{(\sigma/n)}_{n-1}(q^{(i^c_{n}, \overline{i^c}_{n})}_{n \, \bot})|^{2}, \label{eqn:corrected_dipole}
\end{align}
where $\sigma$ is a colour flow and $\sigma/n$ is the same colour flow but with the $n$th colour line removed. We use $i^c_{n}$ to index the (anti-)colour line(s) of parton $i$ in a final state dressed with $n$ soft or collinear partons, i.e. if parton $i$ is a quark it has a single colour line and so $i^c_{n} = i^q_{n}$, if parton $i$ is a gluon it will have a colour and an anti-colour line so $i^c_{n} = i^g_{n}, i^{\bar{g}}_{n}$ respectively. $\overline{i^c}_{n}$ is the (anti-)colour line connected to $i^c_{n}$. Momenta with colour line indices are the momenta of the partons associated to that colour line, i.e. $p_{i^c_{n}}=p_{i_{n}}$. The shower is ordered in dipole $p_{T}$, defined as
\begin{equation}
(q^{(i^c_{n},\overline{i^c}_{n})}_{n \, \bot})^{2} = \frac{2(p_{i^c_{n}}\cdot q_{n})(p_{\, \overline{i^c}_{n}} \cdot q_{n})}{p_{i^c_{n}}\cdot p_{\, \overline{i^c}_{n}}}.
\end{equation}
The dipole splitting functions are 
$$ P_{q q}(z_{n}) = \mathcal{C}_{\TT{F}} \frac{1+ z^{2}_{n}}{1-z_{n}}, \qquad P_{gg}(z_{n}) = \frac{\mathcal{C}_{\TT{A}}}{2} \frac{1+ z^{3}_{n}}{1-z_{n}}.$$
These splitting functions are related to those in the previous section according to \linebreak \mbox{$\mathcal{P}_{gg}(z) = P_{gg}(z) + P_{gg}(1-z)$}, and $\mathcal{P}_{qq}(z) = P_{q q}(z)$.
Note that to simplify Eq.~\eqref{eqn:corrected_dipole} we have omitted the possibility of $g\rightarrow qq$ transitions, which is sub-leading in colour and only contributes a leading logarithm to single-logarithm, collinear-sensitive observables or at NLL for double-logarithmic observables. In Appendix \ref{sec:DipoleEvo} we present Eq.~\eqref{eqn:corrected_dipole} with this splitting included. Being explicit, we would write the squared matrix element as
$$|\mathcal{M}^{(\sigma)}_{n}(q_{\bot} ; \{P_{1},...,P_{n_{\TT{H}}},  (z_{1}, q^{(i^c_{1}, \overline{i^c}_{1})}_{1 \, \bot},\phi_{1}),...,(z_{n},q^{(i^c_{n}, \overline{i^c}_{n})}_{n \, \bot},\phi_{n})\})|^{2}.$$ As for the angular ordered shower, this is the squared matrix element for a hard process dressed with $n$ strongly-ordered partons with a unique branching topology, i.e. each emitted parton is specified by a triplet $(z_{m},q^{(i^c_{m}, \overline{i^c}_{m})}_{m \, \bot},\phi_{m})$ and is emitted from the parton with colour line $i^c_{m}$. The dipole recoil function is given by
\begin{align}
\mathfrak{R}^{\TT{dipole}}_{i^c_{n}} = \left(\frac{1}{2} + \TT{Asym}_{i^c_{n}\overline{i^c}_{n}}(q_{n})\right)\mathfrak{R}_{i^c_{n}}, \label{eqn:dipolerecoilsplit}
\end{align}
where
\begin{align}
\mathfrak{R}_{i^c_{n}} = \delta^{4}(p_{i_{n}} - z^{-1}_{n}\tilde{p}_{i_{n}}) \prod_{i_{n} \neq j_{n}} \delta^{4}(p_{j_{n}} - \tilde{p}_{j_{n}}) + \mathcal{O}(q_{\bot}/Q),
\end{align}
and where
\begin{align}
\TT{Asym}_{i^c_{n}\overline{i^c}_{n}}(q_{n}) = \left[ \frac{T \cdot p_{i^c_{n}}}{4T \cdot q_{n} }\frac{(q^{(i^c_{n}\overline{i^c}_{n})}_{n \, \bot})^{2}}{p_{i^c_{n}} \cdot q_{n}} - \frac{T \cdot p_{\, \overline{i^c}_{n}}}{4T \cdot q_{n} }\frac{(q^{(i^c_{n}\overline{i^c}_{n})}_{n \, \bot})^{2}}{p_{\, \overline{i^c}_{n}} \cdot q_{n}}\right], \quad \TT{and} \quad T = \sum_{i_{n}} p_{i_{n}}.
\end{align}
Note, in the limit that $q_{n}$ is collinear to $p_{i^c_{n}}$, $\TT{Asym}_{i^c_{n}\overline{i^c}_{n}}(q_{n}) = 1/2$. Thus, in this limit $\mathfrak{R}^{\TT{dipole}}_{i^c_{n}} \rightarrow \mathfrak{R}_{i^c_{n}}$, as required. Our expression for $\mathfrak{R}^{\TT{dipole}}_{i^c_{n}}$ should be compared to the recoil function one would find using Catani-Seymour dipole factorisation:
\begin{align}
\mathfrak{R}^{\TT{C.S.}}_{i^c_{n}}(q_{n}) = \left(\frac{(q^{(i^c_{n}\overline{i^c}_{n})}_{n \, \bot})^{2}{p}_{\, \overline{i^c}_{n}}\cdot {p}_{i^c_{n}}}{2p_{i^c_{n}} \cdot q_{n} \, ({p}_{\, \overline{i^c}_{n}}+{p}_{i^c_{n}}) \cdot q_{n}}\right)\mathfrak{R}_{i^{c}_{n}}.
\end{align}
$\mathfrak{R}^{\TT{dipole}}_{i^c_{n}} \to \mathfrak{R}^{\TT{C.S.}}_{i^c_{n}}$ if we were to make the replacement $T \to p_{i^c_{n}} + p_{\bar{i^c}_{n}}$. Observables are computed after summing over emission topologies:
\begin{align}
\Sigma(\mu; \{p\}_{0}, \{v\}) \approx & \int \sum_{n} \sum_{\sigma} \sum_{i^c_{1},...,i^c_{n}}  \left(\prod_{m = 1}^{n} \frac{\td q^{(i^c_{m}, \overline{i^c}_{m})}_{m \, \bot}}{q^{(i^c_{m}, \overline{i^c}_{m})}_{m \, \bot}} \frac{\td z_{i} \td \phi_{i}}{2\pi} \right) |\mathcal{M}^{(\sigma)}_{n}(\mu)|^{2}\, u (\{p\}_{n}, \{v\}). \label{eqn:cross-sectionDipole}
\end{align} 
There are several noteworthy points involved in this derivation:
\begin{itemize}
	\item This shower was built around preserving the beneficial features of an angular ordered shower. In fact, azimuthally averaging the dipole shower reinstates an angular ordering. Angular ordered showers provide a sufficient framework to resum global two-jet dominated observables, such as thrust, up to $\As^{n} L^{2n-1}$ terms with full colour. Radiation consecutively ordered in angle generated by the dipole shower presented here will also achieve this accuracy (radiation unordered in angle will differ at sub-leading $N_{c}$). This reduces the doubly logarithmic NLC errors noted in \cite{Dasgupta:2018nvj}, where the particular example of errors in the thrust observable was given.
	\item Traditional angular ordered showers fail to correctly compute $\As^{n} L^{2n-1}$ logarithms for $n>2$ jet observables. This is because soft, wide-angle physics is miscalculated because of the  \mbox{$\Theta(\theta_{j_{n},i_{n}} - \theta_{n,i_{n}}) \approx 1$} approximation, as previously discussed.\footnote{Modern implementations of angular ordered showers do use colour flow information from the hard process, allowing them to compute $\As^{n} L^{2n-1}$ terms at leading colour for global $n>2$ jet dominated observables by deriving appropriate initial conditions from the respective large-$N$ colour flows of the hard process \cite{Herwig_shower}.} It is never necessary to make this approximation in the dipole shower since we can use the underlying colour flows to define variables in the relevant dipole frame, for which \mbox{$\Theta(\theta_{j_{n},i_{n}} - \theta_{n,i_{n}}) = 1$} is always true. Thus we expect the dipole shower to have the capacity to re-sum $\As^{n} L^{2n-1}$ logarithms at leading colour.\footnote{Eq.~\eqref{eqn:corrected_dipole} as it stands only provides a sufficient framework for this resummation. It is not yet sufficient in itself: one would need to enhance the shower with a running coupling and, possibly, higher order splitting functions.}
	\item In the soft limit the dipole shower generates iterative solutions to the BMS equation \cite{BMSEquation,Larkoski:2016zzc} (the proof is as in Section 3 of \cite{SoftEvolutionAlgorithm}). This demonstrates that the shower computes non-global logarithms at leading colour correctly.
	\item At this point in our theoretical development, the dipole shower does not completely conserve energy and momentum. Rather it only conserves momentum longitudinal to the emitting parton. Accounting for total energy-momentum conservation is not needed to compute some observables to NLL accuracy, e.g. thrust. Regardless, it is an important effect that if handled incorrectly can spoil the NLL accuracy of the shower \cite{Dasgupta:2018nvj}. Addressing this is the focus of the next section.
\end{itemize}

\section{Improving recoil in dipole showers}
\label{sec:Correcting}

\begin{figure}[b]
	\centering
	\includegraphics[width=0.95\textwidth]{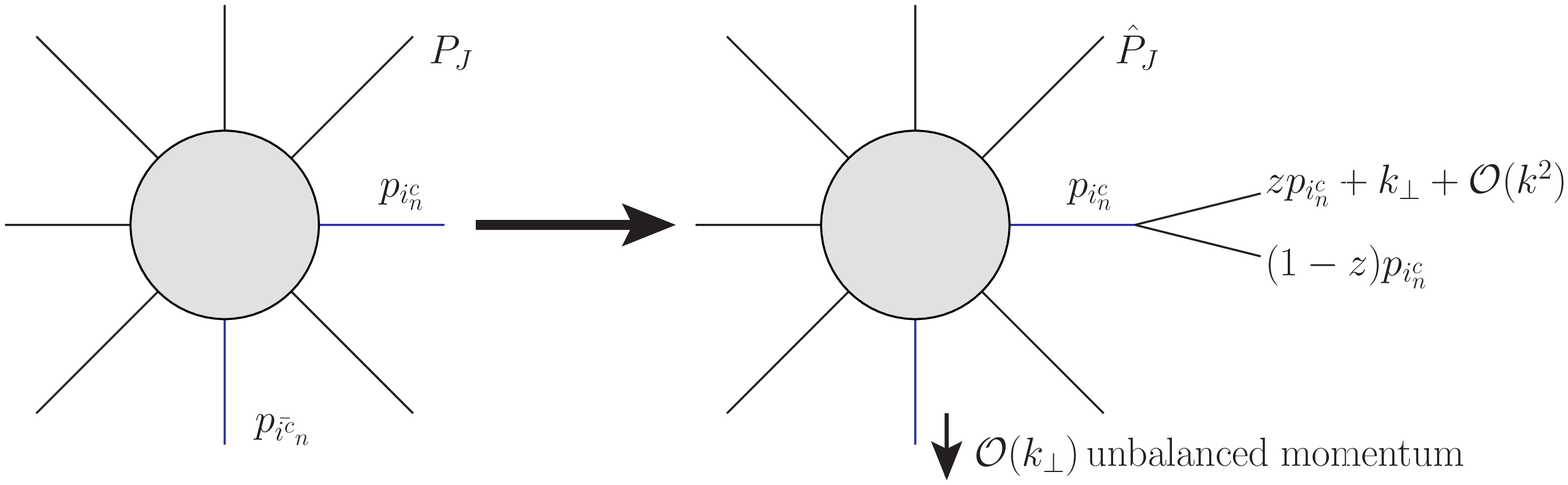} \newline \newline \newline
	\includegraphics[width=0.95\textwidth]{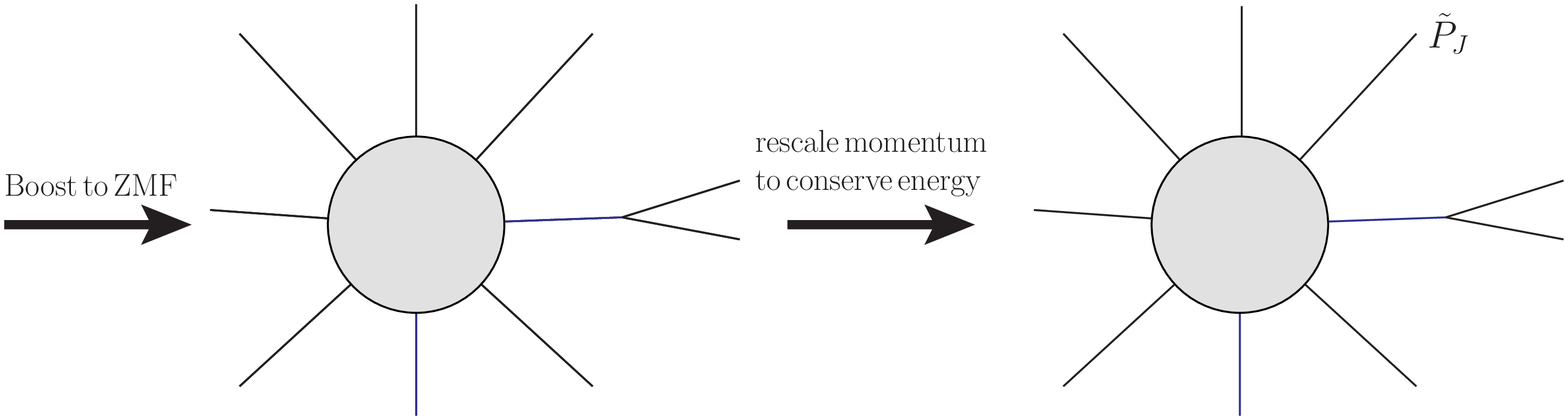}
	\caption{A summary of the dipole shower global recoil scheme (a scheme for energy-momentum conservation). In words: A new particle is emitted which leaves some momentum unbalanced (in the direction of the colour connected parton and in the plane transverse to the dipole); perform a Lorentz boost to the new ZMF, and re-scale the jet momenta in such a way that the rescaling does not change the $k_{\bot}$ of the emission. This leaves an $n$-parton ensemble with the same total energy and total momentum as the $n-1$-parton ensemble.}
	\label{fig:recoil}
\end{figure}

In this section we will address the problem of energy-momentum conservation in a dipole shower, though our approach is simple to map onto an angular ordered shower. The mechanism for energy-momentum conservation (or recoil scheme) we present lacks a formal derivation. Rather it is inspired by the study of recoil by Bewick et al. \cite{Bewick:2019rbu}. Bewick et al. analysed several approaches to recoil in angular ordered showers, reproducing some of the fixed-order checks of \cite{Dasgupta:2018nvj} and performing further numerical checks. They observed that among the better performing recoil schemes are globally defined schemes; schemes that redistribute momentum across an entire jet or event. From our perspective, a global scheme is also preferable, as it is more simply implemented in a dipole shower. Momentum conservation on an emission-by-emission basis is also desirable when it comes to matching to fixed-order and merging of hard processes of different jet multiplicity. In the two-jet limit, our scheme becomes that which is analysed in \cite{Bewick:2019rbu} and implemented in HERWIG's angular ordered shower \cite{Herwig_shower}. For comparison, in Appendix \ref{sec:Spectator} we summarise the implementation and limitations of a spectator recoil scheme, as implemented in \cite{Pythia8,Herwig_dipole_shower,Platzer:recoil}.
	
We start with an observation that is key to all global recoil schemes: when a parton is emitted from another, the parent parton must have been off-shell. We parametrise the amount by which it is off shell by giving it a virtual mass. A parton shower approximates the sum over the multiplicities of QCD radiation dressing a given hard process. Each term in the sum should have the same total energy and the same zero-momentum frame (ZMF). Naively adding a parton to an $n-1$ on-shell parton state changes the total energy and ZMF. We will redistribute parton momenta as simply as possible in order to restore the ZMF and total energy. We will do this using a single global Lorentz boost and a single rescaling that preserves the transverse momentum ordering. This procedure is illustrated in Figure \ref{fig:recoil}. Below we will spell out how to implement this recoil scheme. The simplicity of the scheme can get lost in its mathematical definition and so we encourage the reader to keep Figure \ref{fig:recoil} in mind.

Let us now make Figure \ref{fig:recoil} quantitative. We require that energy is conserved, \\ \mbox{$E_{\TT{before}} = E_{\TT{after}} =Q$} where
\begin{align}
&\sum_{i_{n}}^{n-1} \sqrt{\v{p}^{2}_{i_{n}} + m_{i_{n}}^2} \equiv \sum^{n-1}_{J = 1} \sqrt{\v{P}^{2}_{J} + m^{2}_{J}} = E_{\TT{before}}, \nonumber \\
&\sum_{i_{n}}^{n-1} \sqrt{\tilde{\v{p}}^{2}_{i_{n}} + m_{i_{n}}^2} + \sqrt{ \v{q}^{2}_{n} + m^{2}_{q_{n}}} \equiv \sum^{n-1}_{J = 1} \sqrt{\tilde{\v{P}}^{2}_{J} + \tilde{P}^{2}_{J}} = E_{\TT{after}}, \end{align} 
and that momentum is conserved
\begin{align}
&\sum^{n-1}_{J = 1} \v{P}_{J} = \sum^{n-1}_{J = 1} \tilde{\v{P}}_{J} = 0, \label{eqn:CoMconservation}
\end{align}
where, in the ZMF, $\v{P}_{J}$ is the 3-momentum of $J$th jet amongst the $n-1$ jets constructed from an $n-1$ parton ensemble, i.e. $\v{P}_{J} = \v{p}_{i_{n}}$ for $J= i_{n}$ (recall that $i_n$ labels parton $i$ in an $n$-parton ensemble). We introduce the extra notation because it is the momentum of jets that we particularly focus on conserving. $\tilde{\v{P}}_{J}$ is what we wish to find; it is the momentum of the $J$th jet now constructed from an $n$ parton ensemble after the necessary redistribution of momenta (all jets contain a single parton except for one which contains two partons; the original parton and the newly added parton). $m_{i}$ is the mass of parton $i$, and $m_{i} = 0$ since we consider only massless partons. $\tilde{P}^{2}_{J}$ is the virtual mass squared of the $J$th jet, it also is zero for all jets other than the jet built of two partons. We can achieve our desired redistribution by a Lorentz boost, $\Lambda^{\mu}_{\; \nu}$, from the ZMF of the $n-1$ parton ensemble to the ZMF of the $n$ parton ensemble. Once in this frame, we re-scale all the jet momenta by a global factor $\kappa_{i^c_{n}}$ (the index will prove necessary later on) so as to preserve the centre-of-mass energy. In all, we wish to find $\tilde{P}_{J \, \mu} = \kappa_{i^c_{n}} \Lambda^{\;\nu}_{\mu} \hat{P}_{j \, \nu}$ where $\hat{P}_{j}$ is the four-momentum of the $J$th jet constructed from the $n$ parton ensemble \textit{before} the redistribution of momenta. We place a hat on all intermediary kinematic variables (i.e. those after the emission but before redistribution of momenta). We denote the 3-momentum of $\tilde{P}_{J}$ as $\tilde{\v{P}}_{J} = \kappa_{i^c_{n}}\v{\Lambda}\hat{\v{P}}_{J}$. $\Lambda^{\mu}_{\; \nu}$ is specified by solving Eq.~\eqref{eqn:CoMconservation} and $\kappa_{i^c_{n}}$ is specified by solving
\begin{align}
Q = \sum^{n-1}_{J = 1} \sqrt{\tilde{\v{P}}^{2}_{J} + \tilde{P}^{2}_{J}} = \sum^{n-1}_{J = 1} \kappa_{i^c_{n}}\sqrt{ \,  (\v{\Lambda}\hat{\v{P}}_{J})^{2} + \hat{P}^{2}_{J}}, \label{eqn:kappa}
\end{align}
which comes from requiring $E_{\TT{before}} = E_{\TT{after}} = Q$.
	
We will express this recoil scheme in terms of the shower kinematics and solve for $\tilde{\v{P}}_{J}$. We use the following Sudakov decomposition for a $1 \rightarrow 2$ ($p_{i^c_{n}} \rightarrow \hat{p}_{i^c_{n}} \hat{q}_{n}$) parton transition:
\begin{align}
\hat{q}_{n} &= (1-z_{n}) p_{i^c_{n}}+k_{\bot}+\frac{(q^{(i^c_{n}\overline{i^c}_{n})}_{n \, \bot})^{2}}{1-z_{n}}\frac{p_{\,\overline{i^c}_{n}}}{2p_{i^c_{n}}\cdot p_{\,\overline{i^c}_{n}}},\nonumber \\
\hat{p}_{i^c_{n}} &= z_{n} p_{i^c_{n}}, \qquad (q^{(i^c_{n}\overline{i^c}_{n})}_{n \, \bot})^{2} = - k_{\bot}^{2},  \qquad k_{\bot} \cdot p_{i^c_{n}} = k_{\bot} \cdot p_{\,\overline{i^c}_{n}}=0.
\label{eqn:1_to_2_Sudakov_decomposition}
\end{align}
We label the jet in which the splitting takes place as $P_{J \, \TT{emit}}$, so that $P_{J \, \TT{emit}}= p_{i^c_{n}}$. From Eq.~\eqref{eqn:1_to_2_Sudakov_decomposition}:
$$
\hat{P}^{2}_{J \, \TT{emit}} = \frac{z_{n}(q^{(i^c_{n}\overline{i^c}_{n})}_{n \, \bot})^{2}}{(1- z_{n})}, \quad
\hat{\v{P}}_{J \, \TT{emit}} =  \v{P}_{J \, \TT{emit}} + \v{k}_{\bot} + \frac{(q^{(i^c_{n}\overline{i^c}_{n})}_{n \, \bot})^{2}}{(1-z_{n})\, 2 p_{i^c_{n}}\cdot p_{\,\overline{i^c}_{n}}}\v{p}_{\,\overline{i^c}_{n}}.
$$
For all jets other than ``$J \, \TT{emit}$'' $\hat{\v{P}}_{J} =  \v{P}_{J}$ and $\hat{P}^{2}_{J} = 0$. The Lorentz boost, $\Lambda^{\mu}_{\; \nu}(i^c_{n},\overline{i^c}_{n})$, can now be found. The boost is in the direction of $\v{p}_{\,\overline{i^c}_{n}}$ and is given by the boost velocity
\begin{align}
\boldsymbol{\beta}_{\TT{ZMF}} = \frac{\hat{\v{P}}_{J \, \TT{emit}} - \v{P}_{J \, \TT{emit}} }{ \sum_{J}\sqrt{\hat{\v{P}}^{2}_{J} + \hat{P}^{2}_{J}} + \sqrt{|\hat{\v{P}}_{J \, \TT{emit}} - \v{P}_{J \, \TT{emit}}|^{2} + k_{\bot}^{2}}}.
\end{align}
Finally we must solve for $\kappa_{i^c_{n}}$ using Eq.~\eqref{eqn:kappa},
\begin{align}
\kappa_{i^c_{n}} &=\frac{ \sum^{n-1}_{J = 1} \sqrt{\v{P}^{2}_{J} + P^{2}_{J}} }{\sum^{n-1}_{J = 1}\sqrt{ \,  (\v{\Lambda}\hat{\v{P}}_{J})^{2} + \hat{P}^{2}_{J}}}.
\end{align}
Note that in both the soft and collinear limits $\kappa_{i^c_{n}} \rightarrow 1$.

So now we have everything we need to compute $\tilde{\v{P}}_{J}=\kappa_{i^c_{n}} \v{\Lambda}\hat{\v{P}}_{J}$. We can put this in the dipole shower by introducing a recoil function
\begin{align}
\mathfrak{R}_{i^c_{n}} =& \delta^{4}_{\mathcal{J}}\left(\tilde{p}_{i^c_{n}} - z_{n} \kappa_{i^c_{n}} \, \Lambda(i^c_{n},\overline{i^c}_{n}) p_{i^c_{n}}\right)\prod_{j_{n} \neq i_{n}} \delta^{4}_{\mathcal{J}}\left(\kappa_{i_{n}} \, \Lambda(i^c_{n},\overline{i^c}_{n}) p_{j_{n}} - \tilde{p}_{j_{n}}\right), \label{eqn:recoil}
\end{align}
where 
$\delta^4_{\mathcal{J}}(f(p_{i^c_n}))$ is a delta function normalised against its Jacobi factor: $$\delta^4_{\mathcal{J}}(f(p_{i^c_n})) = \delta^4(p_{i^c_n} - X),$$ where $X$ is the (unique) solution to $f(X)=0$. Note that in an implementation of the algorithm there is never any need to invert the argument of the delta function to solve for $p_{i^c_n}$ since $\tilde{p}_{i^c_n}$ is what is needed going forwards. In Eq.~\eqref{eqn:corrected_dipole}, the delta functions simply kill all of the integrals over $p_{jn}$.
For the sake of being explicit, the emitted parton has momentum
\begin{align}
q_{n} = (1-z_{n}) \kappa_{i^c_{n}} \, \Lambda(i^c_{n},\overline{i^c}_{n}) p_{i^c_{n}}+k_{\bot}+\frac{(q^{(i^c_{n}\overline{i^c}_{n})}_{n \, \bot})^{2}}{\kappa_{i^c_{n}} \,(1-z_{n})}\frac{ \Lambda(i^c_{n},\overline{i^c}_{n})p_{\,\overline{i^c}_{n}}}{2p_{i^c_{n}}\cdot p_{\,\overline{i^c}_{n}}}.
\end{align}
Note that both $z_{n}$ and $\td q^{(i^c_{n}\overline{i^c}_{n})}_{n \, \bot}/q^{(i^c_{n}\overline{i^c}_{n})}_{n \, \bot}$ are Lorentz and jet scaling invariants. This means that all of the emission kernels remain unchanged and so the implementation of this recoil scheme only enters so as to ensure that the real emissions continue to be integrated over the correct phase-space (and through the corresponding $\Theta_{\mathrm{on \, shell}}$ for the virtuals). 

In order to implement the proposed shower computationally we must specify the phase-space boundary for real emissions. In our previous papers, \cite{Forshaw:2019ver,SoftEvolutionAlgorithm,Platzer:recoil}, we gave general formulae for the computation of phase-space boundaries, derived by ensuring the emitted parton is on-shell and has less energy than its parent. Applying these to the recoil prescription we present here, we find that
\begin{align}
z_{n} \in \left(0, 1-\frac{(q^{(i^c_{n}\overline{i^c}_{n})}_{n \, \bot})^{2}}{{2p_{i^c_{n}}\cdot p_{\,\overline{i^c}_{n}}}} \right), \qquad \phi \in [0, 2 \pi), \label{eq:phasespace}
\end{align}
up to terms of the order $(1-\kappa_{i^c_{n}})$; in the following section, we show that these terms are negligible at NLL accuracy. Here $\phi$ is the trivial azimuth in the dipole frame.
Thus, the complete dipole shower is defined by Eq.~\eqref{eqn:corrected_dipole}\footnote{Or better still, Eq.~\eqref{eqn:dipolegqq}, which also includes $g\rightarrow qq$ transitions.}, Eq.~\eqref{eqn:dipolerecoilsplit}, Eq.~\eqref{eqn:recoil}, and Eq.~\eqref{eq:phasespace}.

\subsection{NLC and NLL accuracy of the global recoil}
\label{sec:NLLNLC}
      
In this section we will discuss the colour accuracy of our new dipole shower and test its logarithmic accuracy.

Firstly, the sub-leading colour contained in the shower is inherited from its link to angular ordered showers. In fact, when next-to-leading order splitting functions and the CMW running coupling are introduced the collinear radiation generated by the dipole shower is equivalent (after azimuthal averaging) to that generated by the coherent branching algorithm of \cite{CATANI1991635,CATANI1992419} up to the handling of transverse recoil. We discuss this in more detail in Appendix \ref{sec:Thrust} where we argue that differences due to transverse recoil do not effect next-to-leading logarithmic accuracy in the angular-ordered limit. This means that the dipole shower can be used to compute the leading-colour NLL resummation of thrust, again see Appendix \ref{sec:Thrust}. Correct colour factors will also be assigned to the leading logarithms associated with a broad class of observables that can be computed fully at LL accuracy in the angular-ordered approach (for which radiation unordered in angle generate NLLs). Outside of this limit, only leading colour accuracy is guaranteed. This is an improvement on existing dipole showers, which have been noted to incorrectly compute NLC at LL accuracy \cite{Dasgupta:2018nvj}, even including errors in logarithms originating from radiation ordered in angle. Further improvements on sub-leading colour, for more general observables, require amplitude evolution. We doubt that substantial further improvements in the accuracy of sub-leading-colour effects can be achieved in either the dipole shower or coherent branching frameworks. There is already a body of literature exploring possible resolutions to the NLC errors in dipole showers \cite{Eden:1998ig,Friberg:1996xc,Giele:2011cb}. Our approach of using angular ordering to improve dipole evolution is similar to that of \cite{Eden:1998ig,Friberg:1996xc}, though there it was largely explored only in the context of hadronisation and the computation of jet multiplicity observables. 
We also note that, by partitioning dipoles so as to identify a unique parent, we expect the sub-leading logarithms associated with unresolved soft and collinear radiation to be captured using the CMW scheme for the running coupling \cite{CATANI1991635,Banfi:2004yd}.

We will now proceed to evaluate the logarithmic accuracy of the recoil scheme discussed in the previous section. We do so in two ways. Firstly by re-creating the analysis of Section 4.2 in \cite{Dasgupta:2018nvj}. In this analysis, several event shape observables, defined by functions $V(\{p\})$, are considered at fixed order. The analysis tests the sub-leading contributions from the soft region found in the limit that the transverse momentum of the second emission is of similar magnitude to that of the first but both are small relative to the hard scale. This limit is considered because it is the limit where dipole showers have previously been shown to mishandle recoil. Specifically, we calculate the difference between the $\As^{2}$ LC, NLL contribution to the observable using the fixed-order amplitude, and the shower contribution: \mbox{$\delta \Sigma(L) = \Sigma^{(\As^{2})}_{\TT{shower}}(L) - \Sigma^{(\As^{2})}_{\TT{FO}}(L)$}. As the observables exponentiate, we are looking for differences of the form $\As^{2}\Nc^2L^{2}$ at fixed coupling since these terms contribute to the NLL exponent.

Our second check of logarithmic accuracy is to compare against two known NLL resummations: Thrust and generating functions for jet multiplicity. This is done in Appendix \ref{sec:Checks}.

Let us proceed to compute $\delta \Sigma(L)$ in the doubly-soft limit in $e^{+}e^{-} \rightarrow q \bar{q}$. We label the quark as parton $a$ and the anti-quark as parton $b$. In the same way that we label partons with indices $i_{n}$, each parton label is given a subscript stating the `current' multiplicity of radiated partons (since a parton's momentum changes to conserve momentum as more partons are radiated). From Eq.~\eqref{eqn:corrected_dipole} we can compute the first two soft emissions and find
\begin{align}
\delta \Sigma(L) &= \mathcal{C}^2_{\TT{F}} \sigma_{n_{\TT{H}}}  \int \td \Pi_{2} \, \td \Pi_{1} \int \td q^{(a_{2}, 1_{2})}_{2 \, \bot} \delta (q^{(a_{2}, 1_{2})}_{2 \, \bot} - q_{2 \, \bot}) \int \td q^{(a_{1}, b_{1})}_{1 \, \bot}  \delta(q^{(a_{1}, b_{1})}_{1 \, \bot} - q_{1 \, \bot}) \;  \nonumber \\
& \times \Theta(q_{1 \, \bot} - q_{2 \, \bot})  \bigg[ \int \prod^{2}_{n=1} \prod_{k_{n}} \td^4 p_{k_{n}} \; \mathfrak{R}^{\TT{soft}}_{a_{2} 1_{2}} \, \theta_{a_{2}1_{2}} \; \mathfrak{R}^{\TT{soft}}_{a_{1} b_{1}} \, \theta_{a_{1}b_{1}} \Theta\left(e^{-L} - V(\{p\}_{2})\right) \nonumber \\
& \qquad - \theta^{\TT{correct}}_{a_{2}1_{2}} \theta^{\TT{correct}}_{a_{1}b_{1}} \Theta\left(e^{-L} - V(\{p\}_{\TT{correct}})\right) \bigg], \label{eqn:Fixedorder}
\end{align}
where $\sigma_{n_{\TT{H}}}$ is the hard process cross section. $\theta_{i_{n} j_{n}}$ is the product of theta functions defining the on-shell requirements for emission from dipole $i_{n} j_{n}$ (previously given without indices as $\Theta_{\TT{on} \, \TT{shell}}$). $\{p\}_{\TT{correct}}$ are the momenta used to compute $\Sigma^{(\As^{2})}_{\TT{FO}}(L)$ and $\theta^{\TT{correct}}_{i_{n}j_{n}} = \theta_{i_{n}j_{n}}(\{p\}_{\TT{correct}})$. $\mathfrak{R}^{\TT{soft}}_{i_{n} j_{n}}$ is the combined dipole recoil function, $\mathfrak{R}^{\TT{soft}}_{i_{n} j_{n}} = \mathfrak{R}^{\TT{dipole}}_{i^c_{n}} + \mathfrak{R}^{\TT{dipole}}_{j^c_{n}}$. Before considering any specific event shape, we can simplify our expressions further by using the recoil delta functions to perform some of the integrals. These fix the final state momenta:
\begin{align}
\{p\}_{2} =  \{\tilde{\tilde{p}}_{a}, \tilde{\tilde{p}}_{b}, \tilde{q}_{1}, q_{2}\}, \quad \TT{where} \quad& \tilde{\tilde{p}}_{a} = \kappa_{a_{2}} \kappa_{a_{1}} \, \Lambda(a_{2},1_{2}) \Lambda(a_{1},b_{1}) p_{a}, \nonumber \\
&\tilde{\tilde{p}}_{b} = \kappa_{a_{2}} \kappa_{a_{1}} \, \Lambda(a_{2},1_{2}) \Lambda(a_{1},b_{1}) p_{b}, \nonumber \\
&\tilde{q}_{1} = \kappa_{a_{2}} \, \Lambda(a_{2},1_{2})q_{1}, \qquad q_{2} \; \TT{unmodified}, \nonumber \\
&\tilde{\tilde{Q}} = \kappa_{a_{2}} \kappa_{a_{1}} Q, \quad \tilde{Q} = \kappa_{a_{1}} Q, \quad Q = \mathcal{O}(2 p_{a} \cdot p_{b}).
\end{align}
$q_{1}$ and $q_{2}$ are defined with respect to the rescaled momenta $\tilde{\tilde{p}}_{a},\tilde{\tilde{p}}_{b}$ and so have appropriately modified limits on their phase space. We employ the `equally soft' limit ($Q \gg q_{1 \, \bot}, q_{2 \, \bot}$; $q_{1 \, \bot} \gtrsim q_{2 \, \bot}$) which reduces the complexity of the phase space limits and removes dependence on longitudinal recoil. In total, we find that
\begin{align}
\delta \Sigma(L) \approx &\frac{4\As^{2} \, \mathcal{C}^{2}_{\TT{F}} \, \sigma_{n_{\TT{H}}}}{\pi^{2}} \int^{Q}_{0} \frac{\td q^{(a_{2}, 1_{2})}_{2 \, \bot}}{q^{(a_{2}, 1_{2})}_{2 \, \bot}}  \int^{Q}_{0} \frac{\td q^{(a_{1}, b_{1})}_{1 \, \bot}}{q^{(a_{1}, b_{1})}_{1 \, \bot}} \int^{\ln \tilde{Q} /q^{(a_{1}, b_{1})}_{1 \, \bot}}_{-\ln \tilde{Q} / q^{(a_{1}, b_{1})}_{1 \, \bot}} \td y_{1} \int^{\ln \tilde{\tilde{Q}} /\tilde{q}^{(a_{2}, 1_{2})}_{2 \, \bot}}_{-\ln \tilde{\tilde{Q}} / \tilde{q}^{(a_{2}, 1_{2})}_{2 \, \bot}} \td y_{2} \nonumber \\
& \qquad \times \int^{2\pi}_{0} \frac{\td \phi_{2}}{2 \pi} \Theta\left(e^{-L} - V(\{p\}_{2})\right) \Theta(Q - q^{(a_{1}, b_{1})}_{1 \, \bot})\Theta(\kappa^{-1}_{a_{2}}  q^{(a_{1}, b_{1})}_{1 \, \bot} - q^{(a_{2}, 1_{2})}_{2 \, \bot}) \nonumber \\
&- \frac{4\As^{2} \, \mathcal{C}^{2}_{\TT{F}} \, \sigma_{n_{\TT{H}}}}{\pi^{2}}  \int^{Q}_{0} \frac{\td q^{(a_{1}, b_{1})}_{1 \, \bot}}{q^{(a_{1}, b_{1})}_{1 \, \bot}} \int^{q^{(a_{1}, b_{1})}_{1 \, \bot}}_{0} \frac{\td q^{(a_{2}, 1_{2})}_{2 \, \bot}}{q^{(a_{2}, 1_{2})}_{2 \, \bot}} \int^{\ln Q /q^{(a_{1}, b_{1})}_{1 \, \bot}}_{-\ln Q / q^{(a_{1}, b_{1})}_{1 \, \bot}} \td y_{1} \int^{\ln Q /q^{(a_{2}, 1_{2})}_{2 \, \bot}}_{-\ln Q / q^{(a_{2}, 1_{2})}_{2 \, \bot}} \td y_{2} \nonumber \\
& \qquad \times \int^{2\pi}_{0} \frac{\td \phi_{2}}{2 \pi} \Theta\left(e^{-L} - V(\{p\}_{\TT{correct}}) \right). \label{eqn:difference_global}
\end{align}
In the `equally soft' limit we are considering
\begin{align}
\kappa_{i_{n}} &\approx 1 - \mathcal{O}(q^{2}_{\bot}/2Q^2).
\end{align}
The $\kappa$ dependence in the shower integrals (lines 1 and 2 of Eq.~\eqref{eqn:difference_global}) causes potentially incorrect $\mathcal{O}(q^{2}_{\bot}/2Q^2)$ terms in the phase space limits.\footnote{The algebra to show this is awkward but as $\kappa_{i_{n}}$ is simply a ratio of energies, we can argue that it must be an even polynomial when expanded in small $q_{\bot}$.} These integrate to give dilogarithms in $q^{2}_{\bot}/2Q^2$ which do not contribute $\As^2L^2$ terms but rather $\As^2L^0$ terms that go to zero in both soft and collinear limits.\footnote{The recoil terms in these integrals are reducible to a few general forms. One such form is 
\begin{align}
&\int^{1}_{a} \frac{\td x}{x}\ln^2x \ln \left(x \left(1-\frac{x^2 \epsilon }{2}\right)\right) \nonumber \\
&= \frac{1}{4} \left(\text{Li}_4\left(\frac{a^2 \epsilon }{2}\right)+2 \ln ^2(a) \text{Li}_2\left(\frac{a^2 \epsilon }{2}\right)-2 \ln (a) \text{Li}_3\left(\frac{a^2 \epsilon }{2}\right)-\ln ^4(a)-\text{Li}_4\left(\frac{\epsilon }{2}\right)\right)\nonumber 
\end{align}
where $a$ parametrises the observable, $x \sim q_{\bot}/Q$ and $\epsilon$ parametrises the coefficients to the $\mathcal{O}(q^{2}_{\bot}/2Q^2)$ effects from our recoil scheme; $\epsilon = 0$ gives the leading log result. Note that all terms other than the LL result are not enhanced in the $a \rightarrow 0$ limit. See Appendix \ref{sec:Thrust} for more details.} Thus, with NLL accuracy, Eq.~\eqref{eqn:difference_global} reduces to
\begin{align}
\delta \Sigma(L) &\approx \frac{4\As^{2} \, \mathcal{C}^{2}_{\TT{F}} \, \sigma_{n_{\TT{H}}}}{\pi^{2}}  \int^{Q}_{0} \frac{\td q^{(a_{1}, b_{1})}_{1 \, \bot}}{q^{(a_{1}, b_{1})}_{1 \, \bot}} \int^{\ln Q /q^{(a_{1}, b_{1})}_{1 \, \bot}}_{-\ln Q / q^{(a_{1}, b_{1})}_{1 \, \bot}} \td y_{1} \int^{q^{(a_{1}, b_{1})}_{1 \, \bot}}_{0} \frac{\td q^{(a_{2}, 1_{2})}_{2 \, \bot}}{q^{(a_{2}, 1_{2})}_{2 \, \bot}} \int^{\ln Q /q^{(a_{2}, 1_{2})}_{2 \, \bot}}_{-\ln Q / q^{(a_{2}, 1_{2})}_{2 \, \bot}} \td y_{2} \nonumber \\
&\quad \times \int^{2\pi}_{0} \frac{\td \phi_{2}}{2 \pi} \left[ \Theta\left(e^{-L} - V(\{p\}_{2})\right) - \Theta\left(e^{-L} - V(\{p\}_{\TT{correct}})\right) \right].
\end{align}
Note that $\delta \Sigma(L)$ is only non-zero because $\{p\}_{2} \neq \{p\}_{\TT{correct}}$.
 
We will now consider several specific observables, still following \cite{Dasgupta:2018nvj}. Dasgupta et al. first consider the two-jet rate in the Cambridge algorithm. They argue that for this observable  $V(\{p_{i}\}) = \max_{i} \{p_{i \, \bot}\}$. We notice that $q^{(a_{n}, b_{n})}_{n \, \bot}$ is a Lorentz invariant. As a consequence $q^{(a_{n}, b_{n})}_{n \, \bot}$ is always larger than $q^{(a_{n+1}, b_{n+1})}_{n+1 \, \bot}$ for ourrecoil scheme, up to the neglected dilogarithmic piece. Therefore we find $V(\{p\}_{\TT{correct}}) = V(\{p\}_{2}) = q^{(a_{1}, b_{1})}_{1 \, \bot}$ and that the $\As^2 \Nc^2 L^2$ terms are correctly computed. Similarly, $V(\{p\}_{2})$ is also equal to the correct measurement function (up to neglected poly-logs) for the `fractal moment of energy-energy correlation' ($\TT{FC_{1}}$) which, in the soft-collinear limit, is given by $V(\{p_{i}\}_{\TT{correct}}) = \sum_{i}p_{i \, \bot}$. In the limit we are studying \mbox{$V(\{p_{i}\}_2) = \kappa_{a_{2}}q^{(a_{1}, b_{1})}_{1 \, \bot} + q^{(a_{2}, 1_{2})}_{2 \, \bot} \approx q^{(a_{1}, b_{1})}_{1 \, \bot} + q^{(a_{2}, 1_{2})}_{2 \, \bot} = V(\{p_{i}\}_{\TT{correct}})$}. In fact, it will be the case that for all observables built from Lorentz invariant and jet rescaling insensitive quantities\footnote{Observables not sensitive to the absolute magnitude of energy deposited in a part of a detector.} our recoil scheme is sufficient for the computation of $\As^2 \Nc^2 L^2$ terms. This being because the scheme is constructed by a Lorentz boost and a formally sub-leading re-weighting. We expect that for suitably simple observables this accuracy will also extend to higher orders, see the resummations in Appendix \ref{sec:Checks}. This discussion should be contrasted with that in Appendix \ref{sec:Spectator}, where we perform the same tests with a spectator recoil scheme \cite{Herwig_dipole_shower,Pythia8,DIRE}. In agreement with \cite{Dasgupta:2018nvj}, we find that with such a recoil scheme these observables return $V(\{p_{i}\}_2) \not\approx V(\{p_{i}\}_{\TT{correct}})$. This generates NLL errors.

\section{Conclusions}

Starting from a general algorithm designed to capture both the soft and collinear logarithms associated with the leading infra-red singularities  of scattering amplitudes, we have derived an angular ordered shower and a dipole shower. Our dipole shower is novel in the way that it partitions each dipole in order to account for longitudinal momentum conservation. This partitioning is constructed so as to ensure that the shower implements longitudinal momentum conservation in precisely the same way as the angular ordered shower does. This new dipole partitioning is similar to, but not the same as, Catani-Seymour partitioning. We complete our dipole shower by specifying the transverse recoil and phase-space. The result is a new dipole shower that formally represents an increase in accuracy when compared to the traditional parton shower models employed by many current event generators \cite{Pythia8,Herwig_dipole_shower,DIRE,Lonnblad:1992tz,Herwig_shower,Gleisberg:2008ta,Krauss:2005re,Kuhn:2000dk}. For example it will compute radiation ordered in angle at full-colour, and the leading-colour contribution associated with non-global logarithms, i.e. it will reproduce the correct leading-colour, wide-angle, soft radiation pattern beyond the two, three, and four-jet limits whilst retaining complete leading-colour, global NLLs in the two-jet limit. To our knowledge this is not achieved by other parton shower models.

However, our shower still has substantial limitations. In large part that is because it is based on a cross-section-level, semi-classical picture. Operating at cross-section level necessitates that the shower generally be defined only at leading-colour. General full-colour resummation means a more complicated, amplitude-level, approach \cite{Nagy:2015hwa,Forshaw:2019ver,Nagy:2017ggp,Becher:2016mmh,Caron-Huot:2015bja,Caron-Huot:2016tzz,Neill:2018mmj}. Certainly it would be of considerable interest to compare a parton shower defined at amplitude level, such as the CVolver shower that is currently under construction \cite{QCDTalks,HARPSTalks} or the Deductor shower \cite{Nagy:2019pjp}, with the improved dipole shower we present here.

\section*{Acknowledgements}
The authors want to thank the Erwin Schr\"odinger Institute Vienna for
support while this work was carried out.  This work has received
funding from the UK Science and Technology Facilities Council (grant
no. ST/P000800/1), the European Union’s Horizon 2020 research and
innovation programme as part of the Marie Skłodowska-Curie Innovative
Training Network MCnetITN3 (grant agreement no. 722104), and in part
by the by the COST actions CA16201 ``PARTICLEFACE'' and CA16108
``VBSCAN''. JH thanks the UK Science and Technology Facilities Council
for the award of a studentship. Thanks also to Jack Helliwell, Mike
Seymour and Andy Pilkington for comments and discussions. Finally we would also like to thank the organisers of the ``Taming the accuracy of event generators'' workshop (2020) for facilitating enlightening discussions. We especially want to  thank to Pier Monni and Gavin Salam for useful exchange on the subject.

\section*{Additional comment on PanScales}
Whilst this paper was being finalised, a study of NLL accuracy in parton showers was released by the PanScales collaboration \cite{Dasgupta:2020fwr}. There would seem to be a fair degree of similarity between the dipole shower we derive and the PanGlobal shower with $\beta = 0$ presented in \cite{Dasgupta:2020fwr}. Our recoil schemes in particular are similar. The dipole partitioning they employ also obeys the same basic properties as ours: a rapid rise to 1 in the region that the emitted parton becomes collinear (and a rapid drop to 0 in the anti-collinear region), summing the two halves of the partitioning gives unity at all points in the phase-space of emission, and in the limit that both partons in the dipole have the similar energy the partitioning divides the dipole symmetrically in the event ZMF.
\appendix

\section{The evolution equations supplementary material}
\label{sec:Supp}

\subsection{Amplitude evolution detailed definitions}
\label{sec:AevoDetails}

Before we proceed with the technical details of the PB evolution, it is necessary that we properly introduce the notation we will later be relying on. In these appendices we will often find ourselves manipulating expressions relating states of differing parton multiplicities (for instance Eq.~\eqref{eqn:evo} relates an $n_{\TT{H}}+n-1$ state to a an $n_{\TT{H}}+n$) state. We must label partons and the multiplicity of state they come from carefully since the state's multiplicity determines both the dimension of the colour-helicity space in which the state resides and the momenta of the constituent partons. To this end, we label partons with indices $i_{n},j_{n}, k_{n}, ...$ which run as $i_{n},j_{n},... \in \{1_{\TT{H}}, 2_{\TT{H}}, ... ,n_{\TT{H}}\} \cup \{1, 2, ... ,  n-1 \}$, where $\{1_{\TT{H}}, 2_{\TT{H}}, ... ,n_{\TT{H}}\}$ is the set of hard partons and $\{1, 2, ... ,  n-1 \}$ the set of partons emitted during the evolution. We use $\upsilon_{i_{n}} \in \{q, g\}$ to label the species of a parton $i_{n}$. The momentum of the $i^\text{th}$ parton in a state of multiplicity $n_\text{H}+n-1$ is $p_{i_{n}} \in \{ p \}_{n-1}= \{ P_1, P_2, \cdots P_{n_\text{H}},q_1,\cdots q_{n-1} \}$. The emission operator, $\v{D}_n$, adds a new ($n$th) parton, of four-momentum $q_n$, to the state. After considering energy-momentum conservation, the parton momentum, $q_n$, is added to the set $\{ p \}_{n-1}$, to produce the set $\{ p \}_n$. $\td R_{n}$ acts in conjunction with $\v{D}_n$ to map $\{ p \}_{n-1}$ to a new set, $\{ \tilde{p}\}_{n-1}$. The difference between these two sets is determined by the way we implement energy-momentum conservation (i.e. the recoil prescription). Following this, $\{p\}_n = \{\tilde{p}\}_{n-1} \cup \{q_n\}$ is the set of $n$ momenta including the last emission, $q_n$.

Many of the objects used in this paper carry complicated dependencies. To simplify some lengthy expressions, we will only provide the full list of arguments in an object's definition. In definitions, we will write every object as some $f(x; \{y\})$, where $x$ is the evolution variable on which $f$ depends and the set $\{y\}$ itemises the complete dependences of $f$. In all expressions subsequent to the definition we will drop the $\{y\}$ dependence and only write $f(x)$. We can do this safely as, following the initial definition of an object, each object can always be uniquely determined by the subscripts and superscripts we provide.

In Section \ref{sec:AmpEvo} we gave an overview of the roles of $\v{D}_{n}$, $\int \td R_{n}$ and $\v{\Gamma}_{n}$. Let us now define these operators more carefully\footnote{For pedagogical reviews of the colour-helicity operators relevant in the definition of these operators see \cite{Forshaw:2019ver,SoftEvolutionAlgorithm,QCD_colour_flow}.}
\begin{align}
&\v{D}_{n}(q_{n \, \bot};q_{n} \cup \{\tilde{p}\}_{n-1}) \, \v{O} \, \v{D}^{\dagger}_{n}(q_{n \, \bot}; q_{n} \cup \{\tilde{p}\}_{n-1}) = \nonumber \\
& \sum_{i_{n}, j_{n}} \int  \delta q^{(i_{n}, j_{n})}_{n \, \bot} (q_{n \, \bot}) \, \v{S}^{i_{n}}_{n} \, \v{O} \, \v{S}^{j_{n} \, \dagger}_{n} + \sum_{i_{n}} \int \delta q^{(i_{n}, \vec{n})}_{n \, \bot} (q_{n \, \bot}) \, \v{C}^{i_{n}}_{n} \, \v{O} \, \v{C}^{i_{n} \, \dagger}_{n},
\end{align}
where $\v{O}$ is some operator in the colour-helicity space and where we have used a shorthand notation to help save space
\begin{align}
\delta x(y) \equiv \td x \, \delta(x - y).
\end{align}
Delta functions of this form are used to carry the frame dependence of the ordering variable in a compact form. Physically, $\v{S}^{i_{n}}_{n}$ emits a soft parton from the parton labelled $i_{n}$. These soft partons take the form of interference terms in the evolution. Note that, due to our choice of ordering variable, $\v{S}^{i_{n}}_{n}$ cannot completely factorise from $\v{S}^{j_{n} \, \dagger}_{n}$ as both depend on the momenta $(q^{(i_{n},j_{n})}_{n \, \bot})^{2}$ (defined below). They have been written in this separated form to reflect their operator structure in the colour-helicity space. $\v{C}^{i_{n}}_{n}$ emits a collinear parton from the parton labelled $i_{n}$. The following two definitions for transverse momenta are used as ordering variables for soft and collinear emissions respectively,
\begin{align}
(q^{(i_{n},j_{n})}_{n \, \bot})^{2} = \frac{2(p_{i_{n}}\cdot q_{n})(p_{\, j_{n}} \cdot q_{n})}{p_{i_{n}}\cdot p_{\, j_{n}}}, \quad \TT{and} \quad (q^{(i_{n},\vec{n})}_{n \, \bot})^{2} = \frac{2(p_{i_{n}}\cdot q_{n})(n \cdot q_{n})}{p_{i_{n}}\cdot n},
\end{align}
where $n$ is a light-like reference vector. The choice of $n$ is determined by how recoil is handled in the evolution and is often taken to be in the backwards direction relative to $p_{i_{n}}$. 
Strictly speaking, recoil cannot be entirely factorised from each $\v{D}_{n}$ however the way in which it acts in each $\v{D}_{n}$ follows a simple pattern. Thus we have used the recoil measure $\td R_{n}$ as an abridged notation. It is defined to act by the following rules
\begin{align}
\td R_{n} \; \v{S}^{i_{n}}_{n} \, \v{O} \, \v{S}^{j_{n} \, \dagger}_{n} \equiv \bigg(\prod_{i_{n}} \td^4 p_{i_{n}} \bigg) \; \mathfrak{R}^{\TT{soft}}_{i_{n}j_{n}} \; \v{S}^{i_{n}}_{n} \, \v{O} \, \v{S}^{j_{n} \, \dagger}_{n}, \nonumber \\
\td R_{n} \; \v{C}^{i_{n}}_{n} \, \v{O} \, \v{C}^{i_{n} \, \dagger}_{n} \equiv \bigg(\prod_{i_{n}} \td^4 p_{i_{n}} \bigg) \; \mathfrak{R}^{\TT{coll}}_{i_{n}} \; \v{C}^{i_{n}}_{n} \, \v{O} \, \v{C}^{i_{n} \, \dagger}_{n}.
\end{align}
$\mathfrak{R}^{\TT{soft}}_{i_{n}j_{n}}$ and $\mathfrak{R}^{\TT{coll}}_{i_{n}}$ contain the necessary delta functions and kinematic pre-factors needed to account for energy-momentum conservation. They are discussed in Section \ref{sec:Correcting} and further examples are given in \cite{Forshaw:2019ver}.\footnote{In \cite{Forshaw:2019ver} $\mathfrak{R}^{\TT{soft}}_{i_{n}j_{n}}$ and $\mathfrak{R}^{\TT{coll}}_{i_{n}}$ are written as $\mathfrak{R}^{\TT{soft}\,  *}_{n \, j_{n}}\,\mathfrak{R}^{\TT{soft}}_{n \, i_{n}}$ and $\mathfrak{R}^{\TT{coll}\, *}_{n \, i_{n}}\,\mathfrak{R}^{\TT{coll}}_{n \, i_{n}}$ respectively.} Explicit expressions defining $\v{S}^{i_{n}}_{n}$ and $\v{C}^{i_{n}}_{n}$ are lengthy and can be found in \cite{Forshaw:2019ver}.
Finally,
\begin{align}
&\v{\Gamma}_{n}(q_{\bot}; \{p\}_{n}) = - \frac{\As}{\pi}\int \frac{\td S^{(q)}_{2} }{4\pi} \, \tfrac{1}{2}\v{D}^{2}_{n}(q_{\bot}) \, \Theta_{\TT{on\, shell}} + \frac{\As}{2\pi} \sum_{i_{n+1},j_{n+1}}\mathbb{T}^{g}_{i_{n+1}} \cdot \mathbb{T}^{g}_{j_{n+1}}  \, i\pi \, \tilde{\delta}_{i_{n+1}j_{n+1}}, \nonumber \\
&\tfrac{1}{2}\v{D}^{2}_{n}(q_{\bot}; q \cup \{p\}_{n}) = \int \td R_{n+1} \; \tfrac{1}{2} \TT{Final} \left[ \v{D}_{n+1}(q_{\bot}) \cdot \v{D}_{n+1}(q_{\bot})\right]. \label{eqn:Gamma}
\end{align}
$\TT{Final}[...]$ indicates that the enclosed operators should act on any incoming partons as if they were in the final state (see Eq.~A.1 in \cite{Forshaw:2019ver}, which defines the operators from which $\v{D}_{n+1}$ is constructed, in this context $\TT{Final}[\delta^{\TT{initial}}_j] = 0$ and $\TT{Final}[\delta^{\TT{final}}_j] = 1$ for all $j$). $\Theta_{\TT{on\, shell}}$ is our short-hand notation for the inclusion of the theta functions necessary for restricting the range of integration to the phase-space for an on-shell parton. These are also specified fully in \cite{Forshaw:2019ver} (see functions $\theta_{ij}$ and $\theta_{i}$ in Section 2). $\tilde{\delta}_{i_{n+1}j_{n+1}} = 1$ if both partons $i$, $j$ are incoming or both outgoing and $\tilde{\delta}_{ij} = 0$ otherwise.

We ought to remark on the fact that  $q_{n \, \bot}$ is not equivalent to the dipole transverse momentum derived in \cite{ColoumbGluonsOrdering,ColoumbGluonsOrderingLetter}. The latter was derived using fixed-order perturbation theory and is an amplitude-level object that acts to determine the limits on loop integrals. We have not yet figured out a way to include this physics within our algorithm, though we note that it is a higher-order effect.

\subsubsection{Computing observables \label{sec:hadrons}}
In the main text our focus is on dressing $e^+ e^- \to q \bar{q}$. The formalism is more general and can be used to compute observables in hadron-hadron collisions using
\begin{align}
\td \sigma_{n} = & \left(\prod^{n}_{i = 1} \td \Pi_{i} \right) \Tr \, \v{A}_{n}(\mu; \{p\}_{n}),  \nonumber \\
\Sigma(\mu; \{p\}_{0}, \{v\}) = & \int \sum_{n} \td \sigma_{n} \star \left\{\prod_{i \in \TT{initial}}f_{\upsilon_{i}}\left(\frac{x_{i}}{z_{i_1}z_{i_2}...},\mu\right)\right\}\, u_{n} (\{p\}_{n}, \{v\}), \label{eqn:cross-section}
\end{align}
where $f_{\upsilon_{i}}(x_{i}, \mu)$ are the parton distribution functions (PDFs) with momentum fractions $x_{i}$ and $u_{n} (\{p\}_{n}, \{v\})$ is the $(n_{\TT{H}}+n)$-body measurement function for an observable described by parameters $v_{i} \in \{v\}$. Note that $\Sigma$ is differential in hard process momenta, and that it should be multiplied by the necessary flux factors as necessary. The star operation is defined in Section 4 of \cite{Forshaw:2019ver} but in essence assigns PDF type to a given partonic leg (gluon or quark). In this paper, every concrete use of our formalism concerns the showering of an $e^{+}e^{-}$ hard process and so we will not expand further on the treatment of DGLAP evolution.

\subsection{Derivation of the angular ordered shower}
\label{sec:CBderivation}

This section derives an angular ordered shower from Eq.~\eqref{eqn:evo}. It is split in three parts. Part one forms the main derivation, however it will state some results without proof (when these results are themselves laborious to prove). The subsection following presents the limitations of this derivation. Finally the last subsection fills in the gaps. We will focus on $e^+e^- \rightarrow q \bar{q}$ as the hard processes, and at the end we will sketch the extension to other hard processes.

We begin with the amplitude evolution equation, Eq.~\eqref{eqn:evo}, and introduce an azimuthal averaging operation $\<\>_{1,...,n}$ which averages the lab frame dipole azimuths of partons $1$ to $n$, i.e.
$$ \<f\>_{1,...,n} = \int \frac{\td \phi_{n}}{2\pi} ... \int \frac{\td \phi_{1}}{2\pi} f(\phi_{1},...,\phi_{n}).$$
Implicit in this operation is also spin averaging when acting on spin-dependent operators, as discussed in Appendix \ref{sec:Spin}. 
To keep things simple, we will proceed in this section without discussing any dependence on the observable, which means we are implicitly assuming the observable is not a function of the parton azimuths. We devote the next sub-section to addressing this. After averaging, Eq.~\eqref{eqn:evo} becomes
\begin{align}
&q_{\bot}\pdf{\<\v{A}_{n}(q_{\bot})\>_{1,...,n}}{q_{\bot}} = - \v{\Gamma}_{n}(q_{\bot}) \, \<\v{A}_{n}(q_{\bot})\>_{1,...,n} \nonumber -  \<\v{A}_{n}(q_{\bot})\>_{1,...,n} \, \v{\Gamma}^{\dagger}_{n}(q_{\bot}) \\ 
& + \int \prod_{i_{n}} \td^4 p_{i_{n}} \; \sum_{i_{n},j_{n}} \int \delta q^{(i_{n}j_{n})}_{n \, \bot} (q_{n \, \bot}) \, \<s_{i_{n},j_{n}}\>_{n}  \, \mathbb{T}_{i_{n}} \, \<\v{A}_{n-1} (q_{n \, \bot})\>_{1,...,n-1} \, \mathbb{T}^{\dagger}_{j_{n}} \; q_{\bot} \; \delta(q_{\bot} - q_{n \, \bot}) \nonumber \\ 
& + \int \prod_{i_{n}} \td^4 p_{i_{n}} \; \sum_{j_{n}}\int \delta q^{(j_{n}, \vec{n})}_{n \, \bot}(q_{n \, \bot}) \, \<c_{j_{n}}\>_{n}  \, \mathbb{T}_{j_{n}} \, \<\v{A}_{n-1} (q_{n \, \bot})\>_{1,...,n-1} \, \mathbb{T}^{\dagger}_{j_{n}} \; q_{\bot} \; \delta(q_{\bot} - q_{n \, \bot}), \label{eqn:averaged-evo}
\end{align}
where $s_{i_{n},j_{n}}$ and $c_{j_{n}}$ are the spin-averaged kinematic factors associated with a soft or collinear emission respectively (they will be manipulated into the form of collinear splitting functions shortly). They are defined through the relations
\begin{align}
s_{i_{n},j_{n}} \; \mathbb{T}_{j_{n}} \cdot \mathbb{T}_{i_{n}} &\equiv \frac{1}{2}\sum_{h_{i_{n}}} \< h_{i_{n}} \rkl  \v{S}^{j_{n}}_{n} \cdot \v{S}^{i_{n}}_{n} \lkl h_{i_{n}} \> \; \mathfrak{R}^{\TT{soft}}_{i_{n}j_{n}}, \nonumber \\
c_{j_{n}} \; \mathbb{T}_{i_{n}} \cdot \mathbb{T}_{i_{n}} &\equiv  \frac{1}{2}\sum_{h_{i_{n}}} \< h_{i_{n}} \rkl \v{C}^{i_{n}}_{n} \cdot \v{C}^{i_{n}}_{n} \lkl h_{i_{n}} \>  \; \mathfrak{R}^{\TT{coll}}_{i_{n}}.
\end{align}
We observe that $c_{j_{n}} = \<c_{j_{n}}\>_{n}$ provided $\mathfrak{R}^{\TT{coll}}_{i_{n}}$ is independent of the emission's azimuth (spin correlations provide the only azimuthal dependence for collinear emissions). In Section \ref{sec:aawte} we show that\footnote{Under the assumption that $\mathfrak{R}^{\TT{soft}}_{i_{n} j_{n}}$ is independent of the azimuth up to $\mathcal{O}(1)$ terms, which is true for the two recoil schemes we discuss in this paper.}
\begin{align}
&\int \delta q^{(i_{n}j_{n})}_{n \, \bot} (q_{n \, \bot}) \, \<s_{i_{n},j_{n}}\>_{n} = \nonumber \\
&-\int \prod_{i_{n}} \td^4 p_{i_{n}}  \left(\ \<P_{i_{n}j_{n}} \>_{\phi_{n, i_{n}}} \<\Theta_{\TT{on\, shell}} \>_{\phi_{n, i_{n}}} + \<P_{j_{n}i_{n}} \>_{\phi_{n, j_{n}}} \<\Theta_{\TT{on\, shell}} \>_{\phi_{n, j_{n}}} \right)\mathfrak{R}^{\TT{soft}}_{i_{n} j_{n}} + \mathcal{O}(1),
\end{align}
where
\begin{align}
\<P_{i_{n}j_{n}} \>_{\phi_{n, i_{n}}} = \frac{\Theta(\theta_{j_{n},i_{n}} - \theta_{n,i_{n}})}{1- \cos\theta_{n,i_{n}}}. \label{eqn:P}
\end{align}
The angles in Eq.~\eqref{eqn:P} are defined in Figure \ref{fig:angles}. $\<\Theta_{\TT{on\, shell}} \>_{\phi_{n, i_{n}}}$ contains the necessary theta functions to constrain the phase-space of parton $q_{n}$ so that it is real and on-shell, encoding the phase-space limits for energy conservation. Its lengthy definition can also be found in Section \ref{sec:aawte}. The presence of the functions $\<P_{i_{n}j_{n}} \>_{\phi_{n, i_{n}}}$ enforces an angular ordering, secondary to the $k_{\bot}$ ordering. To bring this ordering to the fore, we now change variables:
$$
q^{2}_{\bot} = E^{2}_{n}\sin^{2} \theta = E^{2}_{n}(1-(1-\zeta)^{2}), \qquad q_{\bot} \pdf{}{q_{\bot}} = \frac{\zeta(2-\zeta)}{1-\zeta}\pdf{}{\zeta}
$$
and define $\zeta_{n, i_{n}} =  1 - \cos \theta_{n, i_{n}}$. In these new variables Eq.~\eqref{eqn:averaged-evo} becomes\footnote{ $\v{\Gamma}_{n}(\zeta)$ is defined as $\v{\Gamma}_{n}(q_{\bot})$ after the change of variables has been made rather than naively swapping out the argument.}
\begin{align}
&\zeta \pdf{\<\v{A}_{n}(\zeta)\>_{1,...,n}}{\zeta} \approx - \v{\Gamma}_{n}(\zeta) \, \<\v{A}_{n}(\zeta)\>_{1,...,n} \nonumber -  \<\v{A}_{n}(\zeta)\>_{1,...,n} \, \v{\Gamma}^{\dagger}_{n}(\zeta) \\ 
& - \int \prod_{i_{n}} \td^4 p_{i_{n}} \; \sum_{i_{n},j_{n}} 2 \<P_{i_{n}j_{n}} \>_{\phi_{n, i_{n}}} \<\Theta_{\TT{on\, shell}} \>_{\phi_{n, i_{n}}} \mathfrak{R}^{\TT{soft}}_{i_{n} j_{n}} \, \mathbb{T}_{i_{n}} \, \<\v{A}_{n-1} (q_{n \, \bot})\>_{1,...,n-1} \, \mathbb{T}^{\dagger}_{j_{n}} \;  \zeta_{n, i_{n}} \; \delta(\zeta - \zeta_{n, i_{n}}) \nonumber \\ 
& + \int \prod_{i_{n}} \td^4 p_{i_{n}} \; \sum_{j_{n}} \, \<P_{j_{n}}\>_{n} \<\Theta_{\TT{on\, shell}} \>_{\phi_{n, j_{n}}} \mathfrak{R}^{\TT{col}}_{j_{n}} \, \mathbb{T}_{j_{n}} \, \<\v{A}_{n-1} (q_{n \, \bot})\>_{1,...,n-1} \, \mathbb{T}^{\dagger}_{j_{n}} \; \zeta_{n, j_{n}} \; \delta(\zeta - \zeta_{n, j_{n}}). \label{eqn:averaged-angle-ordered}
\end{align}
Here we have used 
$$\<c_{j_{n}}\>_{n} \approx \<P_{j_{n}}\>_{n} \<\Theta_{\TT{on\, shell}} \>_{\phi_{n, j_{n}}} \mathfrak{R}^{\TT{col}}_{j_{n}}, $$
where $\<P_{j_{n}}\>_{n}$ is a sum over collinear splitting functions with the soft divergences subtracted away, e.g. when $j_{n}$ is a quark, $\<P_{j_{n}}\>_{n}(z) = (1-z)\overline{\mathcal{P}}_{qq}/2$ where $\overline{\mathcal{P}}_{qq}(z) = - (1+z)$. The details can be found in Appendix A of \cite{Forshaw:2019ver}. We will formulate the evolution in terms of the full splitting functions once equations have been reduced enough that it becomes convenient to do so.

Using the strongly ordered approximation, $\zeta_{1}\gg\zeta_{2}\gg...$\footnote{When working in a frame that ensures $i$ and $j$ are back to back, the theta function is saturated without approximation. In this derivation we are concerned with $e^+e^- \to q \bar{q}$. Thus we can saturate the theta function for emissions from the primary hard partons, so that they are handled without approximation. This means we pick the backwards direction ($n$) (used to define kinematic variables for emissions in a jet) to be in the direction of the other jet. This in turn fixes the definition for the momentum fraction used in later equations: $z_{n} = \frac{\tilde{p}_j \cdot n}{p_{j} \cdot n}$. When working beyond the two-jet limit, tricks can be played to further saturate the theta function using knowledge of the hard process colour flows.}, 
\begin{align}
\<P_{ij} \>_{\phi_{n, i_{n}}} = \frac{\Theta(\theta_{j_{n},i_{n}} - \theta_{n,i_{n}})E_{q}E_{i}}{q \cdot p_{i}} \approx \frac{1}{\zeta_{q, i}}.
\end{align}
Also using strong ordering, the leading part of $\<\Theta_{\TT{on\, shell}} \>_{\phi_{n, i_{n}}}$ does not depend on $j_{n}$ and $\mathfrak{R}^{\TT{soft}}_{i_{n}j_{n}}$ can be chosen so that its leading part can be factorised from the sum over $j_{n}$ as $$\frac{1}{\zeta_{n, i_{n}}} \<\Theta_{\TT{on\, shell}} \>_{\phi_{n, i_{n}}} \mathfrak{R}^{\TT{soft}}_{i_{n}j_{n}} \approx \frac{1}{\zeta_{n, i_{n}}} \<\Theta_{\TT{on\, shell}} \>_{\phi_{n, i_{n}}} \mathfrak{R}^{\TT{col}}_{i_{n}}.$$ Using these simplifications we can apply colour conservation and, by re-labelling indices, write
\begin{align}
\zeta \pdf{\<\v{A}_{n}(\zeta)\>_{1,...,n}}{\zeta} \approx & - \v{\Gamma}_{n}(\zeta) \, \<\v{A}_{n}(\zeta)\>_{1,...,n} -  \<\v{A}_{n}(\zeta)\>_{1,...,n} \, \v{\Gamma}^{\dagger}_{n}(\zeta) + \int \prod_{i_{n}} \td^4 p_{i_{n}} \; \nonumber \\ 
& \times \sum_{j_{n}} \left(\<P_{j_{n}}\>_{n} \<\Theta_{\TT{on\, shell}} \>_{\phi_{n, j_{n}}} + 2 \frac{1}{\zeta_{n, j_{n}}} \<\Theta_{\TT{on\, shell}} \>_{\phi_{n, j_{n}}} \right) \mathfrak{R}^{\TT{col}}_{j_{n}}  \nonumber \\ 
& \times \mathbb{T}_{j_{n}} \, \<\v{A}_{n-1} (q_{n \, \bot})\>_{1,...,n-1} \, \mathbb{T}^{\dagger}_{j_{n}} \; \zeta_{n, j_{n}} \; \delta(\zeta - \zeta_{n, j_{n}}).
\end{align}
By recognising the evolution will become entirely colour-diagonal once the trace is taken, we can diagonalise the colour structures. In turn this allows us to group the soft evolution kernels and the collinear ones into splitting functions. We find  
\begin{align}
\zeta \pdf{\, \Tr\<\v{A}_{n}(\zeta)\>_{1,...,n}}{\zeta} \approx &- 2\Gamma_{n}(\zeta) \, \Tr\<\v{A}_{n}(\zeta)\>_{1,...,n} \nonumber  + \int \prod_{i_{n}} \td^4 p_{i_{n}} \; \frac{(1-z_{n})}{2} \sum_{j_{n}} \sum_{\upsilon \in \{q,g\}} \mathcal{P}_{\upsilon \upsilon_{j_{n}}}(z_{n})\\ 
& \times  \<\Theta_{\TT{on\, shell}}\>_{\phi_{n, j_{n}}} \mathfrak{R}^{\TT{col}}_{j_{n}} \, \Tr\<\v{A}_{n-1} (q_{n \, \bot})\>_{1,...,n-1} \,  \; \zeta_{n, j_{n}} \; \delta(\zeta - \zeta_{n, j_{n}}). \label{eq:dipt}
\end{align}
$\mathcal{P}_{\upsilon \upsilon_{j_{n}}}(z_{n})$ are the usual DGLAP splitting functions, e.g. $\mathcal{P}_{q q}(z_{n}) = \mathcal{C}_{\TT{F}} \frac{1+ z^{2}_{n}}{1-z_{n}}$. Here we have used $\upsilon_{j_{n}}$ to label the species of parton $j_n$ and $\upsilon$ to label the state $j_n$ transitions to; if $\upsilon_{j_{n}} = q$ then $\upsilon=q$ and if $\upsilon_{j_{n}} = g$ then $\upsilon=q,g$. $z_{n}$ is the momentum faction of parton $n$, i.e. if we have a collinear splitting that induces $j_{n-1} \rightarrow j_{n} \, n$ then $p_{j_{n}} \approx z_{n} p_{j_{n-1}}$ and $q_{n} \approx (1- z_{n}) p_{j_{n-1}}$. We specifically require that $z_{n} = \frac{\tilde{p}_{j_{n}} \cdot n}{p_{j_{n}} \cdot n}$ where $n$ is a light-like vector pointing along the primary axis of the jet from which parton ${j_{n}}$ does not stem.

We can make connection to squared matrix elements by letting
\begin{align}
\<|M_{n}|^{2}\>_{1,...,n} = \left(\frac{2\As}{\pi}\right)^{n} \prod_{i=1}^{n}(1-z_{i})^{-1}\Tr \<\v{A}_{n}(\zeta)\>_{1,...,n},
\end{align}
from which we find the evolution equation for a final-state angular ordered shower with a conventional phase-space for a coherent shower in $\td z$. After which, Eq.~\eqref{eq:dipt} can be written as in Eq.~\eqref{eqn:CB} after $\<|M_{n}|^{2}\>_{1,...,n} \rightarrow \sum_{j_{1},...,j_{n}}\<|\mathcal{M}_{n}|^{2}\>_{1,...,n}$.

\subsubsection{Observable dependence and logarithmic accuracy}

In the previous discussion we derived $\<|M_{n}|^{2}\>_{1,...,n}$ from  Eq.~\eqref{eqn:evo}. However, as we highlighted at the beginning, a full treatment should compute $\<|M_{n}|^{2} \, u(\{p\}_{n},\{v\})\>_{1,...,n}$ where $u(\{p\}_{n};\{v\})$ is the measurement function for an observable defined by parameters $v \in \{v\}$. We want to know to what accuracy is 
\begin{align}
	\<|M_{n}|^{2} \, u(\{p\}_{n})\>_{1,...,n} \approx \int \prod^{n}_{i=1} \frac{\td \phi_{i}}{2\pi} \; \<|M_{n}|^{2}\>_{1,...,n} u(\{p\}_{n}) = \<|M_{n}|^{2}\>_{1,...,n} \< u(\{p\}_{n})\>_{1,...,n}.
\end{align}
We can start by considering the effects of only averaging over the $n$th parton and use the following identity 
\begin{align}
	\<|M_{n}|^{2} \, u(\{p\}_{n})\>_{n}= & \<|M_{n}|^{2}\>_{n} \, \< u(\{p\}_{n})\>_{n} \nonumber \\
	& + \sigma_{n}(|M_{n}|^{2}) \, \sigma_{n}(u(\{p\}_{n}) \, \TT{Cor}_{n}(|M_{n}|^{2},u(\{p\}_{n})),
\end{align}
where $\sigma_{n}(x) =  \sqrt{\< x^2 \>_{n} - \< x \>^2_{n}}$ and $\TT{Cor}_{n}(x(\phi_{n}),y(\phi_{n}))$ is the correlation function of $x$ and $y$ under the variation of $\phi_{n}$. Both $|\TT{Cor}_{n}(|M_{n}|^{2},u(\{p\}_{n}))|$ and $\sigma_{n}(u(\{p\}_{n})$ are smaller than unity\footnote{This makes the weak assumption that the measurement function, $u(\{p\}_{n})$ is bounded.}. Next we can consider averaging over both the $n$th and $(n-1)$th partons:
\begin{align}
	\<|M_{n}|^{2} \, u(\{p\}_{n})\>_{n-1,n}= & \<\<|M_{n}|^{2}\>_{n} \< u(\{p\}_{n})\>_{n}\>_{n-1} \nonumber \\
	& + \<\sigma_{n}(|M_{n}|^{2}) \, \sigma_{n}(u(\{p\}_{n}) \, \TT{Cor}_{n}(|M_{n}|^{2},u(\{p\}_{n}))\>_{n-1},
\end{align}
where 
\begin{align}
	&\<\<|M_{n}|^{2}\>_{n} \< u(\{p\}_{n})\>_{n}\>_{n-1}=  \<|M_{n}|^{2}\>_{n-1,n} \< u(\{p\}_{n})\>_{n-1,n} \nonumber \\
	& \qqqquad + \sigma_{n-1}(\<|M_{n}|^{2}\>_{n})\sigma_{n-1}(\< u(\{p\}_{n})\>_{n}) \TT{Cor}_{n}(\<|M_{n}|^{2}\>_{n},\< u(\{p\}_{n})\>_{n}).
\end{align}
This can be iterated to give 
\begin{align}
	&\<|M_{n}|^{2} \, u(\{p\}_{n})\>_{1,...,n} = \<|M_{n}|^{2}\>_{1,...,n} \< u(\{p\}_{n})\>_{1,...,n} \nonumber \\ & \qquad + \sum^{n}_{m = 1} \sigma_{m}(\<|M_{n}|^{2}\>_{1,...,n}) \, \sigma_{m}(\<u(\{p\}_{n})\>_{1,...,n}) \, \TT{Cor}_{m}(\<|M_{n}|^{2}\>_{1,...,n},\<u(\{p\}_{n})\>_{1,...,n}) \nonumber \\ 
	& \qquad + \TT{higher} \; \TT{order} \; \TT{correlations}.
\end{align}
We have been slightly lazy with notation; it is implicit that $$\sigma_{m}(\<x\>_{1,...,n}) \equiv \sigma_{m}(\<x\>_{1,...,m-1,m+1,...,n}).$$ The important question is whether the correlations can provide a logarithmic enhancement to the observable. This is obviously an observable dependent statement. To progress we will place some assumptions on the observable. If the observable is such that the correlation term's contribution to the cross section is suppressed relative to $\<|M_{n}|^{2}\>_{m} \< u(\{p\}_{n})\>_{m}$, we can approximate $\<|M_{n}|^{2} \, u(\{p\}_{n})\>_{1,...,n}$ by only keeping the first order correlations, since second order correlations will necessarily be even further suppressed. The approximation assumed by coherent branching is to neglect correlation terms altogether. Let us look at the $n=m=1$ term for thrust. At this order $u(\{p\}_{n})$ is not a function of the azimuth and so $\sigma_{1}(u(\{p\}_{1})) = 0$. As the observable exponentiates \cite{CATANI1992419,Banfi:2004yd}, this is sufficient to guarantee that it can be computed to NLL using the coherent branching formalism (these last two sentences are an abridged form of the argument in \cite{CATANI1992419}). For contrast, let us look at the $n=m=2$ term in the computation of gaps-between-jets, with the same hard process. The pertinent measurement functions are
\begin{align}
	u_{n} (\{p\}_{n}) = \prod_{m=1}^{n}(\Theta_{\TT{out}}(q_{m})+\Theta_{\TT{in}}(q_{m})\Theta(Q_{0}-q_{m, \bot})),
\end{align} 
where $\Theta_{\TT{in/out}}(q_{m})$ is unity when parton $m$ is in/out the rapidity region between the two highest $p_{\TT{T}}$ jets and zero otherwise. In the following subsection, we compute all the ingredients for $\sigma_{2}(\<|M_{2}|^{2}\>_{1})$. It is reasonably easy to argue (though less easy to compute) that, unless suppressed by multiplicative factors in $\sigma_{2}(\<u(\{p\}_{2})\>_{1})$ and correlation functions, $\sigma_{2}(\<|M_{2}|^{2}\>_{1})$ terms can contribute fourth-order, infra-red poles and with them leading logarithms. By considering the variation of $\phi_{2}$, it is also simple to convince oneself that the correlation function must be finite and positive. So, if angular ordering is to adequately describe this observable, it must be the role of $\sigma_{2}(\<u(\{p\}_{2})\>_{1})$ to screen against contaminating logarithms. This means we only need to test to see if $\sigma_{2}(\<u(\{p\}_{2})\>_{1})$ is non-zero:
\begin{align}
	&\sigma_{2}(\<u(\{p\}_{2})\>_{1}) = \sqrt{\< u(\{p\}_{2})\>_{1,2}\left(1-\frac{\< u(\{p\}_{2})\>_{1,2}}{\< u(\{p\}_{1})\>_{1}}\right)}, \nonumber \\
	&= (\Theta_{\TT{out}}(q_{1})+\Theta_{\TT{in}}(q_{1})\Theta(Q_{0}-q_{1, \bot})) \nonumber \\ 
	& \quad \times \sqrt{\< \Theta_{\TT{out}}(q_{2})+\Theta_{\TT{in}}(q_{2})\Theta(Q_{0}-q_{2, \bot})\>_{2}\left(1-\< \Theta_{\TT{out}}(q_{2})+\Theta_{\TT{in}}(q_{2})\Theta(Q_{0}-q_{2, \bot})\>_{2}\right)} \neq 0.
\end{align}
Furthermore, not only is this non-zero but it contains non-vanishing terms in $\Theta_{\TT{in}}(q_{1})\Theta_{\TT{out}}(q_{2})$. While these terms do screen against fourth order poles and logarithms, they are crucial for the computation of the $\As^2 L^2$ non-global logarithms. As such, a coherence branching algorithm (that makes usage of azimuthal averaging) cannot compute the leading logarithms to gaps-between-jets, as it certainly gets the numerical coefficient to non-global pieces incorrect. This is a general feature: coherent branching will fail to capture leading, non-global logarithms (though in most cases these logarithms are sub-leading in the computation of the overall cross section). This has been previously observed in \cite{Dasgupta:2002bw,Banfi:2006gy}, where the effect of the missing correlations was computed numerically to all-orders. They found that, though the missing correlations are a formally leading effect, phenomenologically their effect is $ <10 \% $. As is widely known, we observe that coherent branching is always capable of calculating logarithms up to $\As^n L^{2n -1}$ in observables for which $\As^n L^{2n}$ is the leading logarithm.

\subsubsection{Azimuthal averaging \label{sec:aawte}}

In this appendix we will fill in the details on the azimuthal averaging of the evolution kernels. The general procedure for azimuthal averaging is well known \cite{CATANI1991635} textbook material \cite{Ellis:1991qj,Dokshitzer:1991wu}. However, the procedure is less widely discussed taking into account phase-space limits and momentum maps. In this section we provide a more careful treatment than the textbook one. We begin by looking at the following integral (which corresponds to the integrated soft emission spectrum),
\begin{align}
&\int \frac{\td S^{(q_{n})}_{2}}{4\pi} \tfrac{1}{2}\v{S}^{j_{n}}_{n} \cdot \v{S}^{i_{n}}_{n} \propto \int \frac{\td S^{(q_{n})}_{2} }{4\pi} \int \frac{\delta q^{(i_{n}, j_{n})}_{n \, \bot} (q_{\bot}) }{q_{\bot}} \,2\, \Theta_{\TT{on} \, \TT{shell}} \nonumber \\ & \qqqquad = \int \frac{\td \Omega_{q_{n}} }{4\pi} \int \frac{\td E_{q_{n}}}{E_{q_{n}}} E^{2}_{q_{n}} \frac{\tilde{p}_{i_{n}} \cdot \tilde{p}_{j_{n}}}{\tilde{p}_{i_{n}}\cdot q_{n} \tilde{p}_{j_{n}} \cdot q_{n}} \, \Theta_{\TT{on} \, \TT{shell}} \; \delta(q^{(i_{n}, j_{n})}_{n \, \bot} - q_{\bot}), \label{eqn:change_to_E}
\end{align}
where $E_{q_{n}}$ is the energy of parton $q$ and $\td \Omega_{q_{n}}$ is solid angle in the frame which $E_{q_{n}}$ is measured. We can regroup the dipole kinematics as
\begin{align}
\text{Eq}.~\eqref{eqn:change_to_E} &= \int \frac{\td \Omega_{q_{n}}}{4\pi} \int \frac{\td E_{q_{n}}}{E_{q_{n}}} \left(P_{i_{n}j_{n}} + P_{j_{n}i_{n}}\right) \, \Theta_{\TT{on} \, \TT{shell}} \; \delta(q^{(i_{n}, j_{n})}_{n \, \bot} - q_{\bot}), \nonumber \\
2 P_{i_{n}j_{n}} &= \frac{n_{i_{n}} \cdot n_{j_{n}}-n_{i_{n}}\cdot n}{n_{i_{n}}\cdot n \; n_{j_{n}} \cdot n} + \frac{1}{n_{i_{n}}\cdot n},
\end{align}
where $n_{i_{n}} = p_{i_{n}}/E_{i_{n}}$. The two terms in this integral are symmetric under the exchange of $i$ and $j$ and so we shall focus only on the first:
\begin{align}
\int \frac{\td E_{q_{n}}}{E_{q_{n}}} &\int \frac{\td \Omega_{q_{n}} }{4\pi} \, P_{i_{n}j_{n}} \, \Theta_{\TT{on} \, \TT{shell}}\; \delta(q^{(i_{n}, j_{n})}_{n \, \bot} - q_{\bot}) \nonumber \\
&= \int \frac{\td E^{2}_{q_{n}}}{2E^{2}_{q_{n}}}  \int \frac{\sin \theta_{n,i_{n}} \, \td \theta_{n,i_{n}} \td \phi_{n, i_{n}} }{4\pi} \, P_{i_{n}j_{n}} \, \Theta_{\TT{on} \, \TT{shell}} \, 2 q_{\bot} \, \delta\left((q^{(i_{n}, j_{n})}_{n \, \bot})^{2} - q^{2}_{\bot}\right).\label{eqn:Pij}
\end{align}
To compute this the integral we take
$n_{i_{n}} = (1, 0 , 0 , 1)$, $n_{j_{n}} = (1, \; \sin \theta_{j_{n},i_{n}}, \; 0, \; \cos \theta_{j_{n},i_{n}})$, and $n = (1, \; \sin \theta_{n,i_{n}} \; \cos \phi_{n, i_{n}} , \; \sin \theta_{n,i_{n}} \; \sin \phi_{n, i_{n}} , \; \cos \theta_{n,i_{n}})$. In this basis
\begin{align}
(q^{(i_{n}, j_{n})}_{n \, \bot})^{2} &= E^{2}_{q_{n}} \frac{2(1 - \cos \theta_{n,i_{n}})(1 - \sin \theta_{n,i_{n}} \cos \phi_{n, i_{n}} \sin \theta_{j_{n},i_{n}} - \cos \theta_{j_{n},i_{n}} \cos \theta_{n,i_{n}})}{1 - \cos \theta_{j_{n},i_{n}}} \nonumber \\ & \equiv E^{2}_{q_{n}} \kappa_{i,j,n},
\end{align}
and
\begin{align}
\TT{Eq}.~\eqref{eqn:Pij} &=  \int \frac{\sin \theta_{n,i_{n}} \, \td \theta_{n,i_{n}} \td \phi_{n, i_{n}} }{4\pi} \int \frac{\td (\kappa_{i,j,n} E^{2}_{q_{n}})}{2 \kappa_{i,j,n}E^{2}_{q_{n}}} \, P_{i_{n}j_{n}} \, \Theta_{\TT{on} \, \TT{shell}} \, 2 q_{\bot} \, \delta\left(E^{2}_{q_{n}}\kappa_{i,j,n} - q^{2}_{\bot}\right) \nonumber \\
&= \frac{1}{q_{\bot}} \int \frac{\sin \theta_{n,i_{n}} \, \td \theta_{n,i_{n}} \td \phi_{n, i_{n}} }{4\pi}  \, P_{i_{n}j_{n}} \, \Theta_{\TT{on} \, \TT{shell}}.
\end{align}
The textbook treatment would set $\Theta_{\TT{on} \, \TT{shell}}=1$ here. For us,
\begin{align}
\Theta_{\TT{on} \, \TT{shell}} &= \Theta(p_{i_{n}} \cdot p_{j_{n}} - q_{n} \cdot (p_{j_{n}} + p_{i_{n}})) \nonumber \\ &= \Theta\left(E_{i_{n}} E_{j_{n}}(1-\cos\theta_{j_{n},i_{n}}) - \frac{q_{\bot} E_{j_{n}}}{\sqrt{\kappa_{i,j,n}}} (1 - \sin \theta_{n,i_{n}} \; \cos \phi_{n, i_{n}} \sin \theta_{j_{n},i_{n}} - \cos \theta_{j_{n},i_{n}} \cos \theta_{n,i_{n}}) \right. \nonumber \\ 
& \left. \quad - \frac{q_{\bot} E_{i_{n}}}{\sqrt{\kappa_{i,j,n}}}(1 - \cos \theta_{n,i_{n}})\right),
\end{align}
which bounds the $\phi_{n, i_{n}}$ integration to the range $|\phi_{n, i_{n}}| \in [\phi^{-}_{q, i},\phi^{+}_{q, i})$. The solutions for the boundaries, $ \phi^{\pm}_{q, i} $ are given by
\begin{align}
&\cos \phi^{\pm}_{q, i} = \pm \, \TT{min}\left(|\alpha^{\pm}|, 1 \right) \quad \TT{for} \; \alpha^{\pm} > 0 \quad \TT{and} \; \cos \phi^{\pm}_{q, i} = 0 \; \TT{otherwise},\nonumber \\
&\alpha^{\pm} = \frac{\pm \sqrt{AF^2(AF^2-2DGH)} + AF^2 - DG(H + CG)}{(\sin \theta_{n,i_{n}} \sin \theta_{j_{n},i_{n}} ) (1 - \cos \theta_{j_{n},i_{n}}) q^2_{\bot} E_{j_{n}}^2} \nonumber \\ 
&F = E_{i_{n}} E_{j_{n}}(1-\cos\theta_{j_{n},i_{n}}) = E_{i_{n}} E_{j_{n}} D, \quad D = 1-\cos\theta_{j_{n},i_{n}}, \nonumber \\
&H = q_{\bot} E_{i_{n}}(1 - \cos \theta_{n,i_{n}}) = q_{\bot} E_{i_{n}} A, \qquad A = 1 - \cos \theta_{n,i_{n}},\nonumber \\
&B = \sin \theta_{n,i_{n}} \, \sin \theta_{j_{n},i_{n}}, \nonumber \\
&C = 1 - \cos \theta_{j_{n},i_{n}} \cos \theta_{n,i_{n}}, \nonumber \\
&G = q_{\bot} E_{j_{n}}.
\end{align}
Note that the expression under the square root is always positive. The usual approach to azimuthal averaging is to employ the soft limit and set $\Theta_{\TT{on} \, \TT{shell}}=1$, after which the $\phi_{n, i_{n}}$ integral can be performed by contour integration. However, in our case this is not viable, due to the boundaries on the $\phi_{n, i_{n}}$ integral. Instead we will write the integral as
\begin{align}
\TT{Eq}.~\eqref{eqn:Pij} &= \frac{1}{q_{\bot}} \int \frac{\sin \theta_{n,i_{n}} \, \td \theta_{n,i_{n}}}{2}  \, \<P_{i_{n}j_{n}} \, \Theta_{\TT{on} \, \TT{shell}} \>_{\phi_{n, i_{n}}} \nonumber \\ 
&= \frac{1}{q_{\bot}} \int \frac{\sin \theta_{n,i_{n}} \, \td \theta_{n,i_{n}}}{2}  \, \big[\<P_{i_{n}j_{n}} \>_{\phi_{n, i_{n}}} \<\Theta_{\TT{on} \, \TT{shell}} \>_{\phi_{n, i_{n}}} \nonumber \\ 
&\qqquad + \sigma_{P_{i_{n}j_{n}}}\sqrt{\<\Theta_{\TT{on} \, \TT{shell}} \>_{\phi_{n, i_{n}}}(1-\<\Theta_{\TT{on} \, \TT{shell}} \>_{\phi_{n, i_{n}}})} \TT{Cor}(P_{i_{n}j_{n}},\Theta_{\TT{on} \, \TT{shell}})\big], \label{eqn:P_Thetasplit_appart}
\end{align}
where $\TT{Cor}(x,y)$ is the correlation function between two variables $x$ and $y$, in context the correlation over variation of the azimuth. Firstly note that 
$$\<P_{i_{n}j_{n}} \>_{\phi_{n, i_{n}}} = \frac{\Theta(\theta_{j_{n},i_{n}} - \theta_{n,i_{n}})}{1- \cos\theta_{n,i_{n}}},$$
the usual result from azimuthal averaging. We can also note that $\<\Theta_{\TT{on} \, \TT{shell}} \>_{\phi_{n, i_{n}}} \in [0,1]$ and $|\TT{Cor}(P_{i_{n}j_{n}},\Theta_{\TT{on} \, \TT{shell}})| \in [0,1]$. By brute-force evaluation and noting $\Theta_{\TT{on} \, \TT{shell}}$ is binomially valued, we find
\begin{align}
&\<\Theta_{\TT{on} \, \TT{shell}} \>_{\phi_{n, i_{n}}} = \frac{|\phi^{+}_{q, i} - \phi^{-}_{q, i}|}{\pi} \, \bar{\theta}_{\TT{on} \, \TT{shell}}, \nonumber \\
&\TT{where} \quad \bar{\theta}_{\TT{on} \, \TT{shell}} =  \Theta_{\TT{on} \, \TT{shell}}\big|_{\phi_{n,i_{n}}=\phi^{\TT{crit}}}, \quad \TT{and} \quad \cos\phi^{\TT{crit}} = \TT{sign}(f)\TT{min}\left(\left|f\right|,1\right), \nonumber \\
&f(\theta_{n,i_{n}},\theta_{j_{n},i_{n}},E_{i_{n}},E_{j_{n}},q_{\bot}) = \frac{\frac{1- (1 - \cos \theta_{n,i_{n}})E_{i_{n}}/E_{j_{n}}}{\sin \theta_{n,i_{n}} \, \sin \theta_{j_{n},i_{n}}}  - 4\frac{1 - \cos \theta_{n,i_{n}}}{1 - \cos \theta_{j_{n},i_{n}}}(1-\cos \theta_{j_{n},i_{n}} \cos \theta_{n,i_{n}})}{1-4\frac{\sin \theta_{n,i_{n}} \, \sin \theta_{j_{n},i_{n}}}{1 - \cos \theta_{n,i_{n}}}}.
\end{align}
The exact angular ordered result is obtained when \mbox{$\<\Theta_{\TT{on} \, \TT{shell}} \>_{\phi_{n, i_{n}}} = \bar{\theta}_{ij} = 1$}, which is the case in the strongly ordered, $q_{\bot}/Q\rightarrow 0$, and collinear, $\theta_{n,i_{n}} \rightarrow 0$, limits (here $Q$ stands in for any other harder invariant). The remainder of this section is used to show that the correlation term can be neglected at least at $\As^n L^{2n-1}$ accuracy (and for NLL thrust). It can be skipped if the reader does not need convincing.

Now we must compute $\sigma^{2}_{P_{i_{n}j_{n}}} = \<P^{2}_{i_{n}j_{n}} \>_{\phi_{n, i_{n}}} - \<P_{i_{n}j_{n}} \>^{2}_{\phi_{n, i_{n}}}$
\begin{align}
\<P^{2}_{i_{n}j_{n}} \>_{\phi_{n, i_{n}}} &= \int \frac{\td \phi_{n, i_{n}} }{2\pi} \, P^{2}_{i_{n}j_{n}} = \int \frac{\td \phi_{n, i_{n}} }{8\pi} \left( \frac{n_{i_{n}} \cdot n_{j_{n}}-n_{i_{n}}\cdot n}{n_{i_{n}}\cdot n \; n_{j_{n}} \cdot n} + \frac{1}{n_{i_{n}}\cdot n}\right)^{2}, \nonumber \\
& = \frac{1}{(n_{i_{n}}\cdot n)^{2}} \int \frac{\td \phi_{n, i_{n}} }{8\pi } \left( \frac{\cos \theta_{n,i_{n}} - \cos \theta_{j_{n},i_{n}}}{1 - \sin \theta_{n,i_{n}} \cos \phi_{n, i_{n}} \sin \theta_{j_{n},i_{n}} - \cos \theta_{j_{n},i_{n}} \cos \theta_{n,i_{n}}} + 1\right)^{2},
\end{align}
using the substitution $z = \exp(i \phi_{n, i_{n}})$ this integral equals
\begin{align}
\<P^{2}_{i_{n}j_{n}} \>_{\phi_{n, i_{n}}} &= \frac{1}{(n_{i_{n}}\cdot n)^{2}} \oint_{S^{1}} \frac{z \, \td z }{2\pi i} \left( \frac{\cos \theta_{n,i_{n}} - \cos \theta_{j_{n},i_{n}}}{2z - \sin \theta_{n,i_{n}} (z^{2} + 1) \sin \theta_{j_{n},i_{n}} - 2z\cos \theta_{j_{n},i_{n}} \cos \theta_{n,i_{n}}} + \frac{1}{2z} \right)^{2}, \nonumber \\
& = \frac{1}{(n_{i_{n}}\cdot n)^{2}} \oint_{S^{1}} \frac{\td z }{2\pi i} \left( \frac{ z ( \cos \theta_{n,i_{n}} - \cos \theta_{j_{n},i_{n}}) }{\sin^{2} \theta_{n,i_{n}}\sin^{2} \theta_{j_{n},i_{n}}(z-z_{+})^{2}(z-z_{-})^{2}} \right. \nonumber \\ 
& \qqqqquad \qquad \left. + \frac{ \cos \theta_{n,i_{n}} - \cos \theta_{j_{n},i_{n}}}{\sin \theta_{n,i_{n}}\sin \theta_{j_{n},i_{n}}(z-z_{+})(z-z_{-})} + \frac{1}{4z}\right),
\end{align}
where
\begin{align}
z_{\pm} = \frac{1-\cos \theta_{j_{n},i_{n}} \cos \theta_{n,i_{n}}}{\sin \theta_{n,i_{n}}\sin \theta_{j_{n},i_{n}} } \pm \sqrt{\left(\frac{1-\cos \theta_{j_{n},i_{n}} \cos \theta_{n,i_{n}}}{\sin \theta_{n,i_{n}}\sin \theta_{j_{n},i_{n}} }\right)^{2}-1}.
\end{align}
Only the $z=z_{-}$ and $z=0$ poles are in the unit circle: 
\begin{align}
&\frac{1}{(n_{i_{n}}\cdot n)^{2}} \oint_{S^{1}} \frac{\td z }{2\pi i} \left( \frac{ \cos \theta_{n,i_{n}} - \cos \theta_{j_{n},i_{n}}}{\sin \theta_{n,i_{n}}\sin \theta_{j_{n},i_{n}}(z-z_{+})(z-z_{-})} + \frac{1}{4z}\right) \nonumber \\
&= \left\{\begin{matrix}
\frac{3}{4 (1 - \cos\theta_{n,i_{n}})^{2}} & \TT{when} \; \theta_{n,i_{n}} < \theta_{j_{n},i_{n}},\\[10pt]
-\frac{1}{4 (1 - \cos\theta_{n,i_{n}})^{2}} & \TT{otherwise},
\end{matrix}\right.
\end{align}
and 
\begin{align}
&\frac{1}{(n_{i_{n}}\cdot n)^{2}} \oint_{S^{1}} \frac{\td z }{2\pi i} \left( \frac{ z ( \cos \theta_{n,i_{n}} - \cos \theta_{j_{n},i_{n}}) }{\sin^{2} \theta_{n,i_{n}}\sin^{2} \theta_{j_{n},i_{n}}(z-z_{+})^{2}(z-z_{-})^{2}}\right) \nonumber \\
&= \frac{1}{(1 - \cos\theta_{n,i_{n}})^{2}(\cos\theta_{n,i_{n}} - \cos\theta_{j_{n},i_{n}})}
\left(1 - \frac{2z_{-}\TT{sign}(\cos\theta_{n,i_{n}} - \cos\theta_{j_{n},i_{n}})}{(\cos\theta_{n,i_{n}} - \cos\theta_{j_{n},i_{n}})^{2}}\right).
\end{align}
Thus
\begin{align}
\<P^{2}_{i_{n}j_{n}} \>_{\phi_{n, i_{n}}} = \left\{\begin{matrix} \frac{1 - \frac{2z_{-}}{(\cos\theta_{n,i_{n}} - \cos\theta_{j_{n},i_{n}})^{2}}}{(1 - \cos\theta_{n,i_{n}})^{2}(\cos\theta_{n,i_{n}} - \cos\theta_{j_{n},i_{n}})}
+ \frac{3}{4 (1 - \cos\theta_{n,i_{n}})^{2}} & \TT{when} \; \theta_{n,i_{n}} < \theta_{j_{n},i_{n}},\\[10pt]
\frac{1 + \frac{2z_{-}}{(\cos\theta_{n,i_{n}} - \cos\theta_{j_{n},i_{n}})^{2}}}{(1 - \cos\theta_{n,i_{n}})^{2}(\cos\theta_{n,i_{n}} - \cos\theta_{j_{n},i_{n}})} -\frac{1}{4 (1 - \cos\theta_{n,i_{n}})^{2}} & \TT{otherwise},
\end{matrix}\right.
\end{align}
and so 
\begin{align}
\sigma^{2}_{P_{i_{n}j_{n}}} = \left\{\begin{matrix} \frac{1 - \frac{2z_{-}}{(\cos\theta_{n,i_{n}} - \cos\theta_{j_{n},i_{n}})^{2}}}{(1 - \cos\theta_{n,i_{n}})^{2}(\cos\theta_{n,i_{n}} - \cos\theta_{j_{n},i_{n}})}
+ \frac{1}{4 (1 - \cos\theta_{n,i_{n}})^{2}} & \TT{when} \; \theta_{n,i_{n}} < \theta_{j_{n},i_{n}},\\[10pt]
\frac{1 + \frac{2z_{-}}{(\cos\theta_{n,i_{n}} - \cos\theta_{j_{n},i_{n}})^{2}}}{(1 - \cos\theta_{n,i_{n}})^{2}(\cos\theta_{n,i_{n}} - \cos\theta_{j_{n},i_{n}})} -\frac{1}{4 (1 - \cos\theta_{n,i_{n}})^{2}} & \TT{otherwise}.
\end{matrix}\right.
\end{align}
This has a collinear divergence that is suitably screened in Eq.~\eqref{eqn:P_Thetasplit_appart} by the accompanying phase space factor, $$\sqrt{\<\Theta_{\TT{on} \, \TT{shell}} \>_{\phi_{n, i_{n}}}(1-\<\Theta_{\TT{on} \, \TT{shell}} \>_{\phi_{n, i_{n}}})},$$ as is the soft divergence from the $q_{\bot}$ pre-factor in Eq.~\eqref{eqn:Pij}. $\TT{Cor}(P_{i_{n}j_{n}},\Theta_{\TT{on} \, \TT{shell}})$, is bounded above and below by $1$ and $-1$ so at most further dampens the effect of the $\sigma^{2}_{P_{i_{n}j_{n}}}$ term. As a result it is a finite non-logarithmic correction at order $\As$ and its contribution is suppressed at higher orders (to be seen explicitly one could repeat the analysis of Appendix \ref{sec:Thrust}). Hence, for $\As^n L^{2n-1}$ accuracy, we need only take the first term on the right hand-side of Eq.~\eqref{eqn:P_Thetasplit_appart}. 

\subsection{Derivation of the dipole shower}
\label{sec:Dipolederivation}

In this section we will derive from Eq.~\eqref{eqn:evo} an evolution equation for a dipole shower for final-state coloured radiation in $e^+e^-$. The extension to an initial state shower does not add complexity but lengthens equations. To derive the dipole shower we will spin average the evolution and make the leading colour approximation. To approximate the colour, we express amplitude density matrices and colour charge operators in the colour-flow basis. We manipulate the colour-flow basis using the mathematical machinery introduced in \cite{SoftEvolutionAlgorithm}.

Before we begin the derivation let us look at Eq.~\eqref{eqn:evo} in more detail and apply some of the knowledge we have gained from deriving an angular ordered shower. Angular ordering is most powerful when applied to the two-jet limit in $e^{+}e^{-}$ , the mono-jet limit of DIS and Drell-Yan. In these cases, angular ordering does not approximate the soft radiation pattern at all. Instead, the soft radiation is colour diagonal. The diagonalisation of soft radiation renders the conservation of momentum longitudinal to a jet unambiguous.  Matching to the angular ordered limit is sufficient to completely constrain the leading component of momentum conservation in Eq.~\eqref{eqn:evo} (it must respect the partitioning defined by $P_{i_{n}j_{n}}$ as given in Appendix \ref{sec:CBderivation}). It is required that 
\begin{align}
\mathfrak{R}^{\TT{soft}}_{i_{n}j_{n}} &= \frac{(q^{(i_{n}j_{n})}_{n \, \bot})^{2}}{2E^{2}_{n}}\left( P_{i_{n}j_{n}} \mathfrak{R}_{i_{n}} + P_{j_{n}i_{n}} \mathfrak{R}_{j_{n}} \right) \nonumber \\
&= \frac{(q^{(i_{n}j_{n})}_{n \, \bot})^{2}}{4}\left( \left[\frac{p_{i_{n}} \cdot p_{j_{n}}}{p_{i_{n}}\cdot q_{n} \; p_{j_{n}} \cdot q_{n}} - \frac{T \cdot p_{j_{n}}}{T \cdot q_{n} }\frac{1}{p_{j_{n}} \cdot q_{n}} + \frac{T \cdot p_{i_{n}}}{T \cdot q_{n} }\frac{1}{p_{i_{n}} \cdot q_{n}} \right]\mathfrak{R}_{i_{n}} + (i\leftrightarrow j)\right),
\end{align}
where $T =  \sum_{i_{n}}p_{i_{n}}$ is a vector for projecting out the energy of a parton in the event ZMF and where $E_{n}$ is the energy of $q_{n}$ in the ZMF. This can be rearranged to give
\begin{align}
\mathfrak{R}^{\TT{soft}}_{i_{n}j_{n}} = \frac{\mathfrak{R}_{i_{n}} + \mathfrak{R}_{j_{n}}}{2} + \TT{Asym}_{i_{n}j_{n}}(q_{n})\mathfrak{R}_{i_{n}} + \TT{Asym}_{j_{n}i_{n}}(q_{n})\mathfrak{R}_{j_{n}},
\end{align}
\begin{align}
\TT{Asym}_{i_{n}j_{n}}(q_{n}) = \left[ \frac{T \cdot p_{i_{n}}}{4T \cdot q_{n} }\frac{(q^{(i_{n}j_{n})}_{n \, \bot})^{2}}{p_{i_{n}} \cdot q_{n}} - \frac{T \cdot p_{j_{n}}}{4T \cdot q_{n} }\frac{(q^{(i_{n}j_{n})}_{n \, \bot})^{2}}{p_{j_{n}} \cdot q_{n}}\right].
\end{align}
 As previously stated in our discussions on angular ordering,
$$\mathfrak{R}_{j_{n}} = \delta^{4}(p_{j_{n}} - z^{-1}_{n}\tilde{p}_{j_{n}}) \prod_{i_{n} \neq j_{n}} \delta^{4}(p_{i_{n}} - \tilde{p}_{i_{n}}) + \mathcal{O}(q_{\bot}/Q).$$
This recoil function is ready to use in Eq.~\eqref{eqn:evo}.

Now, let us begin computing the leading colour evolution of $\v{A}_{n}(q_{\bot})$. We intend to compute
\begin{align}
\Ldg^{(0)}_{\tau \sigma}\left[\v{A}_{n}(q_{\bot})\right] \equiv A^{(0) \; \tau \sigma}_{n}(q_{\bot}) \lkl \tau \> \< \sigma \rkl,
\end{align}
where $A^{(0) \; \tau \sigma}_{n}$ is the leading colour amplitude for
colour flows $\tau$ and $\sigma$, see
\cite{Platzer:2013fha,SoftEvolutionAlgorithm} for details on this
procedure. Term by term in Eq.~\eqref{eqn:evo} we can apply this
operation and find
\begin{align}
&\Ldg^{(0)}_{\tau \sigma}\left[\v{\Gamma}_{n}(q_{\bot}) \, \v{A}_{n}(q_{\bot}) + \v{A}_{n}(q_{\bot}) \, \v{\Gamma}^{\dagger}_{n}(q_{\bot})\right] = 2 \, \gamma^{(\sigma)}_{n}(q_{\bot}) \, \delta_{\tau\sigma} \, \Ldg^{(0)}_{\tau \sigma}\left[\v{A}_{n}(q_{\bot})\right],
\end{align}
where
\begin{align}
&\gamma^{(\sigma)}_{n-1}(q_{\bot}; q_{\bot} \cup \{p\}_{n-1}) =  \frac{\As}{2\pi}\int \frac{\td S^{(q)}_{2} }{4\pi} \,  \Bigg( \sum_{i_{n},j_{n} \, c.c. \, \TT{in} \, \sigma} \lambda_{i_{n}}\bar{\lambda}_{j_{n}} \Nc \int \delta q^{(i_{n}, j_{n})}_{n \, \bot} (q_{\bot}) \nonumber \\
& \times \mathcal{R}^{\TT{soft}}_{i_{n} j_{n}}   + \sum_{i_{n}, \upsilon_{n}} \overline{\mathcal{P}}^{(\TT{final})}_{\upsilon_{i_{n}} \rightarrow \upsilon, \upsilon_{n}} \, (1-z_{n}) \int \delta q^{(i_{n}, \vec{n})}_{n \, \bot} (q_{\bot}) \, \mathcal{R}^{\TT{col}}_{i_{n}}  \Bigg)\, \Theta_{\TT{on\, shell}} \label{eqn:LCgamma}
\end{align}
and where 
\begin{align}
\mathcal{R}^{\TT{soft}}_{i_{n} j_{n}} = \int \prod_{i_{n}} \td^4 p_{i_{n}} \,  \mathfrak{R}^{\TT{soft}}_{i_{n}j_{n}} = 1 + \mathcal{O}(q_{\bot}/Q), \quad
\mathcal{R}^{\TT{col}}_{i_{n}} = \int \prod_{i_{n}} \td^4 p_{i_{n}}  \mathfrak{R}^{\TT{coll}}_{i_{n}} = 1 + \mathcal{O}(q_{\bot}/Q).
\end{align}
The sum over ``$i_{n},j_{n} \, c.c. \, \TT{in} \, \sigma$'' standards for performing the sum over partons dipoles $i_{n}, j_{n}$ which are colour connected in the colour state $\sigma$. $\overline{\mathcal{P}}_{\upsilon_{i_{n}} \rightarrow \upsilon, \upsilon_{n}} \equiv \overline{\mathcal{P}}_{\upsilon, \upsilon_{i_{n}}}$ are the hard-collinear splitting functions defined in Appendix A of \cite{Forshaw:2019ver}. They are the usual collinear splitting functions with soft poles subtracted away, i.e. $\overline{\mathcal{P}}_{qq} = -\mathcal{C}_{\TT{F}}(1+z_{n})$. Note that as we are working in the strict leading colour limit $\mathcal{C}_{\TT{F}} = \Nc/2$. The constants $\lambda_{i_{n}}$ and $\bar{\lambda}_{j_{n}}$ are defined in Table 1 of \cite{SoftEvolutionAlgorithm}, in the situations we will use them (the LC limit) $\lambda_{i_{n}}\bar{\lambda}_{j_{n}} \rightarrow 1/2$. We can observe that the first term on the RHS of Eq.~\eqref{eqn:LCgamma} is of the same form as the standard dipole type term. 
Next we can take the leading colour part of the emission operators. We spin average emission kernels, see Appendix \ref{sec:Spin} for details, and place carats on  objects to remind us that they are spin-averaged. We find
\begin{align}
\, \Ldg^{(0)}_{\tau \sigma}\left[\hat{\v{D}}_{n}(q_{n \, \bot}) \, \hat{\v{A}}_{n-1} (q_{n \, \bot}) \, \hat{\v{D}}^{\dagger}_{n}(q_{n \, \bot})\right] = \hat{W}^{(\sigma)}_{n}(q_{n \, \bot}) \, \delta_{\tau\sigma} \, \Ldg^{(0)}_{\tau \backslash n \,  \sigma\backslash n}\left[\hat{\v{A}}_{n-1}(q_{n \, \bot})\right], \label{eqn:LCDspin}
\end{align}
where
\begin{align}
&\hat{W}^{(\sigma) }_{n}(q_{n \, \bot}; q_{n} \cup \{\tilde{p}\}_{n-1}) = \sum_{i_{n},j_{n} \, c.c. \, \TT{in} \, \sigma} \lambda_{i_{n}}\bar{\lambda}_{j_{n}} \Nc \int \delta q^{(i_{n}, j_{n})}_{n \, \bot} (q_{n \, \bot}) \, \mathfrak{R}^{\TT{soft}}_{i_{n} j_{n}} \nonumber \\
&  + \sum_{\substack{i_{n} \in \TT{final}\\ \upsilon_{n}}} \overline{\mathcal{P}}^{(\TT{final})}_{\upsilon_{i_{n}} \rightarrow \upsilon, \upsilon_{n}} \, (1-z_{n}) \int \delta q^{(i_{n}, \vec{n})}_{n \, \bot} (q_{n \, \bot}) \, \mathfrak{R}^{\TT{col}}_{i_{n}} \,  .
\end{align}
Note that $\hat{\gamma}^{(\sigma)} = \gamma^{(\sigma)}$ as the loops do not depend on spin.

For now we will ignore the single logarithmic, hard-collinear pieces as they are easy to introduce later on (they are uniquely attributed to delta functions of the form $\delta^{4}(p_{j_{n}} - z^{-1}_{n}\tilde{p}_{j_{n}})$ in the recoil). This means that for now our final state will simply be the $q \bar{q}$ pair plus $n$ gluons. It is also typical in the strict LLA to let $\mathcal{R}^{\TT{soft}}_{i_{n} j_{n}}=1$; this will prove to be exact with the recoil scheme given in Section \ref{sec:Correcting} though only approximately so with the spectator scheme in Appendix \ref{sec:Spectator}. Thus the evolution equation is
\begin{align}
&q_{\bot} \Ldg^{(0)}_{\tau \sigma}\left[\pdf{\hat{\v{A}}_{n}(q_{\bot})}{q_{\bot}}\right] \approx  - \, \frac{\As}{\pi}\int \frac{\td S^{(q_{n+1})}_{2} }{4\pi} \sum_{i_{n+1},j_{n+1} \, c.c. \, \TT{in} \, \sigma} \nonumber \\
& \qqqquad \times  4\lambda_{i_{n+1}}\bar{\lambda}_{j_{n+1}} \Nc \int \delta q^{(i_{n+1}, j_{n+1})}_{n+1 \, \bot} (q_{\bot}) \, \Theta_{\TT{on\, shell}} \, \delta_{\tau\sigma} \, \Ldg^{(0)}_{\tau \sigma}\left[\hat{\v{A}}_{n}(q_{\bot})\right] \nonumber \\ 
& \quad + \int \bigg(\prod_{i_{n}} \td^4 p_{i_{n}}\bigg) \;  \sum_{i_{n},j_{n} \, c.c. \, \TT{in} \, \sigma} \lambda_{i}\bar{\lambda}_{j} \Nc \int \delta q^{(i_{n}, j_{n})}_{n \, \bot} (q_{n \, \bot}) \, \mathfrak{R}^{\TT{soft}}_{i_{n} j_{n}} \, \nonumber \\
& \qqqquad \times \delta_{\tau\sigma} \, \Ldg^{(0)}_{\tau \backslash n \,  \sigma\backslash n}\left[\hat{\v{A}}_{n-1}(q_{n \, \bot})\right] \; q_{\bot} \; \delta(q_{\bot} - q_{n \, \bot}).
\end{align}
This is a modified version of the equation for dipole evolution found in \cite{SoftEvolutionAlgorithm} that was shown to reproduce BMS evolution \cite{BMSEquation}. It has been modified to allow for the possibility of kinematic recoil and to account for the phase-space effects from energy conservation.

By taking the leading colour limit, the colour evolution has been made diagonal. We can trivially make the connection with squared spin-averaged matrix elements; for a given colour flow, $\sigma$, 
\begin{align}
|\hat{M}^{(\sigma)}_{n}(q_{\bot})|^{2} \lkl \sigma\> \< \sigma \rkl = \left(\frac{2\As}{\pi}\right)^{n}\Ldg^{(0)}_{\sigma \sigma}\left[\hat{\v{A}}_{n}(q_{\bot})\right],
\end{align}
where $\hat{M}$ is a dimensionless, spin-averaged and leading-colour matrix element, up to global factors of $2$ and $\pi$ which have been absorbed into the definition of our phase-space measure\footnote{The usual dimensionful matrix element is retrieved by multiplying with a factor $\prod_{i_{n+1}}2 \pi^{-1} q^{-2}_{i_{n+1} \, \bot}$.}. Thus 
\begin{align}
&q_{\bot} \pdf{|\hat{M}^{(\sigma)}_{n}(q_{\bot})|^{2}}{q_{\bot}} \nonumber \\  
&\approx - \, \frac{\As}{\pi}\int \frac{\td S^{(q_{n+1})}_{2} }{4\pi} \sum_{i_{n+1},j_{n+1} \, c.c. \, \TT{in} \, \sigma} 4\lambda_{i_{n+1}}\bar{\lambda}_{j_{n+1}} \Nc \int \delta q^{(i_{n+1}, j_{n+1})}_{n+1 \, \bot} (q_{\bot}) \,\Theta_{\TT{on\, shell}} \, |\hat{M}^{(\sigma)}_{n}(q_{\bot})|^{2} \nonumber \\ 
& + \frac{2\As}{\pi} \sum_{i_{n},j_{n} \, c.c. \, \TT{in} \, \sigma} \lambda_{i}\bar{\lambda}_{j} \Nc \int \bigg(\prod_{i_{n}} \td^4 p_{i_{n}} \bigg) \, \delta q^{(i_{n}, j_{n})}_{n \, \bot} (q_{n \, \bot}) \, \mathfrak{R}^{\TT{soft}}_{i_{n} j_{n}} \; |\hat{M}^{(\sigma/n)}_{n-1}(q_{n \, \bot})|^{2} \; q_{\bot} \; \delta(q_{\bot} - q_{n \, \bot}). \label{eqn:dipole_shower}
\end{align}
This is a generalised leading-colour dipole shower evolution equation with fixed coupling. Commonly one would introduce a running coupling with $q_{\bot}$ as its argument. At this point this would be a simple extension. We have omitted the running coupling as it does not effect our discussion. From this point on we drop the carat denoting spin averaging, leaving it implicit that the equations are spin averaged.

To manipulate our new dipole construction into the more usual form we now define a recoil function based on colour flows:
\begin{align}
\mathfrak{R}^{\TT{dipole}}_{i^c_{n}} = \left(\frac{1}{2} + \TT{Asym}_{i^c_{n}\overline{i^c_{n}}_{n}}(q_{n})\right)\mathfrak{R}_{i^c_{n}},
\end{align}
where, just as in Section \ref{sec:DipoleEvo}, we use $i^c_{n}$ to index the (anti-)colour line(s) of parton $i$ in a final state dressed with $n$ soft or collinear partons. Using this we can now return to Eq.~\eqref{eqn:dipole_shower} and manipulate the dipoles so that emissions from each half of a dipole are separated:
\begin{align}
&q_{\bot} \pdf{|M^{(\sigma)}_{n}|^{2}}{q_{\bot}} \nonumber \\  
&\approx - \, \frac{\As}{\pi}\int \frac{\td S^{(q_{n+1})}_{2} }{4\pi} \sum_{i^c_{n+1}} \mathcal{C}_{i^c_{n+1}} \int \delta q^{(i_{n+1}, \overline{i^c}_{n+1})}_{n+1 \, \bot} (q_{\bot}) \,2 \, \Theta_{\TT{on\, shell}} \, |M^{(\sigma)}_{n}|^{2} \nonumber \\ 
& + \frac{\As}{\pi} \sum_{i^c_{n}} \mathcal{C}_{i^c_{n}} \int \bigg(\prod_{j_{n}} \td^4 p_{j_{n}} \bigg) \, \delta q^{(i_{n}, \overline{i^c}_{n})}_{n \, \bot} (q_{n \, \bot}) \, \mathfrak{R}^{\TT{dipole}}_{i^c_{n}} \; |M^{(\sigma/n)}_{n-1}|^{2} \; q_{\bot} \; \delta(q_{\bot} - q_{n \, \bot}).
\end{align}
We can now include the sub-leading logarithms from the hard-collinear limit along with full-colour Casimir invariants. The Casimir invariants and collinear logarithms are each uniquely associated with longitudinal recoil and so a single $\mathfrak{R}^{\TT{dipole}}_{i^c_{n}}$. We note that $\TT{Asym}_{i^c_{n}\overline{i^c_{n}}_{n}}(q_{n})$ gives no logarithmic enhancement in the hard-collinear region, rendering the inclusion of hard-collinear pieces simple (including the re-inclusion of $g\rightarrow qq$ transitions). Thus we arrive at Eq.~\eqref{eqn:corrected_dipole}.\footnote{When constructing Eq.~\eqref{eqn:corrected_dipole} we chose to multiply each matrix element by a phase-space factor so that $|M^{(\sigma)}_{n}|^{2} \rightarrow \prod_{i}1/(1-z_{i})|M^{(\sigma)}_{n}|^{2}$ and separate sums over emission topologies, $|M^{(\sigma)}_{n}|^{2} \rightarrow \sum_{i^{c}_{1},...,i^{c}_{n}}|\mathcal{M}^{(\sigma)}_{n}|^{2}$. This ensures the standard dipole shower phase space can be used \cite{Pythia8,Herwig_dipole_shower,DIRE,Dasgupta:2018nvj}.} We can explicitly include the $g\rightarrow qq$ transitions by extending Eq.~\eqref{eqn:corrected_dipole}:
\begin{align}
&q_{\bot} \pdf{|M^{(\sigma)}_{n}|^{2}}{q_{\bot}} \nonumber \\  
&\approx - \, \frac{\As}{\pi}  \sum_{i^c_{n+1}}\int \td q^{(i^c_{n+1}, \overline{i^c}_{n+1})}_{\bot} \delta (q^{(i^c_{n+1}, \overline{i^c}_{n+1})}_{\bot} - q_{\bot})\int \td z \, \Theta_{\TT{on}\,\TT{shell}} \;  P_{\upsilon_{i_{n}}\upsilon_{i_{n}}}(z)  \, |M^{(\sigma)}_{n}|^{2} \nonumber \\ 
& + \frac{\As}{\pi} \sum_{i^c_{n}} \int \bigg(\prod_{j_{n}} \td^4 p_{j_{n}} \bigg) \,   \mathfrak{R}^{\TT{dipole}}_{i^c_{n}} \, P_{\upsilon_{i_{n}} \upsilon_{i_{n}}}(z_{n})  \; q_{\bot}\delta(q^{(i^c_{n}, \overline{i^c}_{n})}_{n \, \bot} - q_{\bot} ) |M^{(\sigma/n)}_{n-1}|^{2} \nonumber \\ 
& + \frac{\As}{\pi} \sum_{i^c_{n}} \int \bigg(\prod_{j_{n}} \td^4 p_{j_{n}} \bigg) \,   \mathfrak{R}^{\TT{dipole}}_{i^c_{n}} \,\delta_{\upsilon_{i_{n}} g} \, P_{q g}(z_{n})  \; q_{\bot}\delta(q^{(i^c_{n}, \overline{i^c}_{n})}_{n \, \bot} - q_{\bot} ) |M^{(\sigma)}_{n-1}|^{2}, \label{eqn:dipolegqq}
\end{align}
where $P_{q g}(z_{n}) = n_{f}T_{\TT{R}}z^{2}_{n}$. The inclusion of Casimir factors and collinear physics in this fashion ensures our shower correctly computes everything an angular ordered shower can compute, in the angular-ordered limit. There will however be NLC errors for radiation not ordered in angle. At the same time, the usual LC accuracy of a dipole shower is preserved. Also note that at no point in this derivation did we restrict ourselves to a $q\bar{q}$ final state for the hard process. In Section \ref{sec:DipoleEvo} we made this restriction as it allows Eq.~\eqref{eqn:cross-sectionDipole} to be written more simply. For more complex hard-process topologies one should sum over showers originating from each distinct hard-process colour flow (dipole).

So far we have still not constrained the $\mathcal{O}(q_{\bot}/Q)$ pieces in the recoil function associated with recoil in the backwards direction. These pieces are important for the computation of NLLs. Specifying them is the purpose of Section \ref{sec:Correcting} and Appenidx \ref{sec:Spectator}. In these sections we study their effect on NLLs. For contrast, in Section 2 of \cite{Forshaw:2019ver} we considered various recoil functions that specify the $\mathcal{O}(q_{\bot}/Q)$ pieces. We ensured each possible recoil prescription would consistently produce all leading physics, however we did not check sub-leading effects.  One of the prescriptions we considered was based on the spectator recoil commonly employed in modern dipole showers \cite{Platzer:recoil,Pythia8}. This approach involves partitioning the dipole using Catani-Seymour dipole factorisation \cite{Catani:1996vz} and distributing the longitudinal recoil in accordance with this partitioning. The remaining transverse recoil is then given to a third parton, not in the dipole but colour connected to the emitting parton. In \cite{Forshaw:2019ver} we give the functional form of $\mathfrak{R}^{\TT{soft}}_{i_{n} j_{n}}$ necessary to implement this recoil. Using this recoil function instead of the one we present here gives us an evolution equation similar to that governing Pythia8 \cite{Pythia8}.

In \cite{Dasgupta:2018nvj} it was shown that the standard spectator recoil prescriptions used in conjunction with Catani-Seymour dipole type showers are subject to errors computing NLLs and miscalculate next-to-leading colour. The errors in NLC occur because of the misattribution of longitudinal components of recoil and so colour factors. The errors in NLLs occur as unphysical artefacts from the shower construction do not cancel when one properly considers the effects of recoil after multiple emissions. It is for this reason that we have taken so much care to ensure consistency between our dipole shower and angular ordered showers, and why we take great care implementing recoil in Section \ref{sec:Correcting}.

\section{Spin averaging}
\label{sec:Spin}

In the derivation of an angular ordered shower and a dipole shower we had to spin average the evolution from Eq.~\eqref{eqn:evo}. We can introduce spin averaging safe in the knowledge that the spin-correlated evolution can be computed from the spin averaged by re-weighting with the algorithm of Collins, Knowles et al \cite{Collins:1987cp,KNOWLES1990271}. In our previous paper \cite{Forshaw:2019ver} we showed that, given collinear factorisation, the evolution of our algorithm is consistent with that of Collins and Knowles et al. We also showed that complete collinear factorisation can be achieved in the PB algorithm (neglecting Coulomb exchanges, which cancel in the leading colour limit). In this appendix we will summarise the spin averaging procedure. We will do so in the leading colour limit, as this is the limit of interest in the dipole shower case and this limit reduces the number of indices on objects. Real emissions in the leading colour limit without spin averaging give rise to
\begin{align}
\int \td R_{n} \; \Ldg^{(0)}_{\tau \sigma}&\left[\v{D}_{n}(q_{n \, \bot}) \, \v{A}_{n-1} (q_{n \, \bot}) \, \v{D}^{\dagger}_{n}(q_{n \, \bot})\right] = \nonumber \\ & \int \td R_{n} \; W^{(\sigma), \, h^{\TT{L}}_{n}, h^{\TT{R}}_{n}}_{n}(q_{n \, \bot}) \, \delta_{\tau\sigma} \, \Ldg^{(0)}_{\tau \backslash n \,  \sigma\backslash n}\left[\v{A}_{n-1}(q_{n \, \bot})\right], \label{eqn:LCD}
\end{align}
where
\begin{align}
&W^{(\sigma), \, h^{\TT{L}}_{n}, h^{\TT{R}}_{n}}_{n}(q_{n \, \bot}; q_{n} \cup \{\tilde{p}\}_{n-1},\{h^{\TT{L}}\},\{h^{\TT{R}}\}) = \nonumber \\ 
& \qqquad \sum_{i_{n},j_{n} \, c.c. \, \TT{in} \, \sigma} 2\lambda_{i_{n}}\bar{\lambda}_{j_{n}} \Nc \int \delta q^{(i_{n}, j_{n})}_{n \, \bot} (q_{n \, \bot}) \, s^{j_{n}, h^{\TT{R}}_{n} \, \dagger}_{n}s^{i_{n}, h^{\TT{L}}_{n}}_{n} \, \mathfrak{R}^{\TT{soft}}_{i_{n}j_{n}} \nonumber \\
& \qqquad  + \sum_{i_{n}} \int \delta q^{(i_{n}, \vec{n})}_{n \, \bot} (q_{n \, \bot}) \, \mathcal{C}_{i_{n}} \, c^{i_{n},h^{\TT{L}}_{n} \, \dagger}_{n}(h^{\TT{L}}_{i_{n}}) \,  c^{i_{n},h^{\TT{R}}_{n}}_{n}(h^{\TT{R}}_{i_{n}}) \, \mathfrak{R}^{\TT{col}}_{i_{n}}, \label{eqn:LLnotaveraged}
\end{align}
and where $s^{i_{n}, h^{\TT{L}}_{n}}_{n}$ and $c^{i_{n},h^{\TT{L}}_{n}}_{n}(h^{\TT{L}}_{i_{n}})$ are the kinematic factors associated with a soft or collinear emission respectively, for a fixed spin state. We have unpacked some of the recoil factors from $\int \td R_{n}$ and placed them next to the appropriate emission kernels, these are the $\mathfrak{R}^{\TT{soft}}_{i_{n}j_{n}}$ and $\mathfrak{R}^{\TT{col}}_{i_{n}}$ factors. $s^{i_{n}, h^{\TT{L}}_{n}}_{n}$ and $c^{i_{n},h^{\TT{L}}_{n}}_{n}(h^{\TT{L}}_{i_{n}})$ are defined through the relations
\begin{align}
s^{j_{n}, h^{\TT{R}}_{n} \, \dagger}_{n}s^{i_{n}, h^{\TT{L}}_{n}}_{n} \; \mathbb{T}_{j_{n}} \cdot \mathbb{T}_{i_{n}} &= \< h^{\TT{R}}_{j_{n}} \rkl \v{S}^{j_{n} \, \dagger}_{n} \lkl h^{\TT{R}}_{j_{n}}, h^{\TT{R}}_{n} \> \<  h^{\TT{L}}_{i_{n}}, h^{\TT{L}}_{n} \rkl \v{S}^{i_{n}}_{n} \lkl h^{\TT{L}}_{i_{n}}  \>, \nonumber \\
c^{i_{n},h^{\TT{R}}_{n} \, \dagger}_{n}(h^{\TT{R}}_{i_{n}}) \,  c^{i_{n},h^{\TT{L}}_{n}}_{n}(h^{\TT{L}}_{i_{n}}) \; \mathbb{T}_{i_{n}} \cdot \mathbb{T}_{i_{n}} &= \sum_{h'^{\TT{R}}_{i_{n}},h'^{\TT{L}}_{i_{n}}} \< h^{\TT{R}}_{i_{n}} \rkl \v{C}^{i_{n}\, \dagger}_{n} \lkl h'^{\TT{R}}_{i_{n}}, h^{\TT{R}}_{n} \> \<  h'^{\TT{L}}_{i_{n}}, h^{\TT{L}}_{n} \rkl \v{C}^{i_{n}}_{n} \lkl h^{\TT{L}}_{i_{n}}  \> ,
\end{align}
where $ h^{\TT{L/R}}_{i}$ is the helicity of the parton with label $i$ on the left/right hand side of the amplitude. In Eq.~\eqref{eqn:LLnotaveraged} we again used the abbreviation ``$i_{n},j_{n} \, c.c. \, \TT{in} \, \sigma$'' to mean that we sum over pairs $i_{n},j_{n}$ that are colour connected in $\sigma$. Note we have been a little sloppy by omitting sums over trivial spin indices of partons not involved in the splittings induced by $\v{C}^{i_{n}}_{n}$ and $\v{S}^{i_{n}}_{n}$ in Eq.~\eqref{eqn:LCD}. Spin averaging is achieved by setting $\{h^{\TT{L}}\}=\{h^{\TT{R}}\} = \{h\}$ and performing all trivial sums over spin states in Eq.~\eqref{eqn:LCD}. This is equivalent to replacing
\begin{align}
\v{A}_{n} \mapsto \hat{\v{A}}_{n}, & \quad W^{(\sigma), \, h^{\TT{L}}_{n}, h^{\TT{R}}_{n}}_{n}(q_{n \, \bot}) \mapsto \hat{W}^{(\sigma), }_{n}(q_{n \, \bot}),  \nonumber \\
s^{j_{n}, h^{\TT{R}}_{n} \, \dagger}_{n}s^{i_{n}, h^{\TT{L}}_{n}}_{n} \; \mathbb{T}_{j_{n}} \cdot \mathbb{T}_{i_{n}} & \mapsto \hat{s}^{j_{n}i_{n}}_{n} \; \mathbb{T}_{j_{n}} \cdot \mathbb{T}_{i_{n}} = \frac{1}{2}\sum_{h_{i_{n}}} \< h_{i_{n}} \rkl  \v{S}^{j_{n}}_{n} \cdot \v{S}^{i_{n}}_{n} \lkl h_{i_{n}} \>, \nonumber \\
c^{i_{n},h^{\TT{R}}_{n} \, \dagger}_{n}(h^{\TT{R}}_{i_{n}}) \,  c^{i_{n},h^{\TT{L}}_{n}}_{n}(h^{\TT{L}}_{i_{n}}) \; \mathbb{T}_{i_{n}} \cdot \mathbb{T}_{i_{n}} &\mapsto \hat{c}^{i_{n}}_{n} \; \mathbb{T}_{i_{n}} \cdot \mathbb{T}_{i_{n}} = \frac{1}{2}\sum_{h_{i_{n}}} \< h_{i_{n}} \rkl \v{C}^{i_{n}}_{n} \cdot \v{C}^{i_{n}}_{n} \lkl h_{i_{n}} \> ,
\end{align}
where we denoted the spin averaged objects with a carat. We have assumed $\mathfrak{R}^{\TT{soft}}_{i_{n}j_{n}}$ and $\mathfrak{R}^{\TT{col}}_{i_{n}}$ are chosen such that they are not spin dependent, otherwise they too should be averaged in the same fashion.

\section{Dipole shower with spectator recoil}
\label{sec:Spectator}

It is commonplace to use local `spectator' recoils in dipole showers rather than the global approach we have opted for \cite{Pythia8,Herwig_dipole_shower}. In this appendix we introduce one such recoil scheme and show that, despite the other improvements to our dipole shower, it suffers the NLL errors pointed out in \cite{Dasgupta:2018nvj}. 

Following the approach of \cite{Platzer:recoil}, we can treat each transition from an $n-1$ to an $n$ parton matrix element as being generated by a $2 \rightarrow 3$ parton splitting which locally conserves momentum. The splitting is defined such that the parton with colour line $i_{n}$ under goes a primary decay into two partons, the amplitude for which is given by a collinear splitting function. The parton with colour line $\overline{i}_{n}$ acts as a spectator and under goes a secondary $1 \rightarrow 1$ transition where it absorbs the residual recoil from the primary decay. To this end we introduce the following Sudakov decomposition 
\begin{align}
\tilde{p}_{i_{n}} &= z_{n} p_{i_{n}}-k_{\bot}+\frac{(q^{(i_{n}\overline{i}_{n})}_{n \, \bot})^{2}}{z_{n}}\frac{p_{\overline{i}_{n}}}{2p_{i_{n}}\cdot p_{\overline{i}_{n}}}, \qquad (q^{(i_{n}\overline{i}_{n})}_{n \, \bot})^{2} = - k_{\bot}^{2}, \nonumber \\
q_{n} &= (1-z_{n}) p_{i_{n}}+k_{\bot}+\frac{(q^{(i_{n}\overline{i}_{n})}_{n \, \bot})^{2}}{1-z_{n}}\frac{p_{\overline{i}_{n}}}{2p_{i_{n}}\cdot p_{\overline{i}_{n}}}, \nonumber \\
\tilde{p}_{\overline{i}_{n}} &= \left(1-\frac{(q^{(i_{n}\overline{i}_{n})}_{n \, \bot})^{2}}{z_{n}(1-z_{n})}\frac{1}{2p_{i_{n}}\cdot p_{\overline{i}_{n}}}\right)p_{\overline{i}_{n}}, \qquad k_{\bot} \cdot p_{i_{n}} = k_{\bot} \cdot p_{\overline{i}_{n}}=0,
\label{eqn:2_to_3_Sudakov_decomposition}
\end{align}
which conserves momentum as $p_{i_{n}} + p_{\overline{i}_{n}} = \tilde{p}_{i_{n}} + \tilde{p}_{\overline{i}_{n}} + q_{n}$. This decomposition defines the kinematics of the $2 \rightarrow 3$ splitting. Enforcing this local recoil scheme implies that 
\begin{align}
\mathfrak{R}_{i_{n}} = &\left(1 - \frac{(q^{(i_{n}\overline{i}_{n})}_{n \, \bot})^{2}}{z_{n}(1-z_{n})\, 2p_{i_{n}}\cdot p_{\overline{i}_{n}}} \right) \delta^{4}_{\mathcal{J}}\left(\tilde{p}_{\overline{i}_{n}} - p_{\overline{i}_{n}} + \frac{(q^{(i_{n}\overline{i}_{n})}_{n \, \bot})^{2}}{z_{n}(1-z_{n})}\frac{p_{\overline{i}_{n}}}{2p_{i_{n}}\cdot p_{\overline{i}_{n}}}\right) \nonumber \\ & \times \delta^{4}_{\mathcal{J}}\left(\tilde{p}_{i_{n}} - z_{n} p_{i_{n}}+k_{\bot}-\frac{(q^{(i_{n}\overline{i}_{n})}_{n \, \bot})^{2}}{z_{n}}\frac{p_{\overline{i}_{n}}}{2p_{i_{n}}\cdot p_{\overline{i}_{n}}}\right)\prod_{j_{n} \neq i_{n}, \overline{i}_{n}} \delta^{4}(p_{j_{n}} - \tilde{p}_{j_{n}}),
\end{align}
where  $$\delta_{\mathcal{J}}(f(x)) = f'(x_{i})\delta(f(x)) = \delta(x - x_{i}),$$  and $x_{i}$ is the single root of $f(x)$ inside the range of $x$ over which the delta function has support.

\subsection{NLC and NLL accuracy of the spectator recoil}

Let us begin by filling in some of the derivation of Eq.~\eqref{eqn:Fixedorder} with the local dipole recoil specified in previous section. Starting from Eq.~\eqref{eqn:dipole_shower},
\begin{align}
\delta \Sigma(L) =& \sigma_{n_{\TT{H}}} \prod^{2}_{n=1} \bigg( \int \td \Pi_{n} \sum_{i_{n},j_{n} \, c.c. \, \TT{in} \, \sigma} \int \prod_{k_{n}} \td^4 p_{k_{n}} \, \delta q^{(i_{n}, j_{n})}_{n \, \bot} (q_{n \, \bot}) \, \lambda_{i}\bar{\lambda}_{j} \Nc \; \mathfrak{R}^{\TT{soft}}_{i_{n} j_{n}} \, \theta_{i_{n}j_{n}}\bigg) \nonumber \\ 
& \qqquad \times \Theta(q_{1 \, \bot} - q_{2 \, \bot}) \Theta\left(e^{-L} - V(\{p\}_{2})\right)\nonumber \\
& - \sigma_{n_{\TT{H}}} \prod^{2}_{n=1} \bigg( \int \td \Pi_{n} \sum_{i_{n},j_{n} \, c.c. \, \TT{in} \, \sigma} \int \delta q^{(i_{n}, j_{n})}_{n \, \bot} (q_{n \, \bot}) \, \lambda_{i}\bar{\lambda}_{j} \Nc \; \theta^{\TT{correct}}_{i_{n}j_{n}} \bigg) \nonumber \\ 
& \qqquad \times \Theta(q_{1 \, \bot} - q_{2 \, \bot}) \Theta\left(e^{-L} - V(\{p\}_{\TT{correct}})\right), \nonumber
\end{align}
\begin{align}
=&  \mathcal{C}_{\TT{F}} \sigma_{n_{\TT{H}}} \int \td \Pi_{2} \, \td \Pi_{1} \int \delta q^{(a_{2}, 1_{2})}_{2 \, \bot} (q_{2 \, \bot}) \int \delta q^{(a_{1}, b_{1})}_{1 \, \bot} (q_{1 \, \bot}) \;  \Theta(q_{1 \, \bot} - q_{2 \, \bot}) \nonumber \\
& \times \bigg[ \int \prod^{2}_{n=1} \prod_{k_{n}} \td^4 p_{k_{n}} \; \mathfrak{R}^{\TT{soft}}_{a_{2} 1_{2}} \, \theta_{a_{2}1_{2}} \; \mathfrak{R}^{\TT{soft}}_{a_{1} b_{1}} \, \theta_{a_{1}b_{1}} \Theta\left(e^{-L} - V(\{p\}_{2})\right) \nonumber \\
& \qquad - \theta^{\TT{correct}}_{a_{2}1_{2}} \theta^{\TT{correct}}_{a_{1}b_{1}} \Theta\left(e^{-L} - V(\{p\}_{\TT{correct}})\right) \bigg],
\end{align}
where $\{p\}_{\TT{correct}}$ is the set of correct momenta for the 4-body system and where $\theta^{\TT{correct}}_{i_{n}j_{n}} = \theta_{i_{n}j_{n}}(\{p\}_{\TT{correct}})$. From this we find
\begin{align}
\delta \Sigma(L) &\approx \frac{4\As^{2} \, \mathcal{C}^{2}_{\TT{F}} \, \sigma_{n_{\TT{H}}}}{\pi^{2}}  \int^{Q}_{0} \frac{\td q^{(a_{1}, b_{1})}_{1 \, \bot}}{q^{(a_{1}, b_{1})}_{1 \, \bot}} \int^{\ln Q /q^{(a_{1}, b_{1})}_{1 \, \bot}}_{-\ln Q / q^{(a_{1}, b_{1})}_{1 \, \bot}} \td y_{1} \int^{q^{(a_{1}, b_{1})}_{1 \, \bot}}_{0} \frac{\td q^{(a_{2}, 1_{2})}_{2 \, \bot}}{q^{(a_{2}, 1_{2})}_{2 \, \bot}} \int^{\ln Q /q^{(a_{2}, 1_{2})}_{2 \, \bot}}_{-\ln Q / q^{(a_{2}, 1_{2})}_{2 \, \bot}} \td y_{2} \nonumber \\
&\quad \times \int^{2\pi}_{0} \frac{\td \phi_{2}}{2 \pi} \left[ \Theta\left(e^{-L} - V(\{p\}_{2})\right) - \Theta\left(e^{-L} - V(\{p\}_{\TT{correct}})\right) \right].
\end{align}
The kinematics are encapsulated by $\{p\}_{2}$, just as in the global scheme given in Section \ref{sec:Correcting}. They are in fact exactly the same kinematics as those specified in Section 3.3 of \cite{Dasgupta:2018nvj} and we have arrived at the same expression as B.5 of \cite{Dasgupta:2018nvj}. Thus, we can follow their argument from Appendix A and Section 4 and conclude that our local dipole prescription does suffer the same NLL errors as other local dipole prescriptions. For example, we can consider the two-jet rate using the Cambridge algorithm, for which $V(\{p_{i}\}) = \max_{i} \{p_{i \, \bot}\}$. In the limit we have considered, this reduces to \mbox{$V(\{p\}_{\TT{correct}}) = q^{(a_{1}, b_{1})}_{1 \, \bot}$} whereas $V(\{p\}_{\TT{2}}) = \max (q^{(a_{1}, b_{1})}_{1 \, \bot}, q^{(a_{2}, 1_{2})}_{2 \, \bot})$ since the recoil scheme does not ensure that $q^{(a_{1}, b_{1})}_{1 \, \bot} > q^{(a_{2}, 1_{2})}_{2 \, \bot}$ at all points in the phase-space for parton 2's emission. \cite{Dasgupta:2018nvj} show that this error generates a incorrect NLL ($\Nc^{2}\As^{2}L^{2}$). This was expected, as in our local dipole scheme we have only made modifications to fix the NLC of the usual dipole shower procedure. It would be unexpectedly fortuitous if this also fixed the NLL problems.

\section{Further checks}
\label{sec:Checks}

\subsection{Thrust with NLL accuracy using global recoil}
\label{sec:Thrust}

Thrust has a long history. It was first resummed to leading log accuracy in 1980 \cite{Binetruy:1980hd} and then later at next-to-leading in 1993 \cite{CATANI1992419}. More recently, it was resummed to N$^{3}$LL \cite{Becher:2008cf}. In this section we will analyse the consistency of the dipole shower and recoil scheme we present in Sections \ref{sec:DipoleEvo} and \ref{sec:Correcting} with the NLL computation found in \cite{CATANI1992419}. Crucially, the calculation of NLL thrust was performed using a coherent branching algorithm \cite{CATANI1991635} (or equivalently by analytic computation of an angular ordered shower). The coherent branching algorithm employed in the resummation was not strictly momentum conserving and effectively only conserved the momentum longitudinal to the two back-to-back jets. In \cite{CATANI1992419} they show that neglecting the other components is a valid approximation in the computation of NLLs for thrust (see their $\epsilon$ expansion of the correct phase-space). However, in \cite{Dasgupta:2018nvj} it was observed that incorrect handling of transverse momentum in dipole showers can induce NLL errors in thrust from $\mathcal{O}(\As^3)$ onwards. These two papers are not inconsistent with each other, the situation is simply that the incorrect inclusion of momentum conserving terms can induce NLL errors. 

Our dipole shower algorithm was built around consistency with an angular ordered shower. Its collinear radiation pattern reproduces that of an angular ordered shower with the correct longitudinal momentum conservation after azimuthally averaging. At NLL accuracy, it is also consistent at leading-colour with the angular ordered shower (restricted to leading-colour since our dipole shower only has leading-colour accuracy for radiation unordered in angle). Notwithstanding those NLC terms, there is one other main difference between the coherent branching resummed in \cite{CATANI1992419} and our algorithm after azimuthal averaging; ours conserves momentum completely. Thus the only remaining question is whether our approach to momentum conservation breaks the full-colour LL and leading-colour NLL accuracy of our dipole shower. We can compute the difference between our algorithm's computation of thrust and \cite{CATANI1992419}. As thrust is dominated by the two-jet limit, we initially focus on emissions from the primary hard legs (which is sufficient for NLL accuracy in the approach of \cite{CATANI1992419} by assuming inclusivity over jets from secondary jets). Afterwards we will briefly consider the effects of secondary emissions, i.e. possible recoil effects from the multi-jet limit. Firstly note that thrust can be defined as
$$
T(\{p\}_{n}) = \max_{\v{n}}  \frac{\sum_{\forall p \in \{p\}_{n}}|\v{p}\cdot\v{n}|}{\sum_{\forall p\in \{p\}_{n}}|\v{p}|} \stackrel{\TT{NLL}}{\simeq} 1 - \frac{P^{2}_{n} + P^{2}_{\bar{n}}}{Q^2},
$$
where $P_{n}$ ($P_{\bar{n}}$) is the total four-momentum in the hemisphere centred on the forwards (backwards) thrust axis. From this definition, it is clear that thrust is invariant under boosts along the thrust axis and is invariant under global jet energy rescaling. Following the notation of Section \ref{sec:Correcting}, the difference in the two-jet limit between our dipole algorithm and the NLL result due to recoil is of the general form
\begin{align}
\delta \Sigma(L) \sim \sum_{n}& \As^n C_n \Bigg(\int^{Q}_{0} \frac{\td q_{n \, \bot}}{q_{n \, \bot}} ... \int^{Q}_{0} \frac{\td q_{1 \, \bot}}{q_{1 \, \bot}} \; \int^{\ln (\kappa_{n} Q /q_{n \, \bot})}_{-\ln (\kappa_{n} Q /q_{n \, \bot})} \td y_{n} ... \int^{\ln (\kappa_{1} Q /q_{1 \, \bot})}_{-\ln (\kappa_{1} Q /q_{1 \, \bot})} \td y_{1} \nonumber \\
& \qquad \times \Theta(Q - q_{1 \, \bot})...\Theta(\kappa^{-1}_{n}  q_{n-1 \, \bot} - q_{n \, \bot}) \nonumber \\
& -\int^{Q}_{0} \frac{\td q_{n \, \bot}}{q_{n \, \bot}} ... \int^{Q}_{0} \frac{\td q_{1 \, \bot}}{q_{1 \, \bot}} \; \int^{\ln (Q /q_{n \, \bot})}_{-\ln (Q /q_{n \, \bot})} \td y_{n} ... \int^{\ln ( Q /q_{1 \, \bot})}_{-\ln (Q /q_{1 \, \bot})} \td y_{1} \nonumber \\
& \qquad \times \Theta(Q - q_{1 \, \bot})...\Theta( q_{n-1 \, \bot} - q_{n \, \bot})\Bigg)\Theta\left(e^{-L} - (1- T(\{p\}_n)) \right), \label{eqn:thrustdif}
\end{align}
where each transverse momentum is defined relative to the thrust axis and $C_n$ is a constant coefficient. 

It is most beneficial to us if we evaluate the logarithmic order of $\delta \Sigma(L)$ by starting more generally and then applying the result to thrust. As previously stated, each $\kappa_{n} = 1 - \mathcal{O}(q^{2}_{n \bot}/2Q^{2})$. We will parametrise this as $\kappa_{n} = 1 - \epsilon q^{2}_{n \bot}/2Q^{2}$ where $\epsilon$ is order unity. Note that when $\epsilon=0$, $\delta \Sigma(L) = 0$. Eq.~\eqref{eqn:thrustdif} is built from repeated sums over elementary integrals of the following type
\begin{align}
\mathcal{I}_{n} = \int^{1}_{a} \frac{\td x_{n}}{x_{n}}...\int^{1}_{x_{2}} \frac{\td x_{1}}{x_{1}} \left[\prod_{i=1}^{n}\ln\left(x_{i}\left(1 - \frac{\epsilon x^{2}_{i}}{2}\right)\right) - \prod_{i=1}^{n}\ln(x_{i}) \right] \Theta(f(a, \{x_{i}\})), \label{eqn:elint}
\end{align}
where $a$ parametrises the observable dependence (for thrust $a\sim 1-T$), $x_{i} \sim q_{i \, \bot}/Q$ and $\Theta(f(a, \{x_{i}\}))$ parametrises any residual more complex observable dependence. Note that both terms in the square bracket are monotonically decreasing as $x_{i} \rightarrow 0$ and that the second is always of smaller magnitude than the first. Thus $\mathcal{I}$ evaluates to having the largest possible magnitude when $\Theta(f(a, \{x_{i}\})) = 1$, as every point in the domain of the integrand adds constructively to the integral. Therefore we will work assuming $\Theta(f(a, \{x_{i}\})) = 1$ in order to place an upper limit on the order of logarithms produced. With this assumption applied, $\mathcal{I}$ is dominated by the term 
\begin{align}
\mathcal{I}_{n} \approx \int^{1}_{a} \frac{\td x_{n}}{x_{n}}...\int^{1}_{x_{2}} \frac{\td x_{1}}{x_{1}} \left[\sum^{n}_{j=1}\ln\left(x_{j}\left(1 - \frac{\epsilon x^{2}_{j}}{2}\right)\right)\prod_{i\neq j}^{n}\ln (x_{i}) - \prod_{i=1}^{n}\ln(x_{i}) \right],
\end{align}
which is in turn proportional to $g_{2n-2}(a,\epsilon) - g_{2n-2}(a,0)$ where
\begin{align}
g_{n}(a,\epsilon) = \int^{1}_{a} \frac{\td x}{x} \ln\left(x\left(1 - \frac{\epsilon x^{2}}{2}\right)\right)\ln (x)^{n}.
\end{align}
For large $n$, $g_{n}$ is difficult to evaluate. However we can navigate this by constructing a generating function for $g_{n}$, 
\begin{align}
GF(a,\epsilon,\nu) = \int^{1}_{a} \td x \; x^{\nu-1} \ln\left(x\left(1 - \frac{\epsilon x^{2}}{2}\right)\right),
\end{align}
so that $g_{n} = (\partial_{\nu})^{n}GF|_{\nu = 0}$ and
\begin{align}
GF(a,\epsilon,\nu) = & \frac{a^{\nu }-1}{\nu ^2}+\frac{\epsilon \left( \, _2F_1\left(1,\frac{\nu }{2}+1;\frac{\nu }{2}+2;\frac{\epsilon }{2}\right) - a^{\nu +2} \; _2F_1\left(1,\frac{\nu }{2}+1;\frac{\nu }{2}+2;\frac{a^2 \epsilon }{2}\right) \right)}{\nu(\nu +2)} \nonumber \\ &+ \frac{\ln (2) a^{\nu }-\ln (2)+\ln (2-\epsilon ) - a^{\nu } \ln \left(2 a-a^3 \epsilon \right)}{\nu}.
\end{align}
The Taylor series in $\nu$ of $GF(a,\epsilon,\nu)$ can be computed. The series is expressible in the form
\begin{align}
GF(a,\epsilon,\nu) - GF(a,0,\nu) = & \sum^{\infty}_{n = 0} \left(\sum^{n}_{i = 0} A_{i,n}\ln(a)^{n-i}\TT{Li}_{2+i}\left(\frac{a\epsilon}{2}\right)  + B_{n} \TT{Li}_{2+n}\left(\frac{\epsilon}{2}\right) \right) \frac{\nu^{n}}{n!},
\end{align}
where $A_{i,n}$ and $B_{n}$ are order unity constants that we do not need. Thus
\begin{align}
\delta \Sigma(L) \lesssim & \sum^{\infty}_{n = 2} \frac{\As^{n}}{(2n-2)!} \left(\sum^{2n -2}_{i = 0} \tilde{A}_{i,n}\ln(1-T)^{2n -2-i}\TT{Li}_{2+i}\left(\frac{(1-T)\epsilon}{2}\right)  + \tilde{B}_{n} \TT{Li}_{2n}\left(\frac{\epsilon}{2}\right) \right), \nonumber \\
\end{align}
where $L = \ln(1-T)$, and $\tilde{A}_{i,n}$ and $\tilde{B}_{n}$ are  order unity constants. Hence for $T \approx 1$, the limit in which we resum, $\delta \Sigma(L) \ll \sum_{n}\frac{\As^{n} C_{n}}{n!}\ln(1-T)^{2n - 2}$ where $C_{n}$ are also order unity coefficients. Also note that the first logarithmic enhancement from our recoil scheme occurs as $\sim \As^{4} L^{2}$. Finally, we note that this argument applies to recoil distributed along any chain of strongly ordered emissions. Therefore recoil from emissions off secondary legs also contributes terms to $\delta \Sigma(L)$ that are much less than $\sum_{n}\frac{\As^{n} C_{n}}{n!}\ln(1-T)^{2n - 2}$.\footnote{In fact, following the epsilon expansion arguments of \cite{CATANI1992419}, recoil from secondary legs will contribute terms less dominant than $\sum_{n}\frac{\As^{n} C_{n}}{n!}\ln(1-T)^{2n - 4}$.}

We have shown that the recoil scheme for the dipole shower presented in Section \ref{sec:Correcting} does not introduce incorrect next-to-leading logarithms into the resummation of thrust in $e^{+}e^{-}$. We did this using a very general approach, leading us to believe that for other exponentiating two-jet dominated observables the same result will also be found. Thus, one would only need to add a running coupling and the shower could be used to compute the NLL resummation of thrust. In summary, we expect our formalism to be capable of leading-colour NLL accuracy in observables that can be resummed at NLL accuracy using the coherent branching approach and will capture much of the full-colour LL contributions.

\subsection{Generating functions for jet multiplicity using global recoil}

We will now use our algorithm with our new recoil scheme (as presented in Section \ref{sec:Correcting}) to compute the integral equation defining the spin-uncorrelated generating function for the multiplicity of subjets in the final state of $e^{+}e^{-}\rightarrow \text{hadrons}$. The generating function was first computed at NLL accuracy (i.e. including all $\alpha^{n}_{s}L^{2n-1}$ terms) in \cite{Catani:1992tm}. The methodology has since seen a variety of applications \cite{Dokshitzer:1992iy,Forshaw:1999iv} (and references therein) and can be found in graduate texts \cite{Ellis:1991qj,Dokshitzer:1991wu}. We will compute the generating function at LL accuracy, though taking care to include all $\alpha^{n}_{s}L^{2n-1}$ logs from recoil.

The generating function is defined by 
\begin{align}
\phi_{\Sigma}(u,Q) = \sum^{\infty}_{n=0} u^{n}P_{\Sigma}(n,Q) = F \sum^{\infty}_{n=0} u^{n+N} \int \td \Pi_{\TT{Born}}\int \td \sigma_{n}(Q).
\end{align}
It can be used for the computation of the moments of the subjet multiplicity distribution for a process $\Sigma$: 
\begin{align}
\< n_{\Sigma} (n_{\Sigma} - 1) .... (n_{\Sigma}-n+1)\> = \left.\tdf{^{n}\phi_{\Sigma}(u,Q)}{u^{n}}\right|_{u=1}.
\end{align}
Here $F$ is some flux factor for the hard process and $P_{\Sigma}(n,Q)$ is the probability of finding $n$ partons/subjets in the final state of a process with centre-of-mass energy (or hard-scale) $Q$. $N$ is the number of partons in the hard process and $\< n_{\Sigma} \>$ is the mean number of subjets in $\Sigma$.

For $e^{+}e^{-}\rightarrow q\bar{q}$, i.e. computing $\phi_{q \bar{q}}(u,Q)$, it is a textbook result that at our accuracy generating functions factorise as $\phi_{q\bar{q}}(u,Q) = \phi_{q}(u,\tau)\phi_{\bar{q}}(u,\tau)$ where $\phi_{a}(u,\tau)$ is the generating function for subjet multiplicity within the jet from a single parton $a$. $\tau= 2 E \sin(\theta/2)$ is the scale of an individual jet and can be thought of as its maximum transverse momentum, $E$ is the energy of each jet and $\theta$ the opening angle of the jet, e.g. $\phi_{q \bar{q}}(u,Q)=\phi_{q}(u,Q)\phi_{\bar{q}}(u,Q)$ as $\theta=\pi$ and $E=Q/2$ \cite{Dokshitzer:1992iy,Ellis:1991qj}.  

We will now construct an integral equation for $\phi_{a}(u,\tau)$. To do so consider also computing $\phi_{e^{+}e^{-}\rightarrow q \bar{q} [g]}(u, q_{\bot \, 1})$, where the next hardest jet (if one occurs) is a gluon jet of scale $q_{\bot \, 1}$. For the computation of $\phi_{e^{+}e^{-}\rightarrow q \bar{q} [g]}(u, q_{\bot \, 1})$, the hard process can be approximated as $\v{H}^{(e^{+}e^{-}\rightarrow q \bar{q} [g])}(q_{1 \, \bot}) =  \v{A}_{0}(q_{1 \, \bot}) + u\v{A}_{1}(q_{1 \, \bot})$. Hence
\begin{align}
\phi_{e^{+}e^{-}\rightarrow q \bar{q} [g]}(u, q_{\bot \, 1}) =  F \sum^{\infty}_{n=0} u^{n} \int \td \Pi_{\TT{Born}} &\left( u^{2} \int \td \sigma^{(\v{A}_{0})}_{n}(q_{1 \, \bot}) + u^{3} \int \td \Pi_{1} \int \td \sigma^{(\v{A}_{1})}_{n}(q_{1 \, \bot}) \right),
\end{align}
where $\td \Pi_{\TT{Born}} \equiv \td \Pi^{(q)}_{\TT{Born}}\td \Pi^{(\bar{q})}_{\TT{Born}}$ is the Born phase-space for the $q\bar{q}$ pair\footnote{The Born phase-space on the momenta of partons after momentum conservation has been taken into account and includes the momentum conserving delta function $\delta^{4}(P_{\bar{q}} + P_{q})$ as well as a delta function fixing the energy.}. We can rewrite this as
\begin{align}
\phi_{e^{+}e^{-}\rightarrow q \bar{q} [g]}(u, q_{\bot \, 1}) =& \phi_{q}(u,q_{\bot \, 1})\phi_{\bar{q}}(u,q_{\bot \, 1}) \Tr(\v{V}_{q_{\bot \, 1},Q} \cdot\v{V}_{q_{\bot \, 1},Q})  + \int \td \Pi_{\TT{Born}} \int \td \Pi_{1} \nonumber \\ 
& \times \int \td R_{1} \, \int \td^{4}P_{g} \, \tdf{\phi_{q}(u,q_{\bot \, 1})}{^{4}P_{q}}\tdf{\phi_{\bar{q}}(u,q_{\bot \, 1})}{^{4}P_{\bar{q}}}\tdf{\phi_{g}(u,q_{\bot \, 1})}{^{4}P_{g}} \nonumber \\  
& \times \Tr\left(\v{V}_{q_{\bot \, 1},Q}\cdot \v{V}_{q_{\bot \, 1},Q} \<\v{D}^{\dagger}_{1}\cdot \v{D}_{1} \>_{1}\right) \, \delta^{4}(P_{g} - q_{1}), \label{eqn:function_middle_step}
\end{align}
where we have employed azimuthally averaged result of Appendix \ref{sec:CBderivation} since the equation is independent of the azimuth of the gluon. We have also spin averaged at this step. We also note that Eq.~\eqref{eqn:function_middle_step} is equal to $\phi_{q \bar{q}}(u,Q)$ by necessity, i.e. $\phi_{q \bar{q}}(u,Q) = \phi_{e^{+}e^{-}\rightarrow q \bar{q} [g]}(u, q_{\bot \, 1})$ as within the strong ordering approximation the next hardest jet of an $e^{+}e^{-}\rightarrow q \bar{q}$ process must be a gluon jet. After a little work,
\begin{align}
&\phi_{q \bar{q}}(u, Q) = \tfrac{1}{2}\phi_{q}(u,q_{\bot \, 1})\Delta_{q}(q_{\bot \, 1},Q) \, \phi_{\bar{q}}(u,q_{\bot \, 1}) \Delta_{\bar{q}}(q_{\bot \, 1},Q) \nonumber \\ 
& \; + \phi_{\bar{q}}(u,q_{\bot \, 1}) \Delta_{\bar{q}}(q_{\bot \, 1},Q) \frac{\As}{2 \pi} \int^{Q}_{q_{\bot \, 1}} \frac{\td q_{\bot} }{q_{\bot}}  \Delta_{q}(q_{\bot},Q)  \int^{1-\frac{q_{\bot}}{2Q}}_{\frac{q_{\bot}}{2Q}} \td z \, \mathcal{P}_{qq}(z) \, \tilde{\phi}_{q}(u,q_{\bot})\tilde{\phi}_{g}(u,q_{\bot}) \nonumber \\
& \; + (q \leftrightarrow \bar{q}), \label{eqn:generating_function_V_B}
\end{align}
where
\begin{align}
\tilde{\phi}_{q}(u,q_{\bot}) &= \int \td \Pi^{(q)}_{\TT{Born}} \, \td^{4}P_{q} \, \tdf{\phi_{q}(u,q_{\bot})}{^{4} P_{q}} \, \mathfrak{R}^{\TT{primary}}_{q_{1}} \approx \phi_{q}(u,zq_{\bot}), \nonumber \\
\tilde{\phi}_{\bar{q}}(u,q_{\bot}) &= \int \td \Pi^{(\bar{q})}_{\TT{Born}} \, \td^{4}P_{\bar{q}} \, \tdf{\phi_{q}(u,q_{\bot})}{^{4} P_{\bar{q}}} \, \mathfrak{R}^{\TT{secondary}}_{q_{1}} \approx \phi_{\bar{q}}(u,q_{\bot}), \nonumber \\
\tilde{\phi}_{g}(u,q_{\bot}) &= \int \frac{\td \phi_{1}}{2 \pi} \, \td^{4}P_{g} \, \tdf{\phi_{g}(u,q_{\bot \, 1})}{^{4} P_{g}} \, \delta^{4}(P_{g} - q_{1}) \approx \phi_{q}(u,(1-z)q_{\bot}),
\end{align}
and where the recoil functions, using the same definitions as Section \ref{sec:Correcting}, are given by
$$\mathfrak{R}^{\TT{primary}}_{q_{1}} = \delta^{4}_{\mathcal{J}}\left(\tilde{P}_{q_{1}} - z \kappa_{q} \, \Lambda(q,\bar{q}) p_{q}\right), $$
$$\mathfrak{R}^{\TT{secondary}}_{q_{1}} = \delta^{4}_{\mathcal{J}}\left(\kappa_{q_{1}} \, \Lambda(q,\bar{q}) P_{j_{n}} - \tilde{P}_{j_{n}}\right),$$
i.e. each $\tilde{\phi}$ is simply related to each $\phi$ by momentum conservation. At our accuracy, momentum conservation simply maps $E_{q} \rightarrow z_{1} E_{q}$ and $E_{g} = (1 - z_{1})E_{q}$ since $\kappa_{q_{1}}$ and the Lorentz boost are unity at our desired accuracy (noting the argument for neglecting the changes in phase-space due to our recoil scheme given in the previous subsection also holds for this resummation as the measurement function is unity and we are resumming logs up to $\As^n L^{2n-1}$ accuracy). The limits on the $z$ integrals capture angular ordering at NLL accuracy whilst still using a $k_\bot$ ordering variable. $\Delta_{c}(a,b)$ is a Sudakov factor
\begin{align}
\Delta_{c}(a,b) = \exp\left( - \frac{\As}{2\pi}\int^{b}_{a} \frac{\td k^{(c \vec{n})}_{\bot}}{k^{(c \vec{n})}_{\bot}} \int^{1-\frac{k^{(c \vec{n})}_{\bot}}{2Q}}_{\frac{k^{(c \vec{n})}_{\bot}}{2Q}} \td z \, \mathcal{P}_{cc}(z) \right).
\end{align}
We can factorise Eq.~\eqref{eqn:generating_function_V_B} as
\begin{align}
\phi_{q \bar{q}}(u, Q) =& \bigg(\phi_{q}(u,q_{\bot \, 1})\Delta_{q}(q_{\bot \, 1},Q)  \nonumber \\ 
& \; \left. + \frac{\As}{2 \pi} \int^{Q}_{q_{\bot \, 1}} \frac{\td q_{\bot} }{q_{\bot}}  \Delta_{q}(q_{\bot},Q) \int^{1-\frac{q_{\bot}}{2Q}}_{\frac{q_{\bot}}{2Q}} \td z \, \mathcal{P}_{qq}(z) \,  \tilde{\phi}_{q}(u,q_{\bot})\tilde{\phi}_{g}(u,q_{\bot}) \right) \nonumber \\
& \; \times (q \leftrightarrow \bar{q}) + \mathcal{O}(\alpha^{2}_{s}). \label{eqn:factgenerating_function_V_B}
\end{align}
keeping only terms first order in $\As$\footnote{The $\mathcal{O}(\alpha^{2}_{s})$ terms can be computed by instead starting with $\v{H}^{(e^{+}e^{-}\rightarrow q \bar{q} [g] [g])}(q_{2 \, \bot}) = \v{A}_{0}(q_{2 \, \bot}) + u\v{A}_{1}(q_{2 \, \bot}) + u^{2}\v{A}_{2}(q_{2 \, \bot})$ and proceeding as above.}. From this, we can identify
\begin{align}
\phi_{q}(u, Q) =& \phi_{q}(u,q_{\bot \, 1})\Delta_{q}(q_{\bot \, 1},Q)  \nonumber \\ 
& \;  + \frac{\As}{2 \pi} \int^{Q}_{q_{\bot \, 1}} \frac{\td q_{\bot} }{q_{\bot}}  \Delta_{q}(q_{\bot},Q) \int^{1-\frac{q_{\bot}}{2Q}}_{\frac{q_{\bot}}{2Q}} \td z \, \mathcal{P}_{qq}(z) \,  \tilde{\phi}_{q}(u,q_{\bot})\tilde{\phi}_{g}(u,q_{\bot}).
\end{align}
This expression is correct at LL accuracy with complete colour and only requires the coupling to run as $\As(z(1-z)q_{\bot})$ in order to capture the full NLL ($\alpha^{n}_{s}L^{2n-1}$) result. We also can note that the correct NLL resummation might not have been achieved using the local dipole prescription presented in Appendix \ref{sec:Spectator}. This is because the recoil could introduce a correction in the $n>3$ jet limit of the form $\phi_{\bar{q}}(u,q_{\bot \, 1}) \rightsquigarrow \phi_{\bar{q}}(u,|\v{q}_{\bot \, 1} - \v{q}_{\bot \, 2}|)$ (the wavy arrow implying that it will approximately go to). This correction prevents both the usage of naive azimuthal averaging and the factorisation $\phi_{q\bar{q}}\equiv \phi_{q}\phi_{\bar{q}}$ (which naturally emerged between Eq.~\eqref{eqn:generating_function_V_B} and Eq.~\eqref{eqn:factgenerating_function_V_B}), though it is possible that these features could re-emerge once the phase space of each jet has been inclusively integrated over. Due to the other known NLL limitations of this recoil scheme, we did not think it worthwhile further proceeding to evaluate the order of these errors but rather conjecture that NLL errors will also be likely here.

\bibliographystyle{JHEP}
\bibliography{Shower_derivations_paper}

\end{document}